\documentclass[a4paper,12pt]{report}
\usepackage{amsmath,graphicx,epic}
\newlength{\linavst}
\newcommand{\rf}[4]{{\em {#1}} {\bf #2}, #3 (#4)}
\newcommand{\ZP}[3]{\rf{Zeitschr.\ Physik}{#1}{#2}{#3}} 
\newcommand{\PR}[3]{\rf{Phys.\ Rev.}{#1}{#2}{#3}}
\newcommand{\PRL}[3]{\rf{Phys.\ Rev.\ Lett.}{#1}{#2}{#3}}
\newcommand{\alfa}{\mbox{$\alpha$}}

\newcommand{\el}{\ensuremath{e^-}}
\newcommand{\pos}{\mbox{$e^+$}}
\newcommand{\pil}{\rightarrow}
\newcommand{\mi}{\,|\,}

\newcommand{\bra}{\langle}
\newcommand{\ket}{\rangle}
\newcommand{\braket}[1]{\langle#1\rangle}

\newcommand{\One}{1\kern-4.5pt1}

\newcommand{\gam}{\gamma}

\renewcommand{\gam}{\ensuremath{\gamma}}
\begin{document}
\title{Matter and forces in quantum field theory \\
  An attempt at a philosophical elucidation}
\author{Jon-Ivar Skullerud}
\date{May 1991}
\maketitle

\pagestyle{empty}
\section*{Preface}

This thesis arose from one primary motivation. I felt there was little
sense in delving straight into theoretical calculations within e.g.
particle physics without any overarching idea of what the theory was
about and what would be the value of such work --- I had little interest
in doing something I might myself consider next to worthless for my
thesis. But to judge what it could be worth working on it was necessary
to take a step back and evaluate the theory as a whole. This feeling was
further strengthened by the time I spent at CERN in the summer of 1989.
It was an interesting stay and an exciting environment, but I found
myself asking whether much of the activity might not be driven too much
by prestige, and whether perhaps there was not so much scientific insight to be
gained from the large accelerator experiments compared to the
investments. It could be useful to conduct a general evaluation of where
we stood and what was the aim of the experiments.

In addition, I have always (for as long as I have been interested in
physics) been interested in the connections between philosophy and
physics. After also having studied some philosophy in addition to
physics, I believed it could be good to attempt to do some proper work
in this area. It has been good for me --- I have personally got a lot
out of my work with this thesis. I have not managed to do as much as I
had hoped --- there is for example a half-finished section on particle
species which did not make it in because I ran out of time, and there
are several other questions I would have liked to discuss given more
time. I have however got a greater interest in and understanding of both
physics and philosophy. In particular got an idea of which areas of
research within fundamental physics can be of interest --- which was
part of the aim of doing this work in the first place.

I would like to thank my supervisors Audun {\O}fsti and K{\aa}re
Olaussen, who have read through the manuscript and given me good advice
along the way. Thank you also to everyone else who has provided opinions
and encouragement. Finally, thank you to my father, who has given me
access to his computers for writing the thesis, and also given me much
support.

\par \medskip
\begin{center}
Trondheim, 6 May 1991 \par 
jon ivar skullerud
\end{center}
\bigskip

\section*{Preface to the revised edition}

In this revised edition I have made some minor changes to the text in
several places, and included more references. I have also corrected some
typographical errors.

\par \medskip
\begin{center}
Trondheim, 13 July 1991 \\
j.i.s.
\end{center}

\section*{Preface to the English translation}

This thesis should be read as the opinions of a budding physicist in 1991.  I have resisted any temptation to add anything save a small number of footnotes, and have also kept the style, including the idiosyncratic use of punctuation and quote marks, mostly intact.  My opinions (and my writing style) have obviously evolved in the intervening years, but that will have to be for another day.

This translation has been a very long time in the making.  Already in the first few years after it was written several people were asking me if I would translate it into English so that they could read it.  I thought that would be a good idea, but it was never a priority among so many other things to do.  Anyway, what would I do with it?  The thought arose again after I put the original thesis on my web page in 1995, but again time and lack of any urgency meant nothing happened.  Eventually, many years later, I started regularly reading the History and Philosophy of Physics eprints on arXiv [hist-ph], and concluded there would indeed be a potential repository where the thesis could reach a wider audience, justifying the effort that would go into a translation.  Thus started a slow process, in my spare time, not being sure if I could justify this being part of my work or not.

Then, when I went on a one-semester sabbatical to Florence in autumn 2019 I decided this was when I was going to complete the translation, and I stuck to that.  I wish to express my great appreciation to the Galileo Galilei Institute for Theoretical Physics for their hospitality during this time, and especially to the organisers of the mini-workshop ``Beyond Standard Model: Historical-Critical Perspectives'', which was the highlight of my stay at the GGI and gave me an additional spur to finish this work.

I wish to thank M\'aire O'Dwyer for her careful proofreading of the
English translation.  The Feynman diagrams were drawn using
the \emph{FeynGame} package.\footnote{R.~V.~Harlander, S.~Y.~Klein and M.~Lipp,
Comput. Phys. Commun. \textbf{256}, 107465 (2020)
[arXiv:2003.00896].}
\par \medskip
\begin{center}
Firenze, 25 January 2020 / Dublin, 28 November 2020\\
j.i.s.
\end{center}
\pagestyle{plain}

\setcounter{page}{1}
  \tableofcontents
  \chapter{Introduction}

Physics and philosophy are not two independent disciplines.  They
share a common origin; both contribute significantly to our worldview,
and in my opinion there remain several levels of mutual dependence
between them --- in particular in the areas where they impinge on each
other: natural philosophy, fundamental physics and to some extent 
epistemology.\footnote{Physics probably relies more on epistemology
  than vice-versa, but the dependence is not just one way: Kant's
  epistemology, for example, contained among much else a belief that
  Newton's physics (or at least its main features) was the final
  physical theory.}  Not all these relations of mutual dependence can
or should be made explicit, but an awareness of their existence is
necessary for both disciplines.  A philosophical interpretation of
fundamental physical theories is a necessary part of such an
awareness.

Up to the renaissance, religion and philosophy played a far greater
r\^ole in the worldview of Western Europeans than the sciences --- to
the extent that it is possible to talk about a distinction between
religion, philosophy and science.\footnote{Religion probably still
  dominates the worldview of most people on a worldwide basis, but
  this is much less the case in Western Europe.}  The sciences,
including physics, were subordinate to philosophy and religion: the
framework within which physics worked was explicitly drawn up by
philosophy.  The most important characteristic of a good scientific
theory was that it was in accord with the accepted philosophy and
religion of the day.  This has changed.  As the sciences have become
more independent of philosophy, they have played a more independent
and dominant r\^ole in shaping the common worldview.  It has almost
got to the point that the most important criterion for whether a
philosophy is good (acceptable to the public) is whether it accords
with the accepted science of the day.  This can be seen particularly
clearly in the r\^ole classical (Newtonian) physics played in the 19th
century.  The view  of everything as a result of deterministic forces
pervaded not only the understanding of nature, but also of society and
human life.  Newton's physics was transformed into a philosophical
axiom which very many (most people?) took as self-evident.

This is a specific example of a more general phenomenon: any worldview
has a physical aspect, which consists of an understanding of the
ruling paradigm of fundamental physics.  This implies that this
paradigm is tacitly assumed, taken for granted and taken to be
unconditionally true among the general public.  This will also
form part of the basis for our experience of the world, and will thus
also make its imprint on philosophy.  Moreover, if philosophy has as
one of its tasks to comment on our worldview and investigate the
conditions for our experience of the world, it will also have to
comment on physics.  We may say that even though it is accepted that
ontology in the classical sense (as a study of how `being', completely
independently of us humans, must be) is not possible, \emph{an}
ontology (understanding of how the world fundamentally is) is still
necessary, and may consist of an interpretation of fundamental
theories of physics.

A new paradigm of fundamental physics will rarely if ever be
completely understood immediately.  It breaks with several
previously tacit or presumed self-evident assumptions about the
physical world, and will therefore tend to be considered
`incomprehensible'.  Such a situation is, however, untenable in the
long run.  As people become used to using the new conceptual
framework, they will tend to become familiar with and build up a
certain understanding of what it is about.  This understanding will
then (slowly) spread from professional circles to the general public.

There are few physical theories that have been subject to as much
philosophical debate as quantum mechanics.  This debate has however
often started from the premise that quantum mechanics is a
curiosity, far from our ordinary understanding of reality and
remaining for ever the sole domain of experts.  My starting point is
different, namely that quantum mechanics, and not least quantum field
theory, tells us about fundamental features of physical reality, and
that essential features of quantum mechanics will eventually form part
of the common worldview.  There is therefore a need for a
philosophical interpretation which treats the theory as it stands as
fully comprehensible --- a \emph{quantum ontology}, which considers
how quantum mechanics and quantum field theory will fit into a general
worldview.  It will take some time before we reach a `fully mature'
understanding (when it comes to classical mechanics, this may be
represented by Kant, 100 years after Newton).  However, the
understanding does not mature by itself, but is a product of the
interpretations we make.  It is such a sketch of an understanding or
interpretation I wish to contribute to, within the limits allowed by
this dissertation.

Another aspect of the relation between physics, philosophy and common
sense is that none of these, including physics, come without
prerequisites.  Just as philosophy fools itself if it considers
itself independent of any historical context, physics will fool itself
if it considers itself independent of any metaphysical
assumptions.  All our statements about
the world, both in daily life and in physics, presuppose that
something is taken for granted.  These preconditions can be of various
kinds, from conditions for the possibility of talking about an
objective reality (or coming to agreement about matters regarding the
world), via an understanding of what characterises a particular
science, to the complex web of theories that are assumed when carrying
out and interpreting an experiment.  Acknowledging these preconditions
is crucial to the self-understanding of physics, and may also
contribute to guidelines for what kind of further research may be
fertile.  An understanding of and explanation of these preconditions
and their r\^ole is a philosophical problem.

I believe quantum mechanics gives a good illustration of the mutual
interdependence between physics and philosophy.  That philosophy is
dependent on physics is made evident by several ideas that had been
assumed to be necessary features of the world, such as deterministic
causality and the possibility of precisely determining all properties
of things, which are rejected in quantum mechanics.  What had been
considered \emph{a priori} truths turn out only to be valid within
classical physics.  Quantum mechanics has therefore forced us to
reevaluate elements of natural philosophy (and possibly also other
parts of philosophy --- there has been a good deal of work done on
revising logic in the light of quantum mechanics).  The same point is
also illustrated by the fact that much of the opposition to quantum
mechanics was of a philosophical nature, and was voiced by people who
wished to defend the classical worldview.

That physics is dependent on philosophy is of course also illustrated
by all the philosophical debates around quantum mechanics.  Almost
from the outset, the `fathers' of quantum mechanics, like Bohr,
Heisenberg and Schr{\"o}dinger, pointed out the need for a
philosophical interpretation of quantum mechanics.  The connection
with philosophy appears in particular in the r\^ole played by the
subject in the theory or its interpretation.  As an example, I will
look at the `particle--wave-dualism', as it appears in the double-slit
experiment and the view of it within the Copenhagen interpretation.

\begin{figure}[bht]
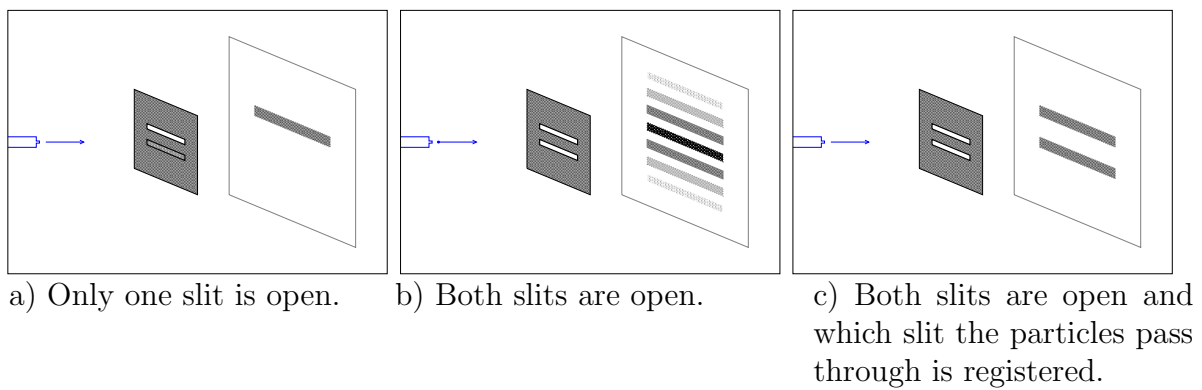

\begin{center}
\includegraphics*[width=0.32\textwidth]{oneslit.eps}
\includegraphics*[width=0.32\textwidth]{twoslit.eps}
\includegraphics*[width=0.32\textwidth]{twoslit-track.eps}\\
\begin{tabular*}{16cm}{p{5cm}@{\extracolsep{\fill}}p{5cm}p{5cm}}
a) Only one slit is open. & b) Both slits are open.
  & c) Both slits are open and which slit the
  particles pass through is registered.
\end{tabular*}
\end{center}
\caption{The double-slit experiment.}
\label{fig:twoslit}
\end{figure}

The experiment is illustrated in figure~\ref{fig:twoslit}.  Single
electrons (or other particles) from a source are sent through two
slits to a screen which is marked when hit by particles.

If only the lower slit is open, a pattern as in fig.~\ref{fig:twoslit}
a) will eventually appear --- the particles have hit with some spread
around one point at the continuation of the line between the source
and the slit --- like classical particles would have done.  If both
slits are open, something strange happens: the pattern formed on the
screen is as in fig.~\ref{fig:twoslit} b) --- a pattern of stripes.
This is what one would have expected in classical physics if a wave
had passed through the slits.  But a wave is something extended, which
should pass through both slits at the same time.  If we now look at
how the electrons pass through the slits (by placing detectors there),
we find that they only pass through one slit at a time --- but at the
same time we obtain a pattern on the screen as in fig.~\ref{fig:twoslit}
c) --- as if it was classical particles passing through the slits.
Depending on how the experiment has been constructed and what we
observe, the electrons behave as particles or as waves, if we insist
on using classical language.

The Copenhagen interpretation (in all its variants) says that the
description of the electron as wave and as particle are equally
valid.\footnote{--- if one is to use such classical concepts at all.
I believe this is unnecessary, and that attempts to describe quantum
mechanical systems in terms of classical concepts (which are not
necessarily better understood than the quantum mechanical ones) hinder
the understanding of quantum mechanics on its own terms.  In
particular, the wave concept is widely misused.  In the early years of
quantum mechanics, on the other hand, it was necessary to make use of
`known' concepts.  Today we are better off formulating both quantum
mechanics itself and the Copenhagen interpretation in purely quantum
mechanical terms.}  Quantum mechanics is free of contradictions
because the two descriptions are \emph{complementary} rather than
contradictory: they cannot be applied simultaneously where they would
have given conflicting results.  What decides which description should
be used is the experiment or what is observed.  The quantum
mechanical phenomenon `in itself' is not fully determined: only when
it is observed (or placed in the context of an experiment) is it
something definite.  Here it is important that the experimental
apparatus according to the Copenhagen interpretation must be described
classically --- i.e., there are no indeterminacies there.  Our classical
world of things is in other words characterised by definiteness and
ordinary causality, while the `quantum world' is characterised by
indeterminacy, complementary descriptions and non-deterministic
behaviour, and depends on our actions to become something definite.

The primary question raised by the Copenhagen interpretation is about
the conditions for observation of physical systems --- which is a
philosophical question.  It turns out that the character of the
physical system depends on both \emph{that} and \emph{how} it is
observed.  More generally, the question arises of what an observation
is as such; this is at the bottom of all the debates about
measurement in quantum mechanics (which I will not go into any
further).  The `naive' view of observation in classical physics (the
observation has no bearing on what is observed) can no longer be
sustained; instead physics becomes dependent on a well-developed
epistemology.

Another reason why I feel a study like the one I
will conduct here is necessary is that very little has so far been written about
what might be called the implicit ontology of quantum field theory ---
the worldview implied by accepting quantum field theory as a
fundamental theory of physics.  What has been written has largely been
limited to popularisations and discussions about specific problems.  I
know of only two wide-ranging expositions: Werner Heisenberg's `Physics
and Philosophy'~\cite{Heisenberg:Phil} and Fritjof Capra's `Tao of
Physics'~\cite{Capra}.  Heisenberg's book gives a brilliant,
probing analysis of the philosophical issues
arising out of quantum mechanics (including quantum field theory),
informed by his variant of the Copenhagen interpretation.  Capra's
book is more directly aimed at advocating a particular interpretation
or presenting one particular view, which is interesting, but would in
any case benefit from being confronted with other possible views.  Some
articles discussing philosophical aspects of quantum field theory
(including its implicit ontology) are collected in \cite{found,PSA}.

The dissertation is aimed at both philosophers with an interest in,
but not a detailed knowledge of, physics, and at physicists with an
interest in philosophy.  It therefore has a different style to what
you might find in e.g. a journal of philosophy of science, where the
readers may be expected to have a good knowledge of both.  It is
however not a `popular' exposition, and demands a fair amount of the
reader.  I have tried to avoid excessive use of physics and philosophy
jargon, but have not been able to get rid of all such `weeds' --- I
hope this is not too annoying.  It should in any case be possible to
follow the main lines of the arguments; in particular, I will
emphasise that this does not depend on having understood the
mathematical details in chapter~\ref{chap:qft}.  However, I do assume some
elementary knowledge of vector arithmetic, calculus (knowing the
meaning of differentiation and integration) and complex numbers.  The required
knowledge of physics is strictly speaking limited (I believe) to not
much more than knowing what energy and momentum refer to, but it is an
advantage to also have a reasonably clear idea of classical fields
(e.g., electric and magnetic fields).  I have also used particle symbols
in a number of places.  A superficial knowledge of the history of
philosophy should be sufficient to follow the philosophical arguments,
if you can look past the jargon.

I have chosen to divide this dissertation into 4 chapters.  Chapter 2
has a dual function.  Firstly, it presents quantum field theory from
both a historical and a systematic point of view.  Secondly, the
historical presentation may serve to give an insight into the process
of active physical research, which is of interest for discussions of
physical research in general.

In chap.~\ref{chap:philos} I wish to sketch a framework of ideas in
epistemology and natural philosophy which may form the basis of a
philosophical discussion of fundamental theories of physics.  An
important point is that this framework should accord with what I find
to be scientific practice: the physicist should recognise himself or
herself in the description of physics.  There are of course many
views, also among physicists, of what physics is, but I assume (or
hope) that my description of aims, methods and preconditions can
be broadly accepted.

Chapter~\ref{chap:critique} contains the most important part of the
dissertation.  The chapter is divided in three.  First there are some
`preliminaries', where I primarily assert what quantum field theory
is not, and say something about why this is so: why classical ideas of
matter would eventually become untenable.  Subsequently, I outline
different possible interpretations of quantum field theory, centred
around four `paradigms'.  The conceptual framework of the theory is
thus also illuminated from different angles.  Finally, I address
some critical issues within the theory, including the question of how
our world may be reconstructed from quantum field theory.

The parameters of the dissertation have not allowed me to discuss
thoroughly all the philosophical issues related to quantum field
theory.  I have concentrated on `ontological' issues --- trying to say
something about what the theory says about the external world ---
rather than issues relating to philosophy of science, which look at
the theory as a human activity.  The questions of renormalisation and
of the validity of the approximations used in the theory are mostly
just mentioned in passing, but this should not be taken as an
indication that they are uninteresting.  I had hoped to include a
section on particle mixing --- how the distinction between different
particle species becomes blurred in quantum field theory --- but time
did not allow this.  Some of the issues I would have addressed are
discussed by Heisenberg \cite{Heis-Einst}, Redhead \cite{Redhead} and
Weingard \cite{Weingard}, as well as in several articles
in \cite{found}.  The Higgs mechanism would also have deserved a
proper analysis, but is not discussed here.

I have avoided many of the typical topics of `quantum philosophy',
such as causality vs determinism, quantum logic, the EPR paradox and
Bell's theorem, the measurement problem etc.  These questions have
been thoroughly debated by others, and I do not expect to contribute
anything significant here.  I will not be able to completely avoid
these issues, and will to some extent express a particular view
without having justified this any further, or merely address the
issues implicitly.  This does not mean that I ignore that these are
real issues (although I consider some of the discussions to be about
non-issues); only that I do not consider a detailed
discussion necessary or appropriate in this context.

In chap.~\ref{chap:future} I try to look into the future, and in
particular consider the positive heuristics of the theory: which
possibilities it contains for future research.  I have not had the
time to make a detailed analysis of the situation (and in any case,
such predictions are notoriously unreliable), but hope it can help
form a basis for a more proper evaluation of potentially fertile
areas of research: in which directions may we expect progress.  In
this chapter I will also take a closer look at physics as a human
activity.

The parameters of the dissertation do not allow the extensive
literature survey that would have been desirable --- and the selection
of literature is perhaps also somewhat `random' and determined largely
by what was to hand.  Hence there is a clear risk that I have missed
significant points and possible previous debates on the topic, and I
may have repeated things that have been said before.  Some articles
which explicitly address philosophical aspects of quantum field
theory \cite{PSA,Teller:1985} are included in the bibliography, although I
had not read them myself at the time of writing.

I hope that this can be a small contribution to a debate I consider
necessary. 

  \chapter{What is quantum field theory?}
\label{chap:qft}

\section{Introduction}
\label{sec:qft-intro}

This chapter is divided into two parts: a historical and a systematic
part.  This is done because I have wanted to present the theory in a
(more or less) logical order, as it appears today, to provide the best
possible starting point for the subsequent discussion.  Since the logical and
the historical order can differ considerably, and since the concepts
have often evolved quite a lot, I found it most appropriate to avoid
confusion by not mixing the two up.  The disadvantage is that the
history is somewhat abbreviated, but I hope this is balanced by
greater clarity later on.

I will also repeat myself to some extent, and the distinction between
history and systematics will not be strictly observed.  In
particular, non-abelian gauge theories will be presented in their
entirety in the historical section.  I found this most natural both
because the logical and the historical developments (at least as I see
them) follow each other closely in this case, and because a systematic
presentation of the theory cannot add much to what is required to
present the historical development --- unless you want to delve into
abstract group theory.

The main point of this chapter is to provide a background for the
reflections I will make in chapters \ref{chap:critique} and \ref{chap:future}.
The main emphasis is therefore on the conceptual aspects of the
theory, and I will attempt as far as possible to avoid using the
mathematical formalism.  Space does not allow me to go into detail,
neither about theoretical contents, nor about history or debates among
physicists regarding the status of the theories.  For anyone who wants
to delve further into these matters, I have included some books
(including textbooks at different levels and with different
approaches) in the references.  I should in particular mention the
book by Abraham Pais \cite{Pais} on the history of particle physics,
which in addition to being written with great insight and giving a
near-encyclopaedic overview, contains almost everything that would be
required of additional references in this area.  I may also mention
Max Jammer's \emph{The Philosophy of Quantum
  Mechanics}~\cite{Jamm:Phil}, which similarly gives a
near-encyclopaedic overview of the debates within nonrelativistic
quantum mechanics.

Attempting to present the physical contents of the theory without
becoming too mathematical is a difficult balance --- especially if (at
this stage) I am to try to take a `neutral' stance in regard to the
various interpretations.  Any (not purely mathematical) concept I
employ will necessarily have its connotations, implying a certain
metaphysical `bias'.  I could have presented many basic concepts in
the theory without using any mathematics, but this would have tied me
to one specific interpretation.  This issue also arises where there
are several (equivalent) mathematical formulations of the same theory:
different formulations make it easier to see different aspects of the
theory.  As long as I make use of physical concepts, and do not
explicitly and consistently explain the equivalence between different
formulations, the presentation will thus have a bias.  The most
obvious expression of this is probably my having put Feynman's path
integral formalism on its own, right at the end of the systematic
presentation.\footnote{Since writing this dissertation, I have worked
  for more than 20 years as a practising particle physicist using path
  integral methods, and these now form the cornerstone of my view of
  the theory.  This is likely to have modified my views and would
  deserve a followup or postscript.}

It would have been tempting to start the historical presentation with
the debates between Newton and Huygens on the nature of light
(particles or waves), since this is the question quantum field theory
(quantum electrodynamics) solves quite brilliantly by removing the
dichotomy.  Feynman writes that Newton was right: light is particles;
and his formulation and visualisation of the theory does on the face
of it lend most support to this point of view.  However, if you look
behind the diagrams, you see that the particles must be some strange
entities which are quite far removed from Newton's ideas and
principles, and Feynman also points this out.\footnote{For example,
  they may go in all possible directions without this being caused directly
  by any external force.  In fact, they take all possible paths,
  including forwards and backwards in time --- at the same time, if
  one may say so.}  I will discuss this in more depth in
section~\ref{sec:Feynman}.

I could also have started with Faraday's and Maxwell's investigations
of electromagnetism, which gave us a completely new understanding of
what light is, and in retrospect must be said to have started the
process leading to the demise of newtonian physics.  This is also the
origin of the concept of fields --- it could have been interesting to
follow the evolution of this concept, first in classical mechanics
from Faraday's description of field lines, via Maxwell's mechanical
model of the aether, to Einstein's special theory of relativity, and
then how it is taken over and developed further in quantum mechanics.
I will however not go any further into this part of the
prehistory.\footnote{I will, however, look at it in a quasi-historical
  perspective, with the benefit of hindsight, in
  section~\ref{sec:Newton}.}  The reader who wants to acquaint herself
further with this topic may read the histories by Whittaker or Meyer
\cite{Whittaker,Meyer}. 

\section{Historical overview}

\subsection{The quantum mechanical revolution (1900--27)}

Quantum field theory may be said to have started with Einstein's
remark in his 1917 article on emission and absorption of radiation:
`The properties of elementary processes make the task of formulating a
genuinely quantised theory of radiation appear
inevitable.'\footnote{A.Einstein:
  \rf{Phys.\ Zeitschr.}{18}{121}{1917}.} In this article he had, using
Planck's law of radiation, Bohr's quantum postulates and thermodynamic
considerations, computed coefficients for the emission and absorption
of radiation by matter --- the so-called Einstein's A- and
B-coefficients.  The theory he had in mind, which could explain these
coefficients from fundamental principles of quantum mechanics (though
not in the way Einstein had in mind), would be formulated by Dirac in
1927: quantum electrodynamics.

But now I have jumped straight into a story which started with
Planck's 1900 article on thermal radiation.\footnote{M.Planck:
\rf{Verh.\ Deutsche Phys.\ Ges.}{2}{237}{1900}; \rf{Ann.\
Physik}{4}{553}{1901}.}  Planck did not understand what he did with his
quantum postulate: originally, his radiation law was a pure
interpolation with no theoretical basis, and the quantum postulate was
introduced purely as a justification for this law.  Initially, it was
thus only applicable to a single, relatively peripheral problem in
physics (black-body radiation, or thermal radiation from a
cavity)\footnote{That it is a peripheral issue in physics does not
  imply that it is of little practical importance: the whole issue of
  the greenhouse effect and global warming is closely connected with
  this.}  --- and he tried long, unsuccessfully, to incorporate it
into classical physics.  Max Planck, who is considered to be the
founder of quantum physics, and after whom its fundamental constant is
named, never accepted quantum mechanics.

It was Einstein who first understood that Planck's work constituted a
revolution in physics.  He soon understood that Planck's \emph{ad hoc}
assumption that the atoms in the cavity wall only emitted and absorbed
radiation in quanta, implied that the radiation (or light) must be
considered as consisting of particles (photons, symbolised by \gam),
each with an energy $h\nu$, where $\nu$ is the frequency of the
radiation.  He used this to carry out detailed investigations of the
interaction between radiation and matter, in studies of the
photoelectric effect and black-body radiation.\footnote{A.Einstein:
  \rf{Ann.\ Physik}{17}{132}{1905}.} Through this work, the famous (or
infamous) wave-particle duality of light was demonstrated for the
first time.

It was also Einstein who showed that quantum physics had applications
beyond issues related to radiation.  In an article from
1907\footnote{A.Einstein: \rf{Ann.\ Physik}{22}{180}{1907}.} he showed
that the energy quantum hypothesis, applied to molecular vibrations,
could explain deviations from the predictions of classical physics
regarding the heat capacity of solids, something which had previously
been a mystery.  As a result, the quantum postulate could be
considered physically relevant, not just as an anomaly, but as a more
general theory.

The last area where quantum theory had an early impact, was the study
of the structure of atoms.  It had been known for quite a while that
different elements have their characteristic spectra: only radiation
with certain frequencies is emitted or absorbed, and certain rules for
these frequencies had been worked out for simple atoms.  In addition,
Rutherford made discoveries in 1911 which could be explained by
thinking of the atom as a mini-`solar system' where the electrons
orbit the nucleus as planets around the sun.\footnote{E.Rutherford:
  \rf{Phil.\ Mag.}{12}{143}{1911}.} Both these observations posed
problems for classical physics.  Firstly, the emission spectrum would
be expected to be continuous; and secondly, an electron orbiting a
nucleus would emit radiation, and hence lose energy and spiral towards
the nucleus.  Bohr attempted to solve these problems with his 1913
model of the atom,\footnote{N.Bohr: \rf{Phil.\ Mag.}{26}{1}{1913}.}
where he postulated that the electrons could only move in certain
orbits with discrete energy values (\emph{stationary states}), where
they did not radiate anything.  Transitions from one state with energy
$E_1$ to another one with a lower energy $E_2$ occurred by the atom
emitting a photon with frequency $\nu=(E_1-E_2)/h$.  Conversely, the
atom could be brought into a higher energy state by absorption of a
photon.  The energy levels were chosen such that the radiation would
satisfy the known rules.  The discrete energy level postulate was
directly confirmed in an experiment by Franck and Hertz in
1914.\footnote{J.Franck and G.Hertz: \rf{Verh.\ Deutsche
Phys.\ Ges.}{16}{457}{1914}.}

Based on this model, a series of further investigations of atomic
structure was carried out, in particular by Bohr, Sommerfeld and their
collaborators.  The result was a set of `quantum postulates' or
quantisation rules, which explained how to find the correct quantum
description of a problem after first having formulated it
classically.  A great help was that the quantum description always
approached the classical one in the limit of large quantum numbers,
and it was hence possible to consider the objects of quantum
physics as analogous to the classical ones.  Bohr formulated this as
the \emph{correspondence principle} in 1923.\footnote{N.Bohr:
  \ZP{13}{117}{1923}.}  However, the quantum postulates were for the
most part unrelated to one another, and quantum physics was thus more
a set of rules for calculation than a unified theory.  In the words of
Abraham Pais, `Quantum physics was not in a crisis.  Quantum physics
\emph{was} a crisis.'~\cite[p217]{Pais}  There was a need for a
`rational quantum mechanics'.  The big breakthrough would happen in
the years 1924--27, originally from two different starting points.

In 1923, Louis de Broglie\footnote{L.de Broglie: \rf{Comptes 
Rendus}{177}{507, 548}{1923}.} managed to show that Bohr's quantum
postulates could be explained by considering the electron as a wave
with wavelength $\lambda = h/p$, where $p=mv$ is the momentum of the
electron.  This line of thought was continued by
Schr{\"o}dinger,\footnote{E.Schr{\"o}dinger:
\rf{Ann.\ Physik}{79}{361,489}{1926}; {\bf 80}, 437 (1926); {\bf 81},
109 (1926).} who considered the electron as an extended charge
with wave characteristics, and constructed a non-relativistic equation
for this wave --- yielding correct results for the energy states of
the hydrogen atom.  At the same time,
Heisenberg\footnote{W.Heisenberg: \ZP{33}{879}{1925}; M.Born and
P.Jordan: \ZP{34}{858}{1925}.} carried out a study of the harmonic
oscillator (an oscillating or vibrating system with only one
frequency), where he sought to describe the system purely in terms of
relations between `directly' observable quantities.  The result of
this was that the values of the observable quantities, which in the
classical theory had been ordinary numbers, now became elements of
matrices --- and the theory was called \emph{matrix mechanics}.  Dirac
and Jordan\footnote{P.A.M.Dirac: \rf{Proc.\ Roy.\ Soc.}{A
109}{642}{1925}; P.Jordan: \ZP{40}{809}{1926}.} showed that the two
forms of quantum mechanics were equivalent, and developed a theory for
transforming betweeen different representations of quantum mechanics.
This transformation theory would thus form a unified theoretical basis
for all of quantum mechanics.

Some important elements of the new physics which was developed during
these years should be mentioned:

\paragraph{\em 1. Quantum statistics.}  Quantum statistics was
developed just before the great breakthrough in quantum mechanics, and
was only fully integrated with the rest of the theory with the advent
of quantum field theory.  It is however essential for calculating
processes and quantum mechanical systems with more than one particle,
it is closely related to symmetries of the systems, and it tells us much
about the concept of a quantum mechanical state.  Two different
statistics were developed: they share the feature that they are based
on the particles being absolutely identical (as opposed to classical
statistical mechanics, where one may imagine `marking' each particle).
They are distinguished by how many particles can be in the same state
at the same time.  In \emph{Bose--Einstein
statistics}\footnote{S.Bose: \ZP{26}{178}{1924}; A.Einstein: {\em
Sitz.\ Ber.\ Preuss.\ Ak.\ Wiss.}  1924, p.\ 261.} (which is valid
for photons and other \emph{bosons}, there are no limits on this.
\emph{Fermi--Dirac statistics},\footnote{E.Fermi:
\rf{Rend.\ Acc.\ Lincei}{3}{145}{1926}, \ZP{36}{902}{1926};
P.A.M.Dirac: \rf{Proc.\ Roy.\ Soc.}{A 112}{661}{1926}.} which holds for
electrons and other \emph{fermions}, requires that no more than one
particle can be in each state.  This is known as the \emph{Pauli
  exclusion principle}, formulated first by Pauli in
1925.\footnote{W.Pauli: \ZP{31}{765}{1925}.}  A connection between the
two statistics was established with \emph{Wigner's sum rule}, which
states that where a system of particles may be considered an
indivisible unit, it behaves as a boson if it consists of an even
number of fermions, and as a fermion if it consists of an odd
number.\label{Wigner}

\paragraph{\em 2. The statistical interpretation.} \label{Born}
With the emergence of the new quantum mechanics it became evident that
it was not possible to predict the exact outcome of an atomic
process.  For example, in matrix mechanics the question of the precise
time of a transition between two energy states had become
meaningless.  It was also not possible to say exactly which stationary
state an electron would end up in after an excitation, even though it
was possible to compute the intensity of the spectral lines.  Nor was
it possible to predict the result of individual scattering processes.
Max Born therefore proposed\footnote{M.Born: \ZP{37}{863}{1926}.} that
the matrix elements and wave functions referred to the distribution of
the results of a series of identically prepared experiments.  This
interpretation was soon accepted by most physicists, and is today
considered one of the basic principles of quantum mechanics.

\paragraph{\em 3. The indeterminacy relation.}
In March 1927, Heisenberg published his indeterminacy
relations,\footnote{W.Heisenberg: \ZP{43}{172}{1927}.} which show that
in quantum mechanics there is a theoretical limit to how precisely
several physical quantities may be measured simultaneously.  It is for
example impossible, as a matter of principle, to simultaneously give
precise values for the position and the momentum of a particle.  This
result emerges both from the mathematical formalism of quantum
mechanics, and from considerations of the impact of the measuring
apparatus on the system to be measured: an impact that cannot be made
arbitrarily small.  This inherent indeterminacy is one of the things
distinguishing quantum mechanics most clearly from classical physics,
and is often considered the first principle of quantum mechanics.  It
also led Bohr to formulate his complementarity
principle.\footnote{N.Bohr: \rf{Nature}{121}{580}{1928}.}

All of this happened in the course of a few years, in an intellectual
climate with few precedents in the history of physics.  A large amount
of the work was done by a group of very young physicists (Pauli,
Jordan, Heisenberg, Dirac, Wigner, Fermi and others were all born
between 1900 and 1902), under the supervision and influence of Niels
Bohr and Max Born in particular.  In this environment, with Bohr as
the most central character, the still-dominant interpretation of
quantum mechanics --- the so-called Copenhagen interpretation --- was
also developed.  This interpretation represented a break with the
realism of classical physics: that things are as they are,
independently of whether we observe them --- and also with the
associated determinism, which had been widespread in the 19th
century.  This was poorly received by many physicists, especially in
the older generation, also among those who had played central roles in
the development of quantum physics: Planck, Einstein, Schr{\"o}dinger,
de Broglie.  Einstein's words, `God does not play dice', have become
famous.  The debate between the two `camps' reached its climax at the
Solvay conference in 1927, with Einstein and Bohr in the lead roles.
This can also be considered the starting point for all later
philosophical discussions of quantum physics, as depicted for example
in \cite{Jamm:Phil}.

\subsection{Quantum electrodynamics}

Since a proper theoretical foundation for quantum mechanics now
existed, it was time to return to the question that Planck had started
with: the interaction between radiation and matter.  It was now
possible to conduct an analysis of radiation phenomena taking into
account the quantum nature of \emph{both} radiation and matter.  The
task was then to attempt to use the theory that had been developed for
the study of matter, to also quantise the radiation field and hence to
explain Planck's quantum postulate.

The idea of \emph{field quantisation} (often called second
quantisation) can be said to have been originated by Jordan, and was first
expressed in a paper by Born, Heisenberg and Jordan in 1925.
A concrete formulation of the idea came in 1927, with Dirac's first
paper on quantum electrodynamics.\footnote{P.A.M.Dirac: \rf{Proc.\ Roy.\
Soc.}{A 114}{243}{1927}, reproduced in \cite{qed}.}  Treating the
field as a system of harmonic oscillators, it was possible to obtain
discrete (quantised) energy states of the field, and these states
(\emph{field quanta}) behaved just like photons (massless particles
with Bose--Einstein statistics).  In this paper he also showed how the
photons were \emph{created} and \emph{destroyed}.  He carried out a
consistent (but still non-relativistic) quantisation of the radiation
field, and obtained the correct values for the emission and absorption
coefficients.  This theory was essentially a \emph{perturbation
  theory}\label{pert}, ie., one first calculated the effects of
processes involving one photon, then added contributions from
processes with two photons (which should be less probable and hence
give smaller contributions), etc.

Dirac's work was followed up by further work on a relativistic formulation
of the theory.  Certain problems emerged with obtaining a consistent
quantisation of all the field components --- Dirac could ignore this
problem because it is not necessary to treat all the components
equally in a non-relativistic theory.  Heisenberg and
Pauli\footnote{W.Heisenberg and W.Pauli: \ZP{59}{160}{1929}.} solved
this problem by exploiting a freedom in the choice of fields ---
\emph{gauge invariance} --- which follows from Maxwell's equations.
This could be used to eliminate the `unphysical' field components and
end up with a theory which was relativistically invariant although it
did not look so.  Other (equivalent) solutions were also proposed, but
these were conceptually more difficult, and were not used in the
1930s.  They would, however, form the basis for the later, explicitly
relativistic theory.

It was shown that field quantisation of the kind carried out by Dirac
could be used to describe bosonic systems in general.  That raised the
question of whether fermions also could be considered in a similar
manner.  This was shown by Jordan and Wigner in 1928.\footnote{P.Jordan and
E.P.Wigner: \ZP{47}{631}{1928}, reproduced in \cite{qed}.}  As in
the bosonic case, variable particle numbers could be taken into
account, suggesting that not only photons, but also electrons, were
field quanta which could be created and destroyed.

The same year, Dirac developed his relativistic equation for the
electron,\footnote{P.A.M.Dirac: \rf{Proc.\ Roy.\ Soc.}{A
    117}{610}{1928}.} which gave the correct results for the structure
of the hydrogen atom, and which also explained the spin and magnetic
properties of the electron, which it had previously been necessary to
introduce as \emph{ad hoc} assumptions.  Combining the Dirac equation
and quantum electrodynamics, it was also possible to calculate
scattering processes to lowest order, with results in very good
accordance with observed values.  The first success was the
Klein--Nishina formula \footnote{O.Klein and Y.Nishina: 
\ZP{52}{853}{1929}.} for Compton scattering ($\el\gamma\to
\el\gamma$).  Later on, Jordan--Wigner quantisation was employed for
the Dirac equation, and a number of processes were calculated on this
basis: pair creation ($\gamma\gamma\to\el\pos$), pair annihilation
($\el\pos \to \gamma\gamma$), bremsstrahlung ($\el\to\el\gamma$ in an
electrostatic field), Bhabha scattering ($\el\pos\to\el\pos$) etc. 
A very important theoretical result from this period is the study by
Paul in 1940,\footnote{W.Pauli: \PR{58}{716}{1940}, reproduced in
\cite{qed}.} where he showed that all particles with spin 1/2 (such as
the electron) \emph{must} obey Fermi--Dirac statistics, while all
particles with integer spin (like the photon) \emph{must} obey
Bose--Einstein statistics.  This conclusion relies fundamentally on
relativistic considerations (the strong requirements of causality in
relativity), and requires that everything is formulated in a
relativistically invariant manner.

So then it seemed everything should be OK.  But the new, relativistic
quantum theory turned out --- despite its successes --- to be a
disaster.  There were in particular 3 or 4 issues that caused despair:

\paragraph{\em 1. The positron.}
Dirac's relativistic equation for the electron was, as we have seen,
an amazing success, as it could be used to derive all the important
properties of the electron, without any additional assumptions.  But
it also contained a paradox: it predicted twice as many quantum states
as expected.  The `superfluous' states corresponded to the
negative-energy solutions of $E^{2} = p^{2}c^{2}+m^{2}c^{4}$.  These
solutions could not be simply rejected as unphysical: there was no
mechanism which could prevent transitions between positive and
negative energies.  Instead, Dirac proposed that the negative energies
could correspond to particles with positive energy and positive
charge: all the states were originally occupied, and an electron left
behind a positively charged `hole' when it jumped to a higher state.
The holes were thus identified with protons, which were the then known
positively charged particles.

This theory was a complete disaster.  Oppenheimer showed that it led
to spontaneous decay of the hydrogen atom, giving a lifetime of
$10^{-10}$ seconds for ordinary matter.  There was also no way of
explaining the mass difference.  Dirac had to postulate the existence
of a new, so far undiscovered particle --- an outrageous idea at that
time.  With the recently discovered neutron, one knew of only 4
particles --- and would rather not have any more.  Luckily, Anderson
discovered the positron by accident in 1932,\footnote{C.D.Anderson:
  \rf{Science}{76}{238}{1932}.} and the disaster turned to triumph.

\paragraph{\em 2. The self-energy of the electron.}
In classical electrodynamics, the electron has an internal
electrostatic energy, which is infinite when considered as a point
particle, and equal to the mass when considered as a hard sphere with
radius $a=e^{2}/4\pi\epsilon_{0}mc^{2}$ (the classical electron
radius).  In quantum mechanics and relativity, the particles are
treated as points, and the infinite self-energy appears.  But in
addition there is an electromagnetic effect, which is purely quantum
mechanical --- and this turned out at first to behave even worse than
the classical self-energy: it gave not just an infinite shift in the
energy levels, but also an infinite relative shift, which would give
an infinite shift of the observed spectral
lines.\footnote{J.R.Oppenheimer: \PR{35}{461}{1930}.}

But this was before the positron theory, and before the Dirac field
was quantised.  When this was included, and a consistent procedure was
developed to subtract quantities which are nonzero in the vacuum, it
turned out that there were indeed still infinities (divergent
integrals), but they were `nicer' (the divergence was weaker) than in
both classical theory and the first calculation of the self-energy.
On the other hand, all correspondence with the classical theory was
lost and with it, apparently, any possibility of obtaining meaningful
results by separating out the infinities --- which appeared in all
attempts at higher order calculations.

\paragraph{\em 3. Vacuum polarisation.} 

The process $\gamma\to\el\pos\to\gamma$, or an electromagnetic field
inducing the creation of an electron--positron pair, with an
associated charge distribution, must be included in the Dirac theory.
If the induced charge and current is calculated, it turns out to be
infinite.  These charges in turn produce an electromagnetic field, and
we get a new quantum effect: an (infinite) polarisation of the vacuum.
This gave both the vacuum and the photon a self-energy, which of
course could not have anything to do with the mass, as was the case
for the electron.  It was suggested that this polarisation might be
\emph{renormalised}\footnote{R.Serber: \PR{49}{545}{1936}.} by
e.g.\ absorbing the infinities in the charge of the electron --- and
Weisskopf observed that `a constant polarisability is in no way
observable.'\footnote{V.F.Weisskopf: {\em Kgl.\ Danske
    Vid.\ Selsk.\ Math.-fys.\ Medd.} {\bf 14}, nr.6 (1936), reproduced
  in \cite{qed}.} But as to \emph{how} the process of eliminating the
infinities was to be carried out, there was no answer.

Weisskopf showed\footnote{V.F.Weisskopf: \PR{56}{72}{1939}, reproduced in
\cite{qed}.} that all self-energies are at most logarithmically
divergent (i.e., they only just diverge) --- a result that would
become important when a theory of renormalisation was developed.  But
this was still some time away.  And in some cases (in particular for
bremsstrahlung) there were also divergences for very low (photon)
energies --- the self-energy and vacuum polarisation diverge at
extremely high energies.  Because of these problems it was concluded
that a consistent theory was still not available, and the theoretical
research became quite dormant.  Dirac distanced himself from quantum
electrodynamics --- the theory he himself had created --- from 1936
on, and devoted the rest of his life until his death in 1984 to trying
to develop an alternative electrodynamics.  This was an extreme, but
not completely untypical, expression of the prevailing mood.  On top
of this came the war.                                                           

Many of the underlying problems were solved by Japanese physicists
during the war, but these results did not become known in the west
until the end of the 1940s, by which time (in particular) US
physicists also had made great progress in the same area.  Much of the
renewed effort was due to new experimental results: in 1947, Lamb and
Retherford\footnote{W.E.Lamb and R.C.Retherford:
\PR{72}{241}{1947}, reproduced in \cite{qed}.} had discovered, with the help of
microwave technology, a small shift of lines in the hydrogen spectrum
--- the so-called Lamb shift.  This was presented at the big Shelter
Island conference the same year, and soon after,
Bethe\footnote{H.A.Bethe: \PR{72}{339}{1947}, reproduced in
  \cite{qed}.} was able to explain the results as a consequence of
radiative corrections.

Bethe's calculation gave the correct result, but was not
relativistically invariant, and could not be used as a more general
procedure. In general the renormalisation, where the infinities were
absorbed into a mass or charge term, appeared arbitrary, but it turned
out that the outcome was unique if explicit Lorentz and gauge
invariance was maintained throughout all stages of the calculation.
Hence, the development of theories exhibiting such invariance became
an essential aim.

Such theories were developed primarily along two lines.  The
first (Tomonaga, Schwinger)\footnote{S.Tomonaga:
  \rf{Progr.\ Theor.\ Phys.}{1}{27}{1946}, reproduced in \cite{qed};
  J.Schwinger: \PR{74}{1439}{1948}.} took field theory as its starting
point, and introduced a new formulation of quantum mechanics, where
the interaction was separated from the remainder (non-interacting
part) of the system.  The second method, developed by
Feynman,\footnote{R.P.Feynman: \PR{76}{749, 769}{1949} \cite{Feynman:1949qed},
  reproduced in \cite{qed}.} was less general, but more intuitive and
easily applicable.  It started from the scattering problem, and
considered the interaction as an action at a distance with a finite
propagation velocity.  The result was a
description of quantum electrodynamics where the motion of the
electron in time and space is fundamental.  A particular feature of
this formulation was that the positron may be interpreted as an
electron moving backwards in time!  Dyson\footnote{F.J.Dyson:
  \PR{75}{486}{1949}, reproduced in \cite{qed}.} showed that the two
formulations were equivalent: Feynman's `rules' may be produced by
integrating Tomonaga's and Schwinger's theory.

As a consequence of this breakthrough, there was fresh interest in
formulating quantum field theory in such a way that it could `stand on
its own legs' --- with strict relativistic invariance and gauge
invariance built into the foundations of the theory.  Until then,
quantum mechanics had been almost exclusively constructed from the
non-relativistic Hamiltonian formulation of classical mechanics; now
many people sought to develop, from quantum mechanical principles, a
theory in line with classical Lagrangian theory, which has a more
invariant formulation.  Such theories could clearly also be applied to
problems other than electrodynamics, and this was now attempted.

\subsection{Strong and weak interactions}

At the first Solvay conference, in 1911, Marie Curie remarked that
`radioactive phenomena form a world of their own', without any
connection with other physical phenomena.  `It looks as if [they] have
their origin in a deeper area of the atom.'\footnote{{\em Th{\'e}orie du
rayonnement et les quanta}, p.\ 385, eds. P.Langevin and M.de Broglie,
Gauthier--Villars, Paris 1912.}  This was before she got to know about
Rutherford's discovery of the atomic nucleus earlier the same year.
In the following years, the perceptiveness of her observation
was confirmed: $\alpha$ (helium nuclei), $\beta$ (electrons) and
$\gamma$ (photons) radiation all turned out to originate in the nucleus, and
it also slowly became evident that completely new forces were needed
to explain these phenomena.

Initially there was full agreement that the nucleus must be made up of
the two particles that were known up to then: the proton (hydrogen
nucleus) and electron, and electromagnetic forces should be sufficient
to keep this system bound.  However, this view soon met with
difficulties.  Rutherford discovered deviations from Coulomb's law in
scattering of $\alpha$ particles off hydrogen at very small
distances.  The $\beta$ particles turned out to be emitted from the
atom with a continuous energy spectrum, in apparent contradiction to
the requirements of energy conservation and the same process being at
work in each case.  The tight binding of protons and electrons was
forbidden according to Bohr's quantum postulates, and also according
to the new quantum mechanics under development.  Also, many nuclei
($\rm {^14}N$ being the best known and most studied) had the wrong
spin and statistics compared with the theoretical expectations.  (If
${}^{14}{\rm N}$ consists of 14 protons and 7 electrons, it should
according to Wigner's sum rule have half-integer spin and behave as a
fermion --- but it has spin 1 and behaves as a boson.)  The most
common response to all these problems was that the known laws of
physics break down at distances of the order of the nuclear radius;
quantum mechanics is no longer valid (just as classical physics breaks
down when Planck's constant no longer can be considered small); the
electron (and perhaps also the proton) completely loses its identity
inside the nucleus --- and Bohr also believed that energy is not
strictly conserved.

By 1931 it had become clear that Coulomb's law would have to be
modified, or new forces introduced, at small distances; and a
qualitative account of radioactivity based on non-relativistic quantum
mechanics had been obtained.  There was however still no proper
theory.  The $\beta$ spectrum and the spin--statistics problem was
still a mystery.  We have seen that Bohr advocated the view that
physics as we know it, including conservation of energy, breaks
down. \label{nu} Pauli\footnote{W.Pauli: \PR{38}{579}{1931}.} had
postulated the existence of an electrically neutral spin 1/2 particle
with a very small ($\approx0$) mass (first called the neutron, but
later renamed by Fermi to the \emph{neutrino} $\nu$) to save energy
conservation and solve these problems --- and became more and more
sarcastic towards Bohr.  There were no conclusive arguments on either
side.  Then, the following year, things again started to happen.

In February 1932, Chadwick discovered the
neutron.\footnote{J.Chadwick: \rf{Nature}{129}{312}{1932}.} He, and
other collaborators of Rutherford, had for 12 years been looking for a
strongly bound system of a proton and an electron --- and now he
believed he have found it.  However, in the course of the following
year, evidence emerged for the neutron being a true elementary
particle, with spin 1/2. This would at least explain the spin and
statistics paradox, but $\beta$ radiation would be a greater mystery
than ever, unless one was prepared to still `hide' some electrons
inside the nucleus.  So the situation was new, but the old fronts more
or less remained in a stalemate.

In the same year, Heisenberg presented the first proper theory of
nuclear forces.\footnote{W.Heisenberg: \ZP{77}{1}{1932}; {\bf 78}, 156
(1932); {\bf 80}, 587 (1933).}  Heisenberg took the side of Bohr
against his good friend Pauli, believing in a composite neutron, but
he wanted to push all the problems onto the neutron and hence forget
them in the description of nuclear forces.  In this way, the old view
could help him: by considering a proton--neutron system in analogy
with a $\rm H_2$ molecule (two protons and one electron), he could see
that the proton and the neutron could be `exchanged' by interchanging
the electron, and the interaction between them could be described by
this exchange.  This was the first step on the way to the insight that
the strong forces do not distinguish between protons and neutrons
(charge independence), and that they may be considered two states of
the same particle, the nucleon.  It is in fact possible to `mix' (or
superposition) the two arbitrarily wihtout this affecting the strong
nuclear forces.  This symmetry is called \emph{isospin}.

In 1934, Fermi\footnote{E.Fermi: \rf{Ric.\ Scient.}{4}{491}{1934};
 \rf{Nouvo Cim.}{11}{1}{1934}; \ZP{88}{161}{1934}.} sorted out the
dispute between Bohr and Pauli, in favour of Pauli.  He did this by
invoking quantum field theory: a $\beta$ decay was the result of an
interaction between the nucleons and electron and neutrino fields, so
that the neutron changes into a proton (Heisenberg's idea!), and at
the same time an electron and a neutrino (or, according to modern
conventions: an antineutrino) are created.  (At the same time, other,
similar, processes were also explained or predicted, e.g., $\el\to
n\nu$ (electron capture); $p\to n\pos\nu$ ($\beta^+$ radiation); $\nu
n\to\el p$.)  Hence there was no longer any need to hide electrons in
the nucleus: they were only created in the decay process.  And the
neutron turned out to be a true elementary particle with spin 1/2, on
a par with the proton.\footnote{It was later discovered that neither
  is elementary.}  And, last but not least, it was seen that $\beta$
decay (and weak interactions in general) cannot even be understood
qualitatively without quantum field theory.  Fermi was in fact the
first person to amply field quantisation of spin 1/2 fields (the
Jordan--Wigner method) --- only later in the same year did Heisenberg
do the same in his positron theory (quantum electrodynamics).

Fermi's theory was of course not complete, but it is a correct
description at low energies.  Over the next few years, Dirac
wavefunctions were also employed for the nucleons, making it
possible to write the interaction in a relativistically invariant
form, and a more general form for the interaction was also soon
developed.  All of this followed almost automatically, and provided
fertile soil for both calculations and experiments.  But it was not
known what happened at high energies.\footnote{This only became clear
  in the 1960s and 1970s, with experimental verification in 1983.  I
  will come back to this in section~\ref{gauge}.}  And the theory did
not work at all as an explanation for the binding of the nucleus: the
coupling between the fields was far to weak, and it was difficult to
obtain charge independence.  Only slowly was it realised that there
are two kinds of nuclear forces: the strong, responsible for binding
the nuclei, and the weak, responsible for $\beta$ decay and similar
processes.

It was Hideki Yukawa who in November the very same
year\footnote{H.Yukawa:
  \rf{Proc.\ Phys.\ Mat.\ Soc.\ Japan}{17}{48}{1935}.} proposed a model
of the forces between the nucleons which could account for the main
features of the strong interaction.  It was based on the following
analogy with electromagnetism.  Electromagnetic interactions may be
viewed as exchange of photons between charged particles.  This should
also be the case for the nuclear forces.  The electromagnetic force
has a long (infinite) range, and this is related to the fact that the
photon has zero rest mass.  The nuclear forces, on the other hand, are
very short ranged.  They should therefore be mediated by a massive (as
yet undiscovered) particle, with a mass $m \approx \hbar/lc$, where
$l$ is the range of the nuclear force.  This particle was (eventually)
called the \emph{meson}.  On the basis of this, Yukawa developed a
field theory in analogy with electromagnetism.

Towards the end of the 1930s, much good theoretical work was done with
different variants of meson theories, especially after the discovery,
in 1937, of what was thought to be the meson.  As it turned out, the
particle discovered in 1937 was not the meson, but a heavier version
of the electron: the muon $\mu$.  In 1947, the `real' meson --- the
$\pi$ meson (pion) --- was discovered, and people then realised why
the previous calculations in the meson theory had given strange or
wrong results.  Now it seemed to work a lot better.

But it did not.  Quantum field theory, which had just enjoyed great
triumphs in electrodynamics, was a complete failure when applied to
meson theory.  No calculations gave any sensible result: the theory
was not useful for anything.  It did not help that it could be shown
that the theory fitting the pion that was discovered was
renormalisable: it was still all wrong.  The explanation is simple:
any process with $n$ photons in quantum electrodynamics has a
probability proportional to $\alpha^{2n}$, where $\alpha\approx1/137$
is the fine structure constant, which is a measure of how strongly the
electromagnetic field couples to matter.  A perturbative expansion
(see p.~\pageref{pert}) works because $\alpha$ is so small.  The equivalent
coupling constant for the nuclear forces, on the other hand, was found
to be $\approx15$.  A perturbative expansion in powers of this number
is clearly meaningless.

Particle physics in the 1950s and early 1960s was chaos.  New
accelerators and new detectors were built, and hordes of new particles
and `resonances' (structures with a lifetime too short to be called
particles) were discovered: K-mesons, $\Lambda$, $\Sigma$ and $\Xi$
particles, $\Delta$ and $\rho$ resonances, etc., etc. --- dashing
hopes that everything was now sorted and the `real' elementary
particles had been found.  There was almost enough to do trying to
find the correct quantum numbers for the new particles and perhaps
bring them into some kind of system.  They were divided into three
main groups: \emph{mesons}, which are strongly interacting bosons;
\emph{baryons}, which are strongly interacting fermions (such as
nucleons); and \emph{leptons}, which are fermions which do not
interact strongly (the electrons, the muon, and the neutrinos).
Mesons and baryons are in turn grouped together and called
\emph{hadrons}.  The quantum field theory of the strong interactions
was, as we have already seen, a disaster, and theorists were limited
to formulating phenomenological laws, and trying to find relations and
rules which were as independent as possible of the dynamical details
of the interaction.

At this point, symmetries and conservation laws turned out to be
invaluable.  It was possible to obtain many results relating the
probabilities of various processes using for example isospin
invariance or spin and parity considerations.  Several new quantum
numbers and abstract symmetries were employed to describe `allowed'
and `forbidden' processes.  Several of these symmetries only hold for
processes involving strong interactions, but since these are much
faster than weak and electromagnetic processes, this yields quite a
lot of information.  One may also find `selection rules' (rules for
which changes in quantum numbers are allowed) for the weak and
electromagnetic processes.

There were several new methods and research programmes which attempted
to go further than this.  Of those, axiomatic field theory and
S-matrix theory were probably the most extensive and general ones.
Axiomatic field theory was, as the name suggests, a programme for
investigating (and making explicit) the (previously implicit)
principles on which quantum field theory rests, and finding out how
much may be derived (strictly mathematically) from the smallest number
of principles.  Some relations which were first discovered
experimentally or derived from plausible assumptions were later
proved within the framework of axiomatic field theory.  Beyond this,
the programme did not contribute many results.

\label{S-theory}
The S matrix programme was, in its most comprehensive version, an
attempt to get rid of almost all of field theory.  It had been
initiated by Heisenberg\footnote{W.Heisenberg: \ZP{120}{513,
673}{1943}.} as an attempt to find out what could be `saved' when
quantum mechanics (in his view) broke down at distances less than 1 fm
($10^{-15}$m).  The S matrix describes the probabilities of
transitions between `asymptotic' states (i.e, states long before and
long after the interaction itself), and the concept is hence
completely independent of whether it is possible to describe what
happens during the interaction itself.  Heisenberg eventually turned
to other interests (a united field theory of all matter), but
supervised several of those who subsequently worked on S matrix
theory.  The study of the S matrix gained renewed interest when it
turned out to be possible, by imposing several generic conditions, to
deduce a number of scattering data (such as relations between
scattering and frequency).  This led some people to suggest that all
relevant physical properties of processes, and the entire spectrum of
particles and resonances, could be explained by imposing conditions on
the S matrix --- the so-called \emph{bootstrap} programme.  Others
were less `starry-eyed', but considered the study of the analytical
properties of the S matrix to be a fertile area of theoretical work.
Several results were obtained, but these became more and more
mathematical and less physical.  Today, this S matrix programme must
be considered to be dead (something which does not necessarily have
any impact on the S matrix interpretation of quantum mechanics, and even
less so on the usefulness of the S matrix in quantum mechanics in general).

Back to weak interactions.  Until the end of the 1940s, the Fermi
theory had been a theory of $\beta$ decay, and that was it.  As the
amount of experimental data increased and the pion was discovered, it
turned out to be possible to construct analogous theories for other
processes: \(p\mu\to n\nu\) (muon capture) and \(\mu^\pm\to
e^\pm\nu\bar\nu\) (muon decay), with approximately the same coupling
constant.  Pion decay could also be described in this way:
\(\pi^+\to\bar{n}p\to\mu^+\nu\), and the ratio of the probabilities
for this process and \(\pi^+\to\pos\nu\) was found to agree with the
predictions.  Similar models could also be constructed for K mesons.
It thus became clear that far from being  some special phenomenon, the
weak interaction is a universal force.

\label{CP}
In 1956, one of the old, well-established symmetries was toppled: it
was discovered that weak processes are not symmetric under spatial
reflections (parity)\footnote{T.D.Lee and C.N.Yang:
  \PR{104}{254}{1956}; C.S.Wu et al.: \PR{105}{1413}{1957}.} --- in
fact, they are maximally unsymmetric, and this is the case for all
weak processes.  This came as a great surprise to theoreticians, but
did not lead to despair: on the contrary, it led to a flourishing of
theories of the weak interactions which aimed to incorporate broken
parity, new theories of the neutrinoes and universal Fermi theory.
This effort came to fruition in 1957--58.  For some years, the belief
was then that all processes were symmetric under the combination (CP)
of reflection and exchange of particles and antiparticles.  This hope
was also dashed: in 1964, it was discovered that this symmetry is also
broken\footnote{J.W.Cronin et al.: \PRL{13}{138}{1964}.} --- but only
very weakly, and only in very special systems.  This symmetry
violation is considerably more difficult than the previous one to
incorporate into the theory in a `natural' way.

The increasing amount of data on the new particles made it possible to
start constructing a kind of `periodic system' of them, which might
lead to the discovery of new symmetries which could yield more
information on the probabilities of various processes. This effort was
closely analogous with the idea of isospin, and was meant as an
extension of this.  In this context, the idea of abstract symmetry
groups became a powerful tool.  Group theory had been introduced in
quantum mechanics by Wigner and Weyl in the 1920s, and had been used
to describe spatial symmetries such as rotations, Lorentz
transformations and reflections, but now the groups started as it were
to live their own lives.  In 1961, Gell-Mann, Ne'eman, Speiser and
Tarski\footnote{M.Gell-Mann: {\em The eightfold way}, Caltech Report
  CTSL--20 (1961), reproduced in \cite{tew}; Y.Ne'eman:
  \rf{Nucl.\ Phys.}{26}{222}{1961}, reproduced in \cite{tew};
  D.Speiser and J.Tarski: \rf{J.\ Math.\ Phys.}{4}{588}{1963}.} managed
to place the 8 most important baryons in a diagram (The Eightfold
Way), and the 8 most important mesons in another diagram --- as
representations of the abstract symmetry group SU(3).

In 1964, Gell-Mann and Zweig\footnote{M.Gell-Mann: \rf{Phys.\
Lett.}{8}{214}{1964}, reproduced in \cite{tew}; G.Zweig: CERN preprint
8182/Th401, 8419/Th412 (1964).} showed that these diagrams and the
associated (approximate) conservation laws emerge naturally if one
assumes that the hadrons are all built up of smaller particles, each
with a charge of $-1/3$ or $+2/3$ of the elementary (electron)
charge.  These particles were called \emph{quarks}, in a nod to a line
from James Joyce's Finnegan's Wake: \emph{Three quarks for Muster
  Mark!}  --- there were three different kinds of quarks: u(up),
d(down) and s(strange), and three quarks were required to make a
baryon.  The quarks were also the simplest (nontrivial) representation
of SU(3).  Mesons consist of a quark and an antiquark.

This model `explained' both the particle spectrum and some properties
of the weak interaction, and must as such be considered a success.
However, a question arose: are quarks real --- or are they only
convenient mathematical and theoretical constructs?  After all, it is
the hadrons that are observed in every reaction, and furthermore, the
quarks appear quite exotic.  Should we not instead consider all
hadrons as equal and elementary, but subject to symmetry laws?  This
position would be in line with the S matrix and bootstrap programme,
but was much more widely shared.  Eventually, however, the quark model
won out, although a free quark has still never been observed.

\subsection{Renewed confidence in quantum field theory}
\label{gauge}

One of the many failed attempts to construct a proper theory of strong
interactions was made by Yang and Mills in 1954.\footnote{C.N.Yang and
  L.R.Mills: \PR{96}{191}{1954}.} Their starting point was a relation
which Weyl had noted between the electromagnetic interaction and
freedom to choose the phase of the quantum mechanical wave functions
(gauge symmetry).  By extending this concept of gauge symmetry to also
include symmetry between different particles (isospin symmetry), they
hoped to be able to generate a useful meson field theory of the strong
interactions.  The results of this first non-abelian gauge theory, as
it was called, were not very encouraging.

The theory did include quanta with the same charge as the $\pi$
mesons, which were assumed to mediate the strong interaction.  It also
included non-linear terms, i.e., the field could interact with itself,
and field quanta could be created and destroyed without any matter in
the vicinity.  This was a completely new phenomenon, but not a great
suprise.  However, the quanta turned out to be massless, and it was
not possible to give them a mass without violating the gauge symmetry
on which the theory was based. Accordingly, it was not possible to
construct a theory of mesons, which definitely were massive particles
--- the theory was thus without (what was believed to be)
reality, and was therefore abandoned.

This failed attempt would however form the basis for the theories that
did emerge, both for strong and weak interactions.  When it came to
the weak interaction, the development of a new theory had two aims.
Firstly, to obtain a renormalisable theory --- Fermi theory was
obviously useless at high energies, and Heisenberg had shown that a
perturbative expansion was impossible in this theory.  This problem
might be solved by introducing new, massive, bosons $\mathrm{W}^\pm$,
which could mediate the weak force in the same way as the photon
mediates the electromagnetic one.  The second aim was to obtain a
theory which could unify weak and electromagnetic interactions.  This
could be done by placing the W bosons, the photon, and possibly a
`new', massive, neutral boson, called $\mathrm{Z}^0$, together in some
symmetry group (and subsequently wonder about the origin of the mass
differences).

\begin{figure}[bht]
\begin{center}
\includegraphics[width=0.65\textwidth]{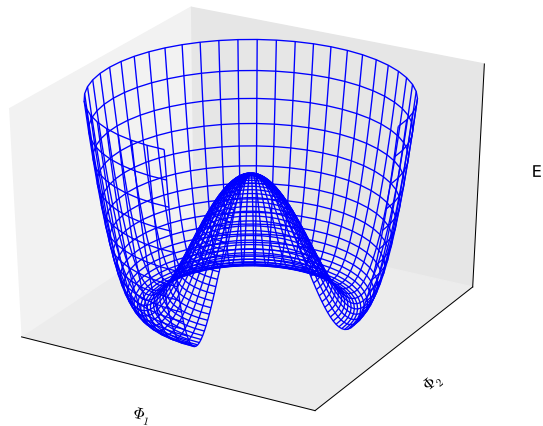}
\caption{A typical Higgs potential}
\label{fig:mexican-hat}
\end{center}
\end{figure}

The problem of particle masses was (partially) solved in the early
1960s, with the introduction of the concept of \emph{spontaneous
  symmetry breaking}.  The essential point here is that the system as a
whole has a symmetry which the ground state does not have.  An example
of this is that the energy (energy density) $E$ depends on the field
$\Phi$ as indicated in fig.~\ref{fig:mexican-hat}.
The system is completely symmetric with regard to `rotations' of the
field components $\Phi_1$ and $\Phi_2$ (this is a kind of gauge
transformation), and the ground state is therefore degenerate: all
states on the dashed circle have the same energy, which is the
lowest achievable.  Any of these may therefore serve as the vacuum.
But when we \emph{define} the vacuum as one of these states, the
symmetry is broken.  When we then express the fields as deviations
from their values in the vacuum, we get one massive and one massless
field.  The massive field can be useful; the massless one (called a
\emph{Goldstone boson}) is a nuisance --- we have never seen such a
creature.

If we now follow Yang and Mills, and let the gauge symmetry be local,
so that the fields may be `rotated' differently at each point in space
and time, we will need to introduce gauge fields (and interactions) to
compensate for this.  These fields are originally massless, as in the
original Yang--Mills model.  But when the symmetry is broken, two
things happen.  Firstly, part of the coupling to the $\Phi$ fields,
specifically the coupling to the value of the field in the vacuum,
appears as a mass term for the gauge fields.  Secondly, a local gauge
transformation may be used to ensure that the Goldstone boson
disappears.  By `mixing' fields, and by also `mixing' `extrinsic'
properties (interactions) and `intrinsic' properties (mass), the
desired result was hence obtained: massive interaction quanta (gauge
bosons).  This is called the \emph{Higgs mechanism}\footnote{Added in translation: or, more  properly, the Brout--Englert--Higgs mechanism.}, after
P.W. Higgs,\footnote{P.W.Higgs: \PRL{12}{132}{1964};
  \PR{145}{1156}{1966}.} and the remaining component of the $\Phi$
field is called the \emph{Higgs boson}.

At this point, Weinberg and Salam (independently) got the idea to try
to unify the weak and electromagnetic interactions in a symmetry group
(SU(2)$\times$U(1)) which had been proposed by Glashow, in a
Yang--Mills type theory using the Higgs
mechanism.\footnote{S.L.Glashow: \rf{Nucl.\ Phys.}{22}{579}{1961};
  S.Weinberg: \PRL{14}{1264}{1967}; A.Salam: 8th Nobel Symposium,
  ed.\ N.Svartholm, Stockholm 1968.}  This worked.  It started with
massless leptons and 4 massless gauge fields, and ended with the three
massive bosons $\rm W^{\pm}$ and $\rm Z^0$ plus the photon --- and
the electron had gained mass through its coupling to the Higgs field.
$\rm Z^0$ and the photon are both `mixtures' (linear combinations) of
two of the original fields.  Broken parity is incorporated into the
theory.  The Fermi theory was recovered in the limit of low energies.

The issue of renormalisability still remained.  But in 1971, 't Hooft
showed that \emph{all} Yang--Mills type theories, with or without
spontaneous symmetry breaking, are renormalisable.\footnote{G.'t Hooft:
\rf{Nucl.\ Phys.}{35}{167}{1971}.} Suddenly quantum field theory was
back in fashion.  And in 1973, the first prediction of the
Weinberg--Salam theory was confirmed: neutral currents (processes
involving exchange of $\rm Z^0$ bosons) were observed in
$\nu_{\mu}$--\el scattering at CERN.

The Weinberg--Salam theory was a good basis on which to proceed; the
next question was to include the quarks (hadrons) in the theory.  In
order to include all possible transitions, the `fundamental' fields
from the point of view of the weak interaction would have to be
combinations of fields corresponding to the `physical' d and s quarks.
In addition, it was necessary to postulate the existence of a `new'
quark, called charm (c), which would serve two purposes.  Firstly, and
most importantly, it would ensure that neutral currents where \(\rm s
\leftrightarrow d \) could not occur (the GIM mechanism).\footnote{S.L.Glashow,
J.Iliopoulos and L.Maiani: \PR{D2}{1285}{1970}.}  Such processes had
never been observed, but would be relatively common if the charm quark
did \emph{not} exist.  Secondly, it would ensure renormalisability.
It was shown that the theory is only renormalisable if the sum of the
charges of all the fermions is 0.  This could be arranged if one
assumes that all quarks appear in three \emph{`colours'} (red, green,
blue), and that a fourth quark exists.

The concept of colour is the starting point for the new theory of
strong interactions: \emph{quantum chromodynamics} (QCD).  The origin
of the model was a concern that the baryon states that came out of the
quark model (with SU(3) and spin) were incompatible with the
assumption that quarks are fermions which obey the Pauli principle.
This problem could be solved by giving the quarks an additional
quantum number --- colour --- and postulating that all hadrons are
colourless, so that the three quarks in a baryon have different
colours.  Furthermore, the colour quantum number was associated with a
new symmetry group $\mathrm{SU(3)_c}$.\footnote{M.Y.Han and Y.Nambu:
  \PR{139B}{1006}{1965}.}  When a Yang--Mills theory (without symmetry
breaking) is then constructed from this group, the result is QCD.

The interactions in QCD are mediated by 8 massless, electrically
neutral \emph{gluons}.  Because the theory is non-abelian, the gluon
fields have self-interactions: they are themselves coloured.  This has
several consequences.  The first one, discovered in 1973, is called
\emph{asymptotic freedom}: if the quarks are very close to each other,
or (equivalently) have very high energies, they will not notice each
other --- they behave approximately like free particles.  This effect
can only occur in a non-abelian theory, where the gluons can be
`dissociated' into gluons as well as into fermions (quark--antiquark
pairs).  In quantum electrodynamics, the effect is the opposite.
Asymptotic freedom can explain experimental results which suggested
that highly energetic leptons scatter off free point particles inside
the hadrons.  It also makes \emph{perturbative QCD} possible: at very
high energies we have a theory of the strong interactions which can be
calculated using conventional methods.

The other consequence of the self-interaction is to be found at the
other end of the energy scale: \emph{confinement}.  At large distances
or small energies, the gluons will multiply into so many gluons of all
possible colours to such an extent that it constitutes a huge force
which will keep captive any quark which tries to `run away': an
`anti-screening' effect.  Only colourless states can have any hope of
escaping. A free quark can never exist!

This provides an explanation both for why quarks always stick together
in colourless states, and for the great problems the strong
interactions had presented.  Until the early 1970s, all experiments
had occurred at relatively low energies, where confinement is
relevant, and perturbative methods break down.  The meson fields are
now considered a residual interaction which can occur over larger
distances than the `free' gluons can reach.  The correct theory of all
these phenomena is now believed to be nonperturbative QCD.

The great breakthrough for the new theories came in the second half of
1974, when charm was discovered.  The context was measurements of the
reaction rate of processes $\pos\el\to$ hadrons, where suddently a
huge, narrow peak was discovered at 3.1 GeV.  This `resonance' was
called $J/\psi$, and was eventually identified as a
$\mathrm{c\bar{c}}$ bound state.  Not only did this confirm the
model with four quarks and three colours: it also turned out that
QCD could be used to \emph{calculate} $\mathrm{c\bar{c}}$ states!  The
reason is that the charm quark is heavy enough that it is bound at
very small distances, so that perturbative QCD may be used, and the
large mass also means that non-relativistic theory may be employed.
With the help of qualified guesswork the problem could then be made to
look like the hydrogen atom.  After this, the combination of the
Weinberg--Salam theory and QCD was called the \emph{Standard Model}.

The particle spectrum received a new and unexpected menber in 1975,
when a heavy lepton, called $\tau$, was discovered.  This disturbed
the balance between leptons and quarks which the Standard Model relies
on, a balance which was restored in 1977 with evidence of a fifth
quark (b: beauty or bottom).  The sixth quark --- t for truth or top
-- was finally discovered in 1995, but its existence was never in
doubt.  This new set of particles provided one important benefit: CP
violation could be introduced in a natural manner by way of mixing
between three quark `families' (u,d), (c,s) and (t,b) --- a mechanism
which had already been proposed by Kobayashi and Maskawa in
1972.\footnote{M.Kobayashi and T.Maskawa:
  \rf{Progr.\ Theor.\ Phys.}{49}{652}{1973}.}  According to the most
recent data from CERN when this was written (1990), this is the end of
the matter in this respect: there are no more than three
families.\footnote{This conclusion has been confirmed by subsequent
  experiments.  Any additional families would have to involve physics
  beyond the Standard Model.}

Further experimental evidence for quantum chromodynamics was provided
in the late 1970s and early 1980s with the observation of `jets':
concentrated showers of hadrons which could be traced back to a single
outgoing quark (or one gluon), in very high energy reactions.  The
angular distributions of these showers were in good agreement with QCD
predictions.

The `final'\footnote{This was originally written in 1990, long before
  the discovery of the Higgs boson in 2012, which may equally be
  called the final confirmation of the theory.  This paragraph has
  been rewritten to take this discovery into account.} confirmation of
the Weinberg--Salam theory came in 1983, when first the W bosons and
later on the Z boson were discovered at CERN, with exactly the
predicted masses.  The only missing piece in the puzzle (after the
discovery of the top quark) was the Higgs boson.  However, in this
case there were no clues as to what its mass should be (although some
limits could be inferred), and it was not even known if there was one, several or
even no Higgs: other, more complicated ways of introducing spontaneous
symmetry braking can be devised.  Only one thing was clear: the
discovery of the Higgs boson would lead to a Nobel Prize.\footnote{The
  1990 text stated that the person who discovers the particle would
  get the Nobel Prize, but this has not happened --- probably because
  the experiments are too large to justify awarding the prize to one
  or two individuals.}

The successful unification of the weak and electromagnetic forces led
many people to attempt to also unify the electroweak and strong forces
in a single theory. The first attempts at such a `Grand Unified
Theory' (GUT) appeared already in 1974.  All these theories predict that
the proton should be unstable and decay, but no proton decay has ever
been observed.  So far, apart from much interesting speculation, such
theories have yielded few if any useful results.  The same is the case
for attempts to integrate gravity into the theories --- but that is a
different story.

\section{Physical principles of quantum field theory}
\label{sec:qft}

Quantum field theory is a relativistic, quantum mechanical
many-particle theory.  It starts from the basic quantum mechanics
concepts of operator and state, and the field theory concepts of field
and Lagrange density, and is furthermore characterised by an extensive
emphasis on symmetries and invariances which is common to field theory
and quantum mechanics.  Using this conceptual framework the theory
treats systems of elementary excitations (`disturbances' in energy
density, charge density, momentum density, etc.), and describes how
these excitations appear and disappear, or relate to each other in
other ways.  If the elementary excitations are identified with
elementary particles, quantum field theory becomes the theory of the
elementary particles and their interactions.  This is how it will be
considered here.\footnote{There also exist non-relativistic quantum
  field theories, which find their applications primarily in condensed
matter theory.  The status of these theories vis-\`a-vis relativistic
quantum field theory and ordinary, non-relativistic quantum mechanics
is not entirely clear to me, and I will not discuss them here.
Methods from quantum field theory are also extensively used in
statistical physics, both classical and quantum, but this version of
`quantum field theory' is definitely completely incommensurate in its
ontological (philosophical) status.}

\subsection{Fields and Lagrangian density.  Symmetries in classical physics}
\label{sec:symmetries}

In classical physics, a \emph{field} is defined as a quantity that has
a specific value (or several, in the case of vector and tensor fields)
at every point in space and time.  Mathematically, it can thus be
defined as a function of the coordinates in 4-dimensional space-time:
$\Phi=\Phi(x)=\Phi(\vec{r},t)$.  In quantum mechanics this will, as we
shall see, be modified because of the indeterminacy principle, but the
fundamental feature remains, that the field exists everywhere: it is
somehow defined everywhere in space-time, and is not localised in one
region. In classical physics, each field represents one physical
phenomenon --- for example, electricity, magnetism, gravity, sound.  A
characteristic feature of quantum field theory is that each field
represents one type of particle.  How this happens will hopefully
shortly become clear.

A \emph{Lagrangian density} \( {\cal L} ={\cal 
L}(\Phi,\partial\Phi/\partial x_\mu ) \) is constructed as a function
of the field and its derivatives; like the field, it is defined
everywhere.  Once the Lagrangian density is given, all the dynamical
properties of the field are also given, and in classical physics this
completely determines the behaviour of the field (when the boundary
conditions are known).  This comes about by requiring that the
\emph{action}, which is the integral of the Lagrange density over all
of 4-dimensional space-time, have a minimum (or a stationary point)
for the actual values of the field.  In other words, those values for
the field and its derivatives are chosen which minimise the action
(or, to be precise: lead to a stationary value for the action with
respect to
variations of the field values).  In mathematical language: \( \delta S =
\delta\int{\cal L}d^4x = 0 \).  This principle, which was first
formulated by William Rowan Hamilton, can be formulated completely
independently of which coordinate system one might choose to use, and
may represent the most invariant (general) form that a physical
principle can possibly take.  The theory is \emph{local}, i.e., the
behaviour of the fields at one point is only determined by the values
of the fields in the vicinity.  This is ensured by $\mathcal{L}(x)$
only depending on the values of the fields and their derivatives at
the point $x$.

If the physical system under consideration has a particular symmetry,
the Lagrangian describing the system must also have the same
symmetry.  For example, if the behaviour of the system is unchanged if
it is rotated, then such a rotation must leave the Lagrangian
unchanged.  This symmetry can also be described by the action
remaining the same after certain kinds of coordinate and field
changes.  It can be shown that for any such symmetry there exists a
corresponding conserved quantity, i.e., a quantity that does not change
with time.  It is thus possible to find out quite a lot about the system
by studying its symmetries.  If the system in addition has a symmetric
initial state, we know that the system will always remain in a
symmetric state.

The symmetry principles may be accorded a somewhat different status
from the `ordinary' laws of nature.  They determine to a certain extent
what form the laws of nature can or must take, or in other words what
kinds of laws are possible, or what must be demanded of a law of
nature.  One may for example associate the symmetries with the concept
of a reproducible experiment, and specifically with the possibility of
repeating an experiment under different conditions (for example in
different locations or with different orientations of the apparatus)
and still claim it to be the same experiment.  They may also be
directly related to the concept of a law of nature: is it possible to
say that the \emph{laws of nature} are different in different
directions?  Even if we should be wary of saying that the symmetry
principles are \emph{a priori} conditions for all science ---
apparently obvious symmetry principles have turned out not to be
satisfied --- then at least a certain amount of symmetry is required.
Wigner has discussed these questions more thoroughly in several
articles in \cite{Wigner:Essay}.  Such questions are also related to the
more philosophical issues of absolute space, absolute time, etc.  The
status of the symmetry principles in cosmology is more problematic,
since there we cannot necessarily assume that all times, places and
directions are equivalent, but to a certain extent we will want to
explain \emph{why} this is so.

The most important symmetries and conservation laws in classical
physics are:

\paragraph{\em 1. Spatial translation.}
It is (we now think) an obvious requirement of a physical theory that
the laws shall be the same wherever we are.  If the laws in a
different part of the universe would differ from ours, then both sets
of laws must be special cases of a more general law, which is valid
everywhere. Moreover, we have no reason to believe that the universe
exhibits irregularities in space which cannot in principle be moved or
smoothed out.  This is formulated as the system being invariant with
respect to a
translation in space, ie., the physics is unchanged if all spatial
coordinates are shifted by a constant: $\vec{r}\to\vec{r}+\vec{a}$,
where $\vec{a}$ is a constant vector.  The conserved quantity
corresponding to this symmetry is momentum: the fundamental
physical principle that all places in the universe are equivalent
(which followed from the break with aristotelian physics, which was
based on the notion of absolute space and a distinction between
terrestrial and celestial dynamics) corresponds to the fundamental law
of \emph{conservation of momentum} (first formulated --- incorrectly
--- by Descartes).

\paragraph{\em 2. Time translation.}
One of the tasks of physics is to find laws which are valid at all
times.  If we find that the laws of physics change with time, we are
forced to search for the laws that govern this change --- and these
laws must of course be time independent.  It must also in principle be
conceivable for any situation to be repeated at any later time: there
is no absolute point in time that we can use to set our time
coordinates.  The Big Bang theory challenges this principle, since it
postulates that the entire universe emerged from a singularity 13
billion years ago.  At that point, all laws of physics, and the
concepts of space and time, may break down.  At the same time, this
theory is based on the principle that the laws themselves do not
change.  One might say that this is a case of a boundary condition
which is not symmetric, while the laws are.  However, when the system
is the entire universe, it may be difficult to distinguish what are
laws and what is a result of boundary conditions.

If we ignore this particular problem (which in any case does not play
a significant role at `normal' scales), we can assert that all
properties of a system are unchanged if we `move' the system forward
or backward in time by a period $t_0$, i.e., if we change our time
coordinate such that $t\to t+t_0$.  This first basic principle of
physics (which has been recognised as long as some kind of physics has
existed) corresponds to the most fundamental law of physics:
\emph{energy conservation} (first formulated, in an incomplete
version, by Leibniz).  After Einstein's demonstration of the
equivalence between mass and energy, the ancient principle of the
constancy of matter has also been absorbed into this law.

\paragraph{\em 3. Rotation.}
A third symmetry principle we assume in our physical considerations
is that the universe looks (more or less) the same in all directions,
and that the laws of physics do not distinguish between directions.
Therefore, if we rotate our coordinate system by a fixed angle about some axis
relative to the physical system, this should not make any
difference.  This symmetry corresponds to the law of
\emph{conservation of angular momentum}, i.e., the sum of the angular
momenta \(\vec{L}=\vec{r}\times\vec{p}\) of all particles in the
system is conserved.\footnote{This holds in classicle particle
  mechanics.  We can also define an angular momentum density
  \(\vec{\ell}=\vec{r}\times\vec{p}\) for fields, where $\vec{p}$ is
  now the momentum density, and the total angular momentum is the
  integral of this over all space.  Conservation of angular momentum
  is hence just as fundamental in field theory.  In addition, in
  quantum mechanics there is an internal angular momentum ---
  \emph{spin} --- which does not have a classical analogue.}

In practice, parts of the system are often taken as given, or as
fixed.  This can be used to define fixed positions and directions in
space, and neither momentum nor angular momentum is defined for the
rest of the system.  For example, when studying the behaviour of the
electrons in a diatomic molecule, the positions of the nuclei are
usually taken to be fixed, and the line between them defines a
direction in space.  In that case, we do not have full rotational
symmetry, but there is still rotational symmetry about this line
(axis).  On the other hand, in a system with a single centre (as an
atom or a solar system), or a system where all particles or subsystems
are on the same footing, we will have full rotational symmetry.  This
information can be used to find out quite a lot about the functional
form any mathematical description of this system must take, beyond the
conservation laws.  In quantum mechanics, this is even more prominent,
both because there are more (abstract) symmetries which may be
exploited, and because the quantum mechanical concepts of operators
and states are very well suited to describing symmetry
transformations: for example, symmetric states may be created from
asymmetric ones.

\paragraph{\em 4. Relativistic invariance.} 
The principle of relativity --- that two systems in uniform,
rectilinear motion relative to each other are equivalent, that the
laws of physics are the same for all such \emph{inertial frames}, and
that there is no way of determining whether a system is at rest ---
was first formulated by Galileo, and is incorporated in Newtonian
physics.  With the special theory of relativity, and the principle
that the speed of light is the same in all inertial frames, relativistic
(Lorentz) invariance becomes an explicit requirement that models may
or may not take into account.  (It is not necessary to take it into
account for systems where all velocities are small.)  This requirement
puts strict demands on the types of quantities that are allowed: for
example, the requirement that the Lagrangian density must be a Lorentz
scalare already tells us much about what kinds of fields are allowed.

Relativity treats space and time in a unified manner, arising from the
insight that it is not possible to introduce an absolute distinction
between time and space, i.e., a distinction that is valid for all
inertial frames.  The 4-dimensional space-time is hence often just
called \emph{space}, which consists of \emph{events}, or points $x =
(t,\vec{r}) = (t,x_1,x_2,x_3)$.  Even though the four coordinates take
different values in different frames, the distance or \emph{interval}
\[   \Delta x^2 \stackrel{\rm def}{=} c^2\Delta t^2 - \Delta \vec{r}^2 
        = c^2(t_2-t_1)^2 - (\vec{r_2}-\vec{r_1})^2 \]
between two points is always the same.  For $\Delta x^2=0$ this
follows directly from the principle that the speed of light is a
universal constant.  If $\Delta x^2<0$, i.e., $\Delta\vec{r}^2 >
c^2\Delta t^2$, we say that the separation between the two points is
\emph{space-like}.  In this case, the two points are completely
separate: a signal from one of them cannot reach the other one.  We
can also talk about space-like (3-dimensional) `planes', which are
such that the separation between all the points on the `plane' is
space-like.  Such a plane will be space-like in all inertial
frames.  An example of a space-like plane is the set of all points in
space at a given time.  If $\Delta x^2>0$, the separation is
\emph{time-like}.  The \emph{lightcone} of a certain point consists of
all the points with separation 0 from this point, or in other words
the points that may be reached with a light signal from this point.
We can then say that separations `within' the lightcone are time-like,
while separations `outside' are space-like.

A physical quantity $A$ characterised by four numbers 
$(A_0,A_1,A_2,A_3) = (A_0,\vec{A})$ such that $A^2\equiv
A_0^2-\vec{A}^2$ is the same in all inertial frames is called a
\emph{4-vector}.  These play a huge role in relativity, not least
because they may be used to construct Lorentz invariant quantities.
Examples of 4-vectors are the position $x=(ct,\vec{r})$ and momentum
$p=(E,c\vec{p})$ of a particle.  Many fields, such as the
electromagnetic and other gauge fields, are described by 4-vectors.

\subsection{States and operators in quantum mechanics}
\label{sec:state-op}

Quantum mechanics is characterised (and differs from classical
mechanics and `common sense') first and foremost by what we may call
the \emph{indeterminacy principle}.  This may be formulated in several
different ways, and is encoded in several of the basic principles of
quantum mechanics.  In brief, it says that it is not possible to
assign definite values to all physical quantities associated with a
system everywhere and at the same time.  The most famous formulation
of this principle is Heisenberg's relation, which gives a limit
to the accuracy with which two physical quantities can be measured
simultaneously.  Before I look more closely at this fundamental
relation, I will explain how the indeterminacy principle is encoded in
the concept of a state in quantum mechanics.

The \emph{state} (the state function or the state vector) comprises
the most complete description of a system that is possible.\footnote{I
  will leave `hidden variables' interpretations of quantum mechanics
  out of this discussion.} Once the total state of a system is
determined, everything that is possible to determine about the system
has been determined.  The state is denoted $\Psi$ or $\mi\Psi\ket$.
The set of all possible states a system can have is called the
\emph{state space}.

The states obey the \emph{superposition principle}: the sum (or a
linear combination) of two states is also a state.  This is where the
indeterminacy principle enters.  If $\Psi_1$ denotes a state where a
particular physical quantity in the system has the value $a_1$, and
$\Psi_2$ one where it has the value $a_2$, then both values will
occur when the system is in the state $\Psi_1+\Psi_2$.  On the other
hand, no other values of this quantity will occur --- the two values
will not be `mixed'.

I can illustrate this with an example that shows how quantum
mechanics in this respect runs counter to common sense.  Let us
imagine that our system is a ball that may change colour, in such a
way that it is always single-coloured.  Each colour it can take
represents a state of the system (i.e., the ball).  Let $\Psi_1$ denote
the state that the ball is red, while $\Psi_2$ denotes that it is
blue.  If our ball is a quantum mechanical system, then
$\Psi_1+\Psi_2$ is an allowed state.  This state will not denote the
ball being purple, but rather that it is \emph{both} red and blue.
Also, it does not denote e.g. half the ball being blue and half of it
red: recall that the ball always remains single-coloured.
$\Psi_1+\Psi_2$ denotes a state where the ball is monochrome red and
blue all over!  That we can say something like this clearly requires
that we do not look to see what colour the ball is --- it is hard to
imagine seeing something monochromatic red and blue.  We can however
not rule out this state having other strange characteristics which
will allow us to conclude that the ball has in fact been monochromatic
red and blue while we did not look at it.

In the case above, since we knew much about the states $\Psi_1$ and
$\Psi_2$, it was natural to express the third state in terms of these
two.  We say that we \emph{decompose} $\Psi$ is $\Psi_1$ and $\Psi_2$
--- $\Psi$ contains a component of $\Psi_1$ and a component of
$\Psi_2$.  This is completely analogous to how we decompose (or add)
vectors: the states can be considered as vectors (but not in our usual
space --- the state space often has an infinite number of
`dimensions').  To continue with the analogy, we can say that only the
`direction' of the state is relevant, and not the `magnitude': no
physical properties change if we multiply the state by an arbitrary
(complex) number.

Starting with $\Psi_1$ and $\Psi_2$, we can construct all possible
states of the ball with different `amounts' of red and blue --- all
states that contain only red and blue depend on (may be expressed in
terms of or decomposed into) $\Psi_1$ and $\Psi_2$.  We can however
never obtain any state containing any amount of yellow --- all states
of the ball containing any yellow are \emph{independent} of $\Psi_1$
and $\Psi_2$.  If we on the other hand have a state $\Psi_3$
containing some yellow (possibly in addition to some red and blue),
then we can decompose all states containing only red, yellow and blue
into $\Psi_1, \Psi_2$ and $\Psi_3$.  $\Psi_3$ can then be taken to
define a third `dimension' in state space relative to $\Psi_1$ and $\Psi_2$.

Instead of decomposing all red and blue states into $\Psi_1$ and
$\Psi_2$, we could have decomposed them into e.g.\ $\Psi_1$ and 
\( \Psi_2' = \frac{1}{2}\Psi_1 + \Psi_2 \)
(if we for some reason had much information about that state).  Our
$\Psi$ would then be written  \( \Psi = \frac{1}{2}\Psi_1 + \Psi_2'
\), and it would look as if it contains less of the red state $\Psi_1$
than in the previous picture.  This appears like an ambiguity in the
description.  We may however get an unambiguous expression for how
much two states $\Psi$ and $\Phi$ contain of each other by looking at
their \emph{product} (scalar product) \( \bra\Phi\mi\Psi\ket \), which
is a (complex) number.  If the product of two states is zero, they are
said to be \emph{orthogonal} to each other, and hence contain nothing
of each other.  (This is completely analogous to the scalar product in
vector algebra.)  The product of a state with itself gives the
`length' of the state; physical states all have the same length
(usually 1).

A set of states $\mi n\ket$ which are mutually orthogonal, have length
1, and are such that all possible states may be decomposed into states
from this set,

\[     \mid\Psi\ket = \sum_{n} \alpha_n \mi n\ket
  \qquad\text{where}\qquad
  \bra m\mi n\ket = \delta_{mn} = 
  \begin{cases}
  1 &; m=n\\ 0 &; m\neq n
  \end{cases}\, \]
is called a \emph{basis} (a set of basis states), and this
decomposition is called a \emph{representation} of the state.  There
are always several possible choices for such basis states: there is no
preferred representation.  Which representation is chosen usually
depends on which properties of the system one is most interested in.

It is important to note that the state is something that characterises
the system as a whole, and not its individual parts.  In some cases
--- when considering a system consisting of independent
(non-interacting)\footnote{Independent and non-interacting are not the
  same thing: as we shall see, two subsystems may not be independent
  even if they do not interact.} parts --- the state may be divided
into substates, each of which can be ascribed to the separate parts.
Very many systems may also be conveniently described as approximations
to such systems of independent parts.  From a strictly quantum
mechanical perspective this is a purely mathematical technique, without
any preferred physical content --- in particular since the state may
always be represented in other ways which do not involve such a
separation.  Strictly speaking, two subsystems cannot be considered to
have any kind of independence if they are interacting, or if they are
in some other kind of non-separable (entangled)
state.\footnote{Non-separable or entangled states may easily be
  constructed from separable states.  If $\Psi_n(1)$ and $\Psi_n(2)$
  are states of subsystems 1 and 2 separately, the `tensor product'
  (which is a state, not a scalar) \(\Psi_n(1)\Psi_m(2)\) with varying
$m$ and $n$ will denote separable states of the combined system of 1 and 2.
  Linear combinations of these states, which according to the
  superposition principle are possible states of the combined system,
  will however be entangled.  Consider for example the state \(
  \Psi_a(1)\Psi_b(2) + \Psi_c(1)\Psi_d(2) \).}  This is most clearly
exhibited by properties of the individual subsystems being
indeterminate.  But usually this is of course an impractical approach.

Since the state is supposed to characterise all properties of the
system, it should be expected that in a system of many particles, not
only the energy, momentum, relative location, etc, of the particles,
but also their number and kind, should be determined by the state.  In
a non-relativistic theory this can in general be ignored, and the
particles can be taken as given --- except for photons, which are
massless and hence can be created in large numbers also at low
energies.\footnote{This `problem' may be avoided by treating the
  photons, or radiation, not as particles, but as
  classical fields.  But this is not a quantum mechanical treatment of
  the radiation, and quantum phenomena such as the photoelectric
  effect cannot be described this way.}  In a relativistic theory
(quantum field theory), on the other hand, the high energies involved
imply that all particles can be created or disappear, and (the free)
particles must be considered as features of the state of the system.

So far, I have treated the state as a purely abstract entity --- and
the state `in itself' is a completely abstract and almost empty
concept.  The state `in itself' contains no \emph{direct} reference to
physical quantities, nor to points in space and time.  (Since the
state characterises the whole system, it cannot itself be anywhere,
neither in space nor in time.)  Such a reference can only be provided
by choosing a particular \emph{representation} of the state, where the
basis states are linked to certain physical quantities or properties.

In quantum mechanics, any property is associated with an
\emph{operator}.  An operator can be defined as a linear function of
states in state space, which returns states.  We write $F\Psi=\Psi'$,
where $\Psi$ and $\Psi'$ are states, and $F$ is the operator.  We
say that the operator acts on the state $\Psi$, and that $\Psi'$ is
the effect of the operator.  (But this is not something that `happens
to' $\Psi$ at some point in time, it is merely the \emph{definition}
of the operator.)  That it is linear means that the effect of an
operator on the sum of two states equals the sum of the effect on each
of the states:
\[     F(\alpha\Psi_1+\beta\Psi_2)=\alpha F\Psi_1+\beta F\Psi_2 \]
This is what makes the superposition principle relevant, since
it implies that addition of states may be connected with addition of
properties. 

If $F\Psi=f\Psi$, where $f$ is a number, $\Psi$ is called an
\emph{eigenstate} of $F$, with \emph{eigenvalue} $f$.  In this case,
the property $F$ has the unique value $f$ in this state.  This will
not be the case in general, and then the operator cannot be said to be
defined \emph{within} the state, since $F\Psi$ contains states which
are orthogonal to $\Psi$.  But for any operator it is possible to find
at least some eigenstates.  Finding eigenvalues and eigenstates is
indeed one of the main topics of quantum mechanics.

In the general case, we may define the \emph{expectation value}
$\braket{F}$ and \emph{matrix elements} $F_{ab}$ of the operator $F$ as
\[     \braket{F}_\Psi = \braket{\Psi}{F\Psi}\,, \qquad
       F_{ab} = \braket{\Psi_a}{F\Psi_b}\,.   \]
If $\Psi$ is an eigenstate of $F$, then the expectation value of $F$
in $\Psi$ is equal to the eigenvalue.  In any given representation, an
operator will be completely determined by the matrix elements of the
operator between the basis states: if all these are known, then the
action of the operator on any state is given.

A \emph{measurable quantity} (often called an \emph{observable}), such
as energy, charge, position, etc., is represented by a real operator, i.e.,
\( \braket{\Phi}{F\Psi} = \braket{F\Phi}{\Psi} = \braket{\Phi|F|\Psi} \) for
all $\Phi, \Psi$.  The eigenvalues (and expectation values) of such
operators will always be real.

When we measure a quantity, we will never obtain anything but an
eigenvalue of the corresponding operator --- even if the system is in
a state that is \emph{not} an eigenstate of that operator.  This can
be related to the decomposition of the state into eigenstates of the
operator (i.e., it can always be written as a linear combination of such
states), where the operator acts on these separately, returning the
eigenvalue for each eigenstate.  In states which are not eigenstates,
we can therefore \emph{not} predict the outcome of each individual
measurement, since the quantity does not have any specific value in
this state --- it is \emph{in principle undetermined}.  This, and the
fact that measurements in general cannot return all possible values,
is something that distinguishes quantum mechanics clearly from
classical physics.  The latter is what is often denoted by many
quantities being \emph{quantised} (and violates the principle that
`nature does not make any leaps').\footnote{I will here not go into
  the discussion about what actually happens during a measurement.
  Fortunately, this is not crucial to the remainder of this
  presentation.} 

Since observables are related to real operators, all possible
measurement results will be real numbers.  The eigenstates of
observables also have the properties required of basis states.  That
is, if $F$ represents an observable, then every state can be written
as 
\[     
     \mi\Psi\ket = \sum_{n}\alpha_n\mi f_n\ket 
        \qquad\text{with}\qquad
        F \mi f_n\ket = f_n \mi f_n\ket \,,
\] 
where\footnote{In general, we do not necessarily have $f_n\neq f_m$
  for $n\neq m$.  In that case, the expansion is not quite unique.  It
  is possible to choose linear combinations of eigenstates with the
  same eigenvalue, satisfying the same requirements.  In such cases it
  will however be possible to find another observable G, which shares
  eigenstates with F, but where states with the same eigenvalue of F
  may not have the same eigenvalue for G.  If we then use the states
  which are eigenstates of \emph{both} F and G (and, if necessary,
  other operators with the same properties), the expansion will be
  unique.  For any system one may determine how many operators are
  required to obtain a unique expansion.  This number can be computed
  on the basis of the number of classical degrees of freedom in the
  system, and its `internal degrees of freedom' (spin, colour, etc.).}
\[ \braket{f_n}{f_m} = \delta_{nm}\,. \]

If $\Psi$ is a physical state, we have
\[     
     \sum_{n} |\alpha_n|^2 = 1 \quad\text{and}\quad
        \bra F\ket_\Psi = \sum_{n} f_n|\alpha_n|^2 \,.
\]
Upon measurement, $F$ can take any of the values $f_1, f_2,\ldots$,
while $|\alpha_n|^2$ gives the probability of each of these results.
We can easily see that this obeys all criteria for probabilities, and
the coefficients $\alpha_n$ are often called \emph{probability
  amplitudes}.  This association of the expansion coeffcients with the
distribution of results in a large number of identically prepared
experiments, is Born's statistical interpretation of quantum
mechanics.

As we have just seen, we can use the eigenstates of observables to
construct representations of the states.  One such representation ---
the position representation --- is obtained by expanding the state in
eigenstates $\vec{r'}\ket$ of the position vector $\vec{r}$ of a
particle (assuming here that the system consists of a single
particle).  In such an eigenstate, the particle will with certainty be
at the position $\vec{r'}$.  A general state can then be written
as\footnote{The position can take a continuous set of values, and
  hence we get an integral rather than a sum.  In other words: not all
  observables are quantised.}
\[     \mi\Psi\ket = \int \psi(\vec r) \mi\vec r\ket d^3r \] 
where \( \psi(\vec r) = \bra\vec r\mi\Psi\ket \) is Schr{\"o}dinger's
wave function, which characterises the state in this representation.  
$|\psi(\vec r)|^2$ gives the probability of the particle being found
at the position $\vec{r}$ as result of a measurement.

Another commonly used representation of the state is the \emph{energy
  representation}.  A quantum mechanical problem is often considered
solved once the energy eigenvalues (and other characteristics of the
energy eigenstates) are found.  This is related to the energy being a
conserved quantity, so that if a system at some point in time is in a
particular energy eigenstate, it will remain in this state forever
(or: a state which is an eigenstate of the energy at one time, will
also be an eigenstate of the energy at any other time).  Examples of
energy eigenstates are stationary states in atoms, and particles with
a given momentum in quantum field theory.

The fact that a given quantity cannot be given a definite value in an
arbitrary state, also implies that \emph{two} quantities cannot always
be given definite values (or be measured precisely) at the same time
or in the same state: an eigenstate of one is not necessarily an
eigenstate of the other.  On the contrary, there are quantities that
do not have \emph{any} common eigenstates.  These can therefore
\emph{never} be given precises values simultaneously.  Other
quantities may have some common eigenstates, and may therefore in
some, but not all cases be given precise values simultaneously.  I
will call all such sets of quantities \emph{incommensurable}.  If two
quantities always can be simultaneously given precise values, i.e., if
all eigenstates $|n\ket$ are common, it follows that
\begin{align*}
    FG\mi\Psi\ket & = \sum_{n} FG\alpha_n\mi n\ket 
        = \sum_{n} \alpha_nFg_n\mi n\ket \\
             & =  \sum_{n} \alpha_nf_ng_n\mi n\ket 
        = \sum_{n} \alpha_nGF\mi n\ket \\
             & =  GF \mi\Psi\ket
\end{align*}
for all states $\mi\Psi\ket$.  This is a necessary and sufficient
condition for the quantities being compatible.  This implies that
\[     [F,G] \stackrel{\rm def}{=} FG-GF = 0\,. \]
The operator $[F,G]$ is called the \emph{commutator} of $F$ and $G$.
If it is not equal to zero, the two quantities are incompatible.

We may often express the commutator of two operators in terms of known
operators, with no reference to any choice of representation.  Such
commutation relations form the most general starting point for
describing a system: most of the physics of the system may be
expressed in terms of commutation relations.  For example, the
dynamics of quantum mechanical systems may be expressed as commutation
relations between known quantities and their derivatives.  We may also
say that a system is given by the commutation relations between the
operators characterising the system.  They define the `shape' of state
space, i.e., which kinds of states are allowed or forbidden.  It is often
also possible to derive information about possible eigenvalues more or
less directly from the commutation relations.\footnote{Textbooks in
  quantum mechanics often describe how this may be done for angular
  momentum operators.}

The expression for the commutator [F,G] can tell us about the
distribution of eigenvalues of G in a state with a given distribution
of eigenvalues of F.  This gives us \emph{indeterminacy relations} ---
lower limits for how precisely it is possible to simultaneously
determine the values of incompatible quantities.  The general form is
that if
\[     [F,G] = i\hbar K \quad\text{where $K$ is a real operator,} \]
then
\[     \Delta F \Delta G \geq \frac{\hbar}{2}|\bra K\ket|  \]
where $\Delta F$ ($\Delta G$) is the spread (indeterminacy) in the
eigenvalues of $F$ ($G$) in the given state.\footnote{More precisely: 
\( \Delta F = \sqrt{\bra (F - \bra F \ket)^2 \ket} \).}
For the momentum and position of the same particle, $[x,p]=-i\hbar$
and we obtain
\[     \Delta x \Delta p \geq \frac{\hbar}{2}  \]
This is the original Heisenberg indeterminacy relation.

In quantum field theory we no longer talk about the positions of the
particles as operators, as is the case in `traditional' quantum
mechanics.  Since the number of particles can vary, it is difficult to
make such a quantity well-defined.  Moreover, it would violate the
requirement of relativistic invariance, since time would have to be a
parameter: if any talk of the position of a particle as a measurable
quantity (or of measurable quantities in general) is to be meaningful,
we must assume that we are talking of the position (or energy or
angular momentum etc) \emph{at some time $t$}.  A measurement will
necessarily have to take place at some specific time (or, rather, over
a particular period of time).  This approach is however obviously not
satisfactory in a relativistic theory, where time and space
coordinates should be treated on an equal footing.\footnote{One
  possible way out would be to make both the time and space
  coordinates of a particle depend on a proper time parameter.  This
  is the starting point for Feynman's path integral formalism, which I
  shall look at in section~\ref{sec:pathint}.}

This problem is solved  by also making the spatial coordinates into
parameters.  A \emph{quantum field} is just like a classical field,
except that the `values' at each point are not numbers, but
operators; i.e., the system is characterised by a set of \emph{field
  operators} which are defined everywhere in space-time.  The
operators need not be real --- most of the fields are indeed complex.
The Lagrangian density and the field equations now describe relations
between operators.  On top of this, there are commutation relations
between the fields, which give rise to typical quantum mechanical
indeterminacies or fluctuations.

An important requirement for the field operators is
\emph{microcausality}: if no signal can travel between two points, ie
if the interval between them is space-like, the fields at these points
cannot depend on each other in any way --- the fields must be
separated, and all measurable quantities that can be constructed from
the field operators at the two points must be compatible.  A
measurement at one point cannot have any impact on a measurement at
the other point (unless the system was initially prepared in a state
where the values at the two points are correlated).  In mathematical
language,
\[      (x-y)^2 < 0 \Rightarrow [A(x),B(y)] = 0\,,   \]
where $(x-y)^2$ is the relativistic interval between the points $x$
and $y$, and \(A(x), B(y)\) are measurable (real) quantities
constructed from the field operators.

Only one of the fields has a classical analogue: the electromagnetic
field.  Every field corresponds to a particular type of particle, and
from the field operators we may construct operators for
energy-momentum density, charge density, etc, for this particle
species.  Particles appear as states with well-defined values of mass,
charge and other quantum numbers.  There is however no reference to
the position of an individual particle.  This is, as we will see, as
it should be.

\subsection{Transformations and symmetries}
\label{sec:transf}

A \emph{transformation} (of the most general kind) in quantum
mechanics is a change in representation,
\[     \mi\Psi\ket = \sum_{n} \alpha_n\mi f_n\ket 
        = \sum_{m} \beta_m\mi g_m\ket\,.  \]
The transformation can be characterised by a transformation matrix 
$S_{mn}$,\footnote{This matrix is \emph{unitary}, which means that the
  inverse transformation $(S^{-1})_{mn} = S^*_{nm} =
  (S^{\dag})_{mn}$.} which relates the expansion coefficients to each other,
\[     \beta_m = \sum_{n} S_{mn}\alpha_n 
        \quad\text{with}\quad
       S_{mn} = \bra g_m \mi f_n\ket\,.   \]

A very important class of transformations is \emph{coordinate
  transformations}, where we consider the system in a coordinate
system which is rotated, shifted or transformed in some other way
relative to the original coordinates.  The classical symmetry
transformations are examples of these.  They are a subset of a larger
group of transformations where the operators that form the basis of
the representations have the same eigenvalue spectrum.  Hence,
coordinate transformations in quantum field theory can also be
included, even though the space coordinates are not operators: the real
operators constructed from the field operators $\Phi(x')$ have the
same eigenvalues as those constructed from $\Phi(x)$.  All such
transformations may be viewed in two ways:

\begin{enumerate}
\item \emph{Passive:} The state (system) is unchanged, but is viewed
  from a different perspective.  We may express the operator $G$,
  which forms a basis for the new representation, in terms of $F$,
  which forms a basis for the old one.  For example, $F=x, \;\;\;\;\; G=x+a$
--- translation.  The expression (matrix element) of all operators is
changed to $A'_{mn} = \sum_{k,l} S_{mk}A_{kl}(S^{-1})_{kn}$.

\item \emph{Active:} We transform the state, e.g. $\psi(x) \to
\psi(x+a)$.  In this case, the transformation matrix may be considered
an operator in state space,
\[     S\Psi = S \sum_{n} \alpha_n\mi n\ket = \Psi' 
        = \sum_{n} \beta_n\mi n\ket\,. \]
This operator is also well-defined independently of the
representation.  An arbitrary operator will in general not have the
same effect on the transformed state as on the original one, but we
can define a \emph{transformed operator} $A'=SAS^{-1}$ which has the
same matrix elements.  The transformed operator is the same as the
original one if it commutes with the transformation operator, $[A,S]=0$.
\end{enumerate}

We often consider a group of similar transformations (e.g., the group
of all possible space translations) together.  We may then express the
transformation in terms of one or more parameters $\alpha$, e.g.
$S(\alpha): x\to x+\alpha$.  The combination of two such
transformations will then be another transformation of the same kind: 
\( S(\alpha)S(\beta)=S(\gamma)\,. \)  The parameters may take a
discrete or a continuous set of values.  In the latter case we may
always write
\[     S(\alfa) = e^{i\alpha G/\hbar}\, \]
where $G$ is a real operator which is called the \emph{generator} of
the transformation.

One transformation that often has an exceptional status is time
evolution.  Here, the bases of the two representations are formed by
operators for the same quantities at different times.\footnote{This may be
  made relativistically invariant by considering field operators on
  spacelike hypersurfaces (foliations) rather than at one point in
  time.}  This is a continuous transformation which is generated by
the energy operator (the Hamiltonian) $H$ (the transformation operator
is now often called $U^{-1}$ instead of $S$),
\[     U(t,t_0) = e^{-i(t-t_0)H/\hbar}  \]

As before, this may be pictured in two ways.  If we choose the
`active' picture --- that the operators remain the same (we measure
the same quantities), but the states change --- we arrive at the
\emph{Schr{\"o}dinger equation}\footnote{
\(     \Psi(t) = U(t,t_0)\Psi(t_0) = e^{-i(t-t_0)H/\hbar}\Psi(t_0) 
     \Rightarrow d\Psi(t)/dt = (-iH/\hbar) e^{-i(t-t_0)H/\hbar}\Psi(t_0)  
                = (-iH/\hbar) \Psi(t) \\
     \Rightarrow i\hbar d\Psi(t)/dt = H\Psi(t)        \) }
\[     i\hbar \frac{d\Psi(t)}{dt} = H\Psi(t)\,.  \]

The `passive' picture --- we are dealing with the same state, but the
operators change with time --- gives rise to \emph{Heisenberg's equation
  of motion},\footnote{
\(     A(t) = U^{-1}(t,t_0)A(t_0)U(t,t_0) 
            = e^{i(t-t_0)H/\hbar}A(t_0)e^{-i(t-t_0)H/\hbar} 
   \Rightarrow\\ 
        dA(t)/dt 
            = (iH/\hbar) e^{i(t-t_0)H/\hbar}A(t_0)e^{-i(t-t_0)H/\hbar}
             - e^{i(t-t_0)H/\hbar}A(t_0)e^{-i(t-t_0)H/\hbar} (iH/\hbar)\\
   = (i/\hbar) (HA(t)-A(t)H)  \Rightarrow i\hbar dA(t)/dt = [A(t),H]  \) }
\[     i\hbar\frac{dA}{dt} = [A,H] \,.   \]

This has an important corollary: any quantity that commutes with the
Hamiltonian is conserved (does not change with time).  This may also
be easily derived from the Schr\"odinger equation.  Depending on your
point of view, the Schr\"odinger equation or Heisenberg's equation of
motion may be considered the equations of motion for the system.

If the transformed system is physically the same as the untransformed
one, we call $S$ a \emph{symmetry transformation}.  It is not always
clear what is meant by `physically the same'; usually we require that
the equations of motion look the same, so that the same boundary
conditions will lead to the same evolution.  In field theory this may
be straightforwardly expressed by the Lagrangian or the action being
the same.  We see that the quantum mechanical equations of motion are
the same if the expression for $H$ is the same and hence $[S,H]=0$.
Since $H$ is derived from $\mathcal{L}$, this does not introduce
anything new, but it shows more clearly the relation between
symmetries and conserved quantitites.  If
\[     S(\alpha) = e^{i\alpha G/\hbar} \quad\text{and}\quad
        [S(\alpha),H] = 0 \quad\text{for all } \alpha\,,  \]
we also have $[G,H] = 0$ -- and $G$ is conserved.

If we compare with what we know about classical symmetries, we find
that space translations are generated by the total momentum, while
rotations are generated by the angular momentum.  In fact, a general
consideration of rotation transformations will show that we have to
take into account not only the classical orbital angular momentum, but
also the spin, which has no classical analogy.

Another important class of transformations is \emph{gauge
  transformations}, where the fields are changed without changing the
coordinates.  The simplest gauge transformation is the global
\[     \Phi(x) \pil e^{i \chi}\Phi(x)\,,  \]
where $\Phi$ is a charged field.  This transformation turns out to be
related to charge conservation.  If we now allow the parameter $\chi$
to depend on $x$ (a \emph{local} gauge transformation), it turns out
that if the Lagrangian is to be unaffected by the transformation, we
are forced to introduce an additional field $A_\mu$, which couples to
$\Phi$ and transforms together with $\Psi$.  The equations of motion
of this field and its interactions with the charges turn out to be
identical to those of the electromagnetic field.  If we postulate that
$\mathcal{L}$ is invariant under the local gauge transformation above,
then all of electromagnetism follows!

We may also construct more complicated gauge transformations of the kind
\[      \Phi_{\alpha}(x) \to
           \sum_{\beta} a_{\alpha\beta}(x) \Phi_{\beta}(x)\,,      \]
where the coefficients $a_{\alpha\beta}$ should satisfy certain
requirements, and where the different field indices may well represent
different kinds of particles.  A similar mechanism to the one above
leads to non-abelian gauge theories (non-abelian refers to the
mathematical properties of the coefficients $a_{\alpha\beta}$), which
form the basis of theories of the strong and weak interactions.

The most important discrete transformations are \emph{reflection}
(parity transformation, represented by $P$\/), \emph{time reversion}
($T$\/) and \emph{particle--antiparticle exchange} or \emph{charge
  conjugation} ($C$\/).  The first of these also has a classical
description and use (it was first introduced by Kant), but becomes
really important only in quantum mechanics.  When a system is
reflected twice, it is obvious that we end up with the same system as
we started with.  Hence, $P^2=1$, and $P$ has eigenvalues
$\pm1$.\footnote{The operator $P$ is real, and $\psi(-x)=\pm\psi(x)$
  if $\psi$ is an eigenstate of $P$.}  Furthermore, if $P$ is a
symmetry transformation (i.e., there is no difference between left
and right in this system), then $P$ is conserved --- a symmetric state
can never turn into an antisymmetric state, and vice-versa.  We can
therefore classify states (and particles, in quantum field theory!) as
symmetric or antisymmetric under reflection.  Moreover, because of the
superposition principle, any state can be written as the sum of a
symmetric and an antisymmetric state, which will evolve separately.
We see that it is possible to find out quite a lot about which kinds
of processes are possible or impossible by assuming that there is no
difference between left and right.

Time reversal has a somewhat special status.  It is easy to imagine
how a system can be reflected, or how we can exchange particles and
antiparticles, but it is more difficult to imagine what is meant by
reversing time.\footnote{This problem is encountered in Feynman's path
  integral formulation, where particles can go backwards in time.  The
  question of the arrow of time --- the difference between past and
  future --- does of course have a number of other aspects, both
  physical and philosophical, than those that are relevant in
  particle physics.  Stephen Hawking~\cite{Hawking} has a popular
  exposition of these questions.}  This operation is however
well-defined in quantum field theory, and can be used to classify
states and processes.  It is difficult --- but not impossible --- to
test time reversal symmetry experimentally.

To describe charge conjugation, it is necessary to employ the
apparatus of quantum field theory, but there it has a status
completely equivalent to the two other discrete transformations.  All
three are very similar in that $P^2=T^2=C^2=1.$  It was a big surprise
when it was discovered that none of them are exact symmetries: they
are all violated in weak interactions (as described on
page~\pageref{CP}).  However, this does not violate any principle of
quantum field theory.  On the other hand, it has been proven that in
any Lorentz invariant, local quantum field theory obeying the
principle of microcausality, their combination must be an absolute
symmetry.  This \emph{CPT theorem} is one of the few real results from
axiomatic field theory.  No violations of this symmetry have so far
been observed.

\subsubsection{Identical particles}
\label{sec:ident}

In quantum mechanics, we often consider systems consisting of a number
of identical particles (particles without any individuating
characteristics, e.g.\ electrons in an atom).  In such a system, it is
possible to define an operation where all coordinates or quantum
numbers of two (or more) particles are exchanged, or (equivalently)
two particles are exchanged.  If the particles (which we may call
`particle 1' and `particle 2') really are identical, it will not be
possible to distinguish the state $\Psi(1,2)$ `before' and $\Psi(2,1)$
`after' this operation --- for example, all operators must have the
same expectation value in both states.  In that case, $\Psi(1,2)$ must
be a constant times $\Psi(2,1)$, and since we obviously must get the
same state back when we repeat the operation, this implies that
$\Psi(2,1)=\pm\Psi(1,2)$.\footnote{This has obvious parallels to space
reflection.  Consider the position representation of two identical
particles without any `internal quantum numbers'.  Instead of using
the coordinates $\vec{r}_1, \vec{r}_2$, one may describe the system
using \( \vec R = \frac{1}{2}(\vec r_1+ \vec r_2) \) and \( \vec r = \vec
r_2- \vec r_1 \).  Particle exchange is then equivalent to 
 \( \vec r \to - \vec r \).}

Particles where $\Psi(2,1)=-\Psi(1,2)$ are called \emph{fermions},
while particles where $\Psi(2,1)=\Psi(1,2)$ are called \emph{bosons}.
This can be generalised to systems with more than two particles: the
state of a system will always be symmetric under exchange of two
identical bosons, and antisymmetric under exchange of two identical
fermions.  We see that there are constraints on the state that arise
exclusively out of the fact that the particles are identical.

If the particles do not interact, we may determine (measure) all
quantum numbers for all the particles simultaneously.  For fermions
we find that two particles can never have the same set of quantum
numbers, since this would give $ \Psi(1,2) =
\Psi(2,1) = -\Psi(1,2)$, which is impossible for a physical state.
This is the \emph{Pauli exclusion principle}: two fermions can never
occupy the same (single-particle) state at the same time.

This is a necessary, but not sufficient condition for a fermion.  To
see this, it is sufficient to note that if particle 1 is in the
single-particle state $\Psi_a$ (has quantum numbers $a$), while
particle 2 is in state $\Psi_b$, the total state can be written as
$\Psi(1,2)=\Psi_a(1)\Psi_b(2)$.  This state does not satisfy the
symmetry requirements for either fermions or bosons.

On the other hand, the states
\begin{align*}
     \Psi_-(1,2) & = \frac{1}{\sqrt{2}}
        ( \Psi_a(1)\Psi_b(2) - \Psi_b(1)\Psi_a(2) )  \\
     \Psi_+(1,2) & = \frac{1}{\sqrt{2}}
        ( \Psi_a(1)\Psi_b(2) + \Psi_b(1)\Psi_a(2) )
\end{align*}
satisfy the requirements for fermions and bosons respectively.  If
particles 1 and 2 are fermions, $\Psi_-$ is an allowed state for the
system consisting of the two particles, while $\Psi_+$ is an allowed
state if the two particles are bosons.  But this means that the
particles do not take on their quantum numbers independently: they are
non-interacting (at least not interacting in the usual sense), but not
independent.  The states cannot be separated into one part that only
concerns one of the particles and another part that only concerns the
other; a change in the state of one will also be a change in the state
of the other.  This gives rise to observable effects: the expectation
value of a quantity $A$ in these states will contain an `interference
term'
\[     \mp {\rm Re} (\Psi_a\Psi_b,A\Psi_b\Psi_a)\,,   \]
which is a manifestation of the system `being' in both states
$\Psi_a\Psi_b$ and $\Psi_b\Psi_a$.  If $A$ is the product of two
quantities, each pertaining to one of the particles, $A =
A_1(1)A_2(2)$ (e.g., the spin of particle 1 in the $z$ direction and
the spin of particle 2 in a different direction), then $\bra A\ket$
expresses a correlation between the values of $A_1$ for particle 1 and
$A_2$ for particle 2.  This correlation will also contain such a
`strange' term.  The positions of the particles is also correlated in
this way, with the result that \emph{bosons like each other, while
fermions detest each other}: the probability of finding two fermions
close to each other is small, while it is considerably larger than
`expected' for bosons.

We may also note that this is the case regardless of how far apart the
two particles are --- the symmetry of the state is independent of
distances (although the size of the correlation term decreases with
distance).  This means that we are in fact correlated with something
in the Andromeda galaxy (or behind the Moon, if we want to be somewhat
more down-to-earth).  This and similar phenomena are part of the
essence of Bell's theorem, which states that local realistic theories
are incompatible with quantum mechanics.

An important result in quantum statistics is Wigner's `summation
formula', which states that where a collection of several fermions can
be considered an indivisible unit (as e.g.\ in an atomic nucleus), it
behaves like a boson if it consists of an even number of fermions, and
like a fermion if it consists of an odd number.  This can be seen
directly by noting that exchange of two units each consisting of an
even number of fermions corresponds to an even number of fermion
exchanges, which gives an even number of sign changes, and hence no
net change in the state.  Conversely, for units consisting of an odd
number, we get an overall minus sign.  Alternatively we may use the
fact (which originally was an empirical discovery) that all particles
with a half-integer spin are fermions, while all particles with
integer spin are bosons.  Using the rules for addition of angular
momentum, we find that aggregates of an even number of particles with
half-integer spin have even `internal angular momentum' (which in this
context may be taken to be the spin of the aggregate), while an odd
number gives a half-integer `internal angular momentum'.  (The orbital
angular momentum is always integer, so it actually makes no difference
to this argument.)

In quantum field theory, all these results follow directly from the
basic principles, without any further assumptions.  The field operator
$\Phi$ can be written as a (linear) combination\footnote{The field is
normally expanded in Fourier modes, although expansions in other basis
functions could also be considered.} of a set of operators  $a(p)$,
$a^{\dag}(p)$ --- one for each possible value of a parameter
$\vec{p}$.  When an energy--momentum density and an energy operator $H$
are constructed from the Lagrangian density, and the operators are
required to satisfy the Heisenberg equation of motion,
\[     
        i\hbar \frac{\partial\Phi}{\partial t} = [\Phi,H]   
\]
(or other, equivalent quantisation conditions), this can be
reformulated in terms of commutation relations between the $a$ and
$a^{\dag}$ operators, so that they appear as \emph{annihilation
operators} and \emph{creation operators} respectively, as follows.

With a definition of the vacuum $\mi0\ket$ and the physical operators
for total energy, $H$, and total momentum, $\vec{P}$, so that
$H\mi0\ket=0, \vec{P}\mi0\ket=0$, $a^{\dag}(p)\mi0\ket$ will become
eigenstates of $H$ and $\vec{P}$ with eigenvalues
$E(\vec{p})=\sqrt{\vec{p}^2c^2+m^2c^4}$ and
$\vec p$ respectively.  The total energy and momentum will in general
be written as
\[
       H = \int \!d^3\!p E(\vec p)\,a^{\dag}(p)a(p) \,,
        \quad
        \vec P = \int \!\!d^3\!p\, \vec p\: a^{\dag}(p)a(p) \,.
\]
\label{particles}
This may be taken to define a \emph{number operator}
$n(p)=a^\dag(p)a(p)$ which represents the number of particles with
momentum $\vec{p}$ and energy $E(\vec{p})$, where the state $|mi
p\ket=a^{\dag}(p)$ is a particle: $a^{\dag}(p)$ creates a particle
with a definite energy and momentum.  We also find that
$a(p)\mi p\ket = \mi0\ket$: $a(p)$ destroys a particle.  We may also
define a \emph{total number operator}
\[     N = \int \!d^3\!p \:a^{\dag}(p)a(p)\,,   \]
which has integer eigenvalues.  An eigenstate of $N$ with eigenvalue
$n$ is then $n$ particles.  These particles need not have definite
energy and momentum, but can be combinations of states with different
energies and momenta.

This formalism makes it impossible to maintain any form of
individuality for the particles.  Considering the states as particles
(or superpositions of particles) is only one possible representation.
States that are superpositions of states with different particle
numbers are not only allowed, but may also be physically relevant.
For example, the eigenstates of electric and magnetic fields $\vec{E}$
and $\vec{B}$ have indeterminate particle numbers --- electromagnetic
field strength and photon number are incompatible quantities.  And
in cases where there is a definite particle number, it is only in
certain limits that they can be individually identified, for example
by their spatial separation.  The state does not give the individual
particle any special status or `personality', and above all the
formalism ensures that particles of the same kind are completely
identical -- they all emerge from the same field.  The states are
states of the fields, and are not assigned to individual particles,
but at most to a particle number.

The distinction between fermions and bosons appears in how the fields
are quantised.  The requirements of microcausality and the existence
of a lowest energy level imply that for a field with integer spin
(scalar, vector or tensor field) we must (essentially)
have\footnote{Additional indices on the creation operators referring
to e.g.\ spin orientation or polarisation give additional
$\delta_{mn}$-type prefactors.}
\[
     [a(p),a(p')] = [a^{\dag}(p),a^{\dag}(p')] = 0 \,,
\]
while for fields with half-integer spin (spinor fields) we must have
\[
       \{a(p),a(p')\} = \{a^{\dag}(p),a^{\dag}(p')\} = 0  
        \qquad \text{where}\qquad
       \{a,b\} \stackrel{def}{=} ab+ba
\]
The former gives rise to Bose--Einstein statistics (bosons), while the
latter gives rise to Fermi--Dirac statistics (fermions).  In
particular we see that the Pauli principle is satisfied: if we try to
create a state consisting of two particles with the same energy and
momentum, we get
\[
       (a^{\dag}(p))^2 \mi\Psi\ket = 
        \frac{1}{2} \{a^{\dag}(p),a^{\dag}(p)\} \mi\Psi\ket = 0  
\]
This \emph{spin--statistics theorem} has been proved on general
grounds within the framework of axiomatic field theory.

Charged particles (or particles that have an antiparticle) are
represented by a field $\Phi$ that is not real.  It has an associated
conjugate field $\Phi^{\dag}$ and two sets of creation and
annihilation operators: $a, a^{\dag}$ annihilate and create particles,
while $b, b^{\dag}$ annihilate and create antiparticles (particles
with opposite charge).  The field $\Phi$ can be written in terms of
the operators $a$ and $b^\dag$, and will hence destroy a particle or
create an antiparticle.  $\Phi^{\dag}$ contains $b$ and $a^{\dag}$
operators, and will therefore create a particle or destroy an
antiparticle.

\subsubsection{Interactions}
\label{sec:interactions}

If two subsystems do not interact, the energy of the total system can
be written as a sum of two terms, each of which only depends on one of
the subsystems, and the subsystems will evolve independently.  The
interaction is written as a term that depends on both subsystems, and
which cannot be separated.  The simplest example is perhaps the
interaction between two massive bodies in classical (Newtonian)
theory,
\[     E = \frac{1}{2} m_1v_1^2 + \frac{1}{2} m_2v_2^2 - 
        G\frac{m_1m_2}{|\vec r_1 - \vec r_2|}\,.    \]

Here we can see where the idea of fields comes from: this may
alternatively be viewed as body 1 establishing a field around it,
which is proportional to the mass $m_1$ and inversely proportional to
the distance from the body.  The second body will then have an
additional energy due to a coupling to this field, which depends on
the mass of the body and the strength of the field.  If we allow the
field to spread out with a finite speed from one body to the other, we
see that the energy will also depend on whether the field has yet
reached the other body.  We may also allow the field to have its own
energy, which depends only on the field strength.

This is the situation in classical, dualist theories --- the typical
(and only?) example is classical electrodynamics.  Here, both the
charged particles and the field have their own intrinsic motion and
their own energy; the coupling between them is added to this, and
gives rise to mutual modifications.

All of this can be rewritten in terms of a Lagrange theory, where the
interactions appear as additions to the `free' Lagrangians and
Lagrangian densities.  In a non-dualist field theory the particles are
viewed merely as `lumps' of fields; each field can be considered a
subsystem; and the interaction is manifested through a local
additional term in the Lagrangian density,
\[     {\cal L}(\Phi_1,\Phi_2) = {\cal L}_1(\Phi_1) + {\cal L}_2(\Phi_2) 
        + {\cal L}_{i}(\Phi_1,\Phi_2)    \]

In my treatment of particles in sec.~\ref{sec:ident} I ignored their
interaction.  It turns out that interactions influence both the
stability of particles and to what extent it makes sense to talk about
particles.  The particles in sec.~\ref{sec:ident} are states of the
system of free fields, and move freely.  When the fields or particles
interact, the picture becomes less clear.  Heisenberg's equation of
motion (or whatever we use as our quantisation condition) looks
different, and the commutation relations are changed.  It can be
difficult to recognise the operators, and the energy eigenstates are not
necessarily eigenstates of the number operators --- in short, the
states have different properties.  Hence, the particles lose even more
of their identity.  And obviously the time evolution is different
(otherwise there would be no point in talking about an interaction).

The situation is however not quite as hopeless as it might seem.  It
is possible to choose a representation of the total state of the
system --- called the \emph{interaction picture} --- such that the
(expression for) the states had been time-independent \emph{if the
interaction were zero}, and the field commutation relations are the
same as for the free fields, whose properties we know.  This is in
other words a third way of viewing the time evolution transformation,
in addition to the Heisenberg and Schr\"odinger pictures described in
section~\ref{sec:transf}.  In this way we can start from the solutions
of a known problem when solving the more complicated interaction
problem. 

The interaction problem can be divided into two main categories:
\begin{enumerate}
\item \emph{Bound states.}  Here we want to find stable states with a
definite energy (which is lower than the energy of free particles),
with a limited and constant spatial extent (i.e.\, the region where
the average energy density is significantly different from zero is
finite and constant).  Examples of such states are atoms (bound states
of a nucleus and one or more electrons) and hadrons (bound states of
quarks and gluons).  There is nothing in the conceptual framework of
quantum field theory which should make it impossible in principle to
compute this; it is however a fact that nobody has succeeded in
developing rigorous methods to tackle these problems.\footnote{\emph{Note
added in translation:} This is evidently not the case, but when I
wrote this, I was only vaguely aware of Bethe--Salpeter equations
(whose rigour in practical applications can be disputed as they rely
on approximations or truncations), and even less aware of lattice
gauge theory, which has since been my main area of research.}  What is
normally done is to transform the problem to a non-relativistic
potential problem with a definite particle number, which may be solved
using established methods.  Quantum field theory may subsequently be
employed to calculate relativistic corrections, which are hopefully
small.  The range of validity of such an approach cannot be known with
certainty.

The corrections to a known (approximate) solution are computed using
the interaction picture.  The known, bound system is now considered as
a `free' (known) subsystem, and the correction terms are the
`interaction'.  If the correction terms are small, perturbative
methods can be used.

\item \emph{Scattering problems.}
I am using the term scattering in its widest sense: the interaction is
significantly different from zero only in a finite time interval;
outside this interval the subsystems can be considered as free (and
independent).  This is in particular the case where several particles
approach each other, react, and the products of the reaction
subsequently go off in different directions and become separated in
space.  Quantum field theory has been shown to be particularly well
suited to solving such problems.

In the interaction picture we can treat the system as if the fields
were free.  We also know that the states long before and long after
the interaction (the asymptotic states) are free particles, and these
can be taken as the starting point of the calculation.  Furthermore,
we are rarely interested in the details of the interaction, but only
in the probability of transitions between different asymptotic
states.  These probabilities are encoded in the \emph{S matrix}, which
is the asymptotic form of the time evolution operator in the
interaction picture.  With
\begin{align*}
\Psi(t) &= U(t,t_0)\Psi(t_0)\,,       \\
\intertext{we have}
     S &\stackrel{\rm def}{=} 
        \lim_{t\to\infty,t_0\to-\infty} U(t,t_0)\,. 
\end{align*}

We can write down an explicit expression for the S-matrix in terms of
the interaction term $\mathcal{L}_i$ in the Lagrangian density,
\[     S = e^{(i/\hbar) \int\mathcal{L}_{i}(x)d^4x}   \]
It is very difficult to solve this operator equation exactly.  If
$\mathcal{L}_{i}$ may be assumed to be small, the exponential may be
expanded in a power series,
\begin{align*}
  e^x &= 1 + x + x^2/2! + x^3/3! + \ldots \text{\,, i.e.,}  \\
   S &= 1 + (i/\hbar) \int {\cal L}_{i}(x)d^4x
            + \frac{1}{2}(i/\hbar)^2 \int {\cal L}_{i}(x_1)d^4x_1 
                \int {\cal L}_{i}(x_2)d^4x_2 + \ldots\,,
\end{align*}
and assume that the higher order terms give only small contributions
to the S-matrix.\footnote{To be precise, the integrands are also
  required to be time-ordered in the power series.}

As a result, the processes can be viewed as composed of `elementary
interactions' (elementary processes) given by $\mathcal{L}_i$, which
may occur (or do occur) anywhere and at any time, and may occur one or
more times.  Processes consisting of more elementary processes are
less likely than those consisting of fewer.  The total S matrix
element for a transition between two given states is found by adding
up the contribution from all possible (elementary or composite)
processes that take the system from one state to the other.  The
transition probability is (proportional to) the absolute square of the
S matrix element.
\end{enumerate}

At this point it can be appropriate to say something about the form
the interactions may take.  Relativistic invariance requires that all
interaction terms which include fermion (spin-1/2) fields must be of
the form\footnote{Relativistic invariance also allows terms like
\( (\Psi_a^{\dag}(x)\Psi_b(x)) (\Psi_c^{\dag}(x)\Psi_d(x)) \).  These
interactions (which Fermi used in his description of weak
interactions) also conserve the fermion number.  This theory is
however not renormalisable, and such terms are therefore excluded.}
\[
       \mathcal{L}_{i}(x) \sim \Psi_a^{\dag}(x)\Phi(x)\Psi_b(x)  \,,
\]
where $\Psi_a$ and $\Psi_b$ are fermion fields (the letter $\Psi$ is
used for historical reasons) and $\Phi$ is a boson field.  If we now
define the total fermion number as the total number of particles minus
the total number of antifermions, we find that \emph{the total fermion
number is conserved}: $\Psi_b$ must \emph{either} destroy a
particle \emph{or} create an antiparticle, and \emph{at the same time}
$\Psi_a^{\dag}$ must \emph{either} create a particle \emph{or} destroy
an antiparticle.  At the same time, a boson is either created
(emitted) or destroyed (absorbed); the boson appears as a `mediator'
of the process.

In quantum electrodynamics these are the only possible interactions.
In non-abelian gauge theories there are also `self-interactions' where
the boson (gauge) fields couple to each other, in terms of the kind
$\Phi_1\Phi_2\Phi_3$ and $\Phi_1\Phi_2\Phi_3\Phi_4$ (the fields may be
the same or different).

Feynman has invented a method to represent the elementary processes in
terms of simple diagrams, where each element of the diagram is related
to a factor in the S-matrix element.  This also gives us a
visualisation of the processes and simple rules for the calculation.
The diagrams are space--time illustrations with a time axis (which is
usually vertical) and a `space axis' (which is supposed to represent
all three space coordinates, and is usually drawn horizontally).  The
particle states are drawn as lines which are supposed to represent the
`world line' of the particle, i.e., its position in space as a
function of time.

We have the following simple diagram elements:

\setlength{\unitlength}{.05cm}
\begin{picture}(240,40)
  \put(5,20){fermion}
  \put(5,12){(spin-1/2 particle)}
  \put(75,0){\vector(1,2){10}}
  \put(85,20){\line(1,2){10}}
  \put(150,20){antifermion}
  \put(190,40){\vector(1,-2){10}}
  \put(200,20){\line(1,-2){10}}
\end{picture}

\begin{picture}(260,40)
  \put(2,30){vector boson}
  \put(2,20){(photon,}
  \put(3,10){ $W^{\pm},Z^0$)}
  \put(50,0){\includegraphics*[width=2cm]{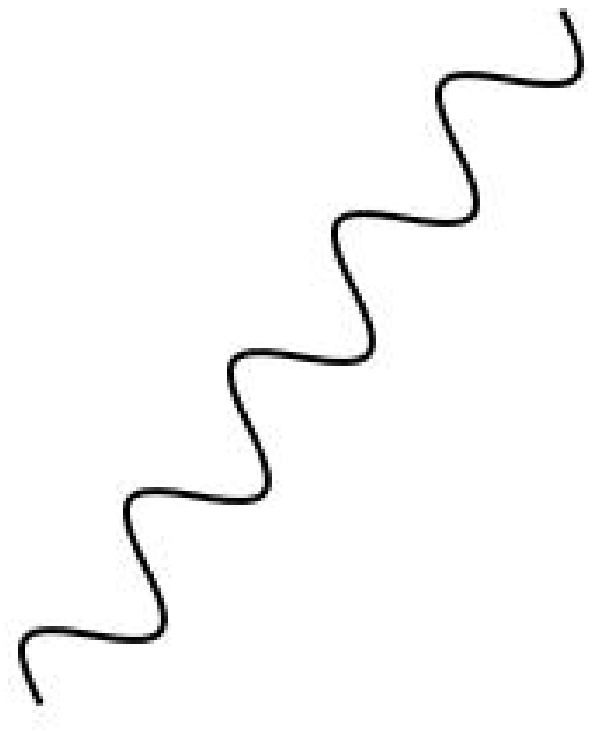}}
  \put(110,20){gluon}
  \put(135,0){\includegraphics*[width=2cm]{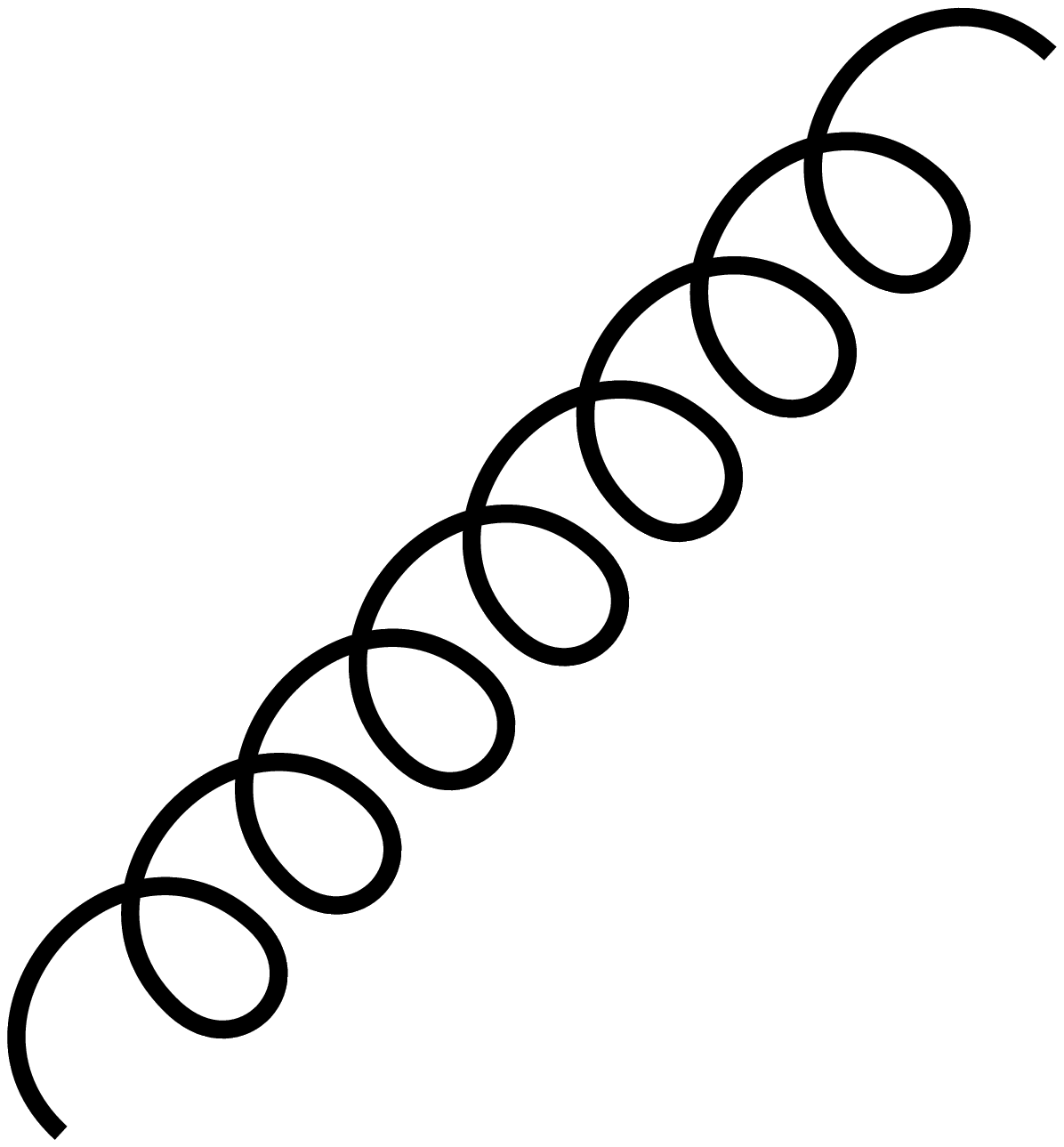}}
  \put(190,25){scalar boson}
  \put(190,15){(Higgs etc.)}
  \dashline{4}(235,40)(260,5)
\end{picture}

\hspace{1em} interactions\\
\includegraphics[width=\textwidth]{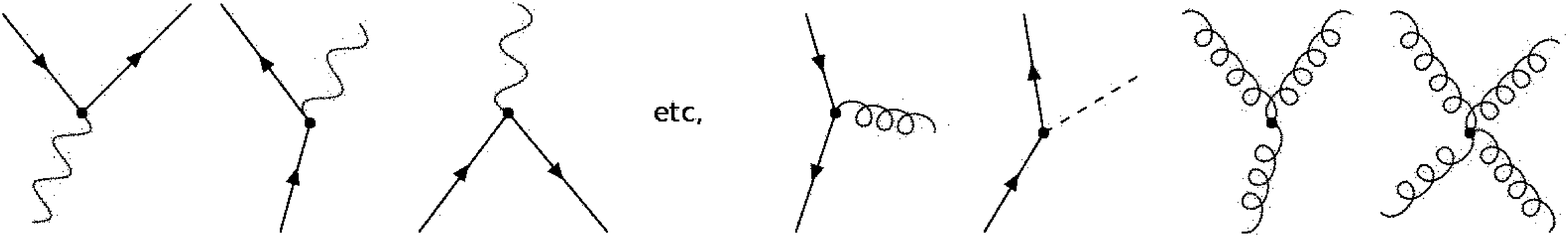}

\begin{figure}[bht]
\begin{center}
\includegraphics*[width=0.36\textwidth]{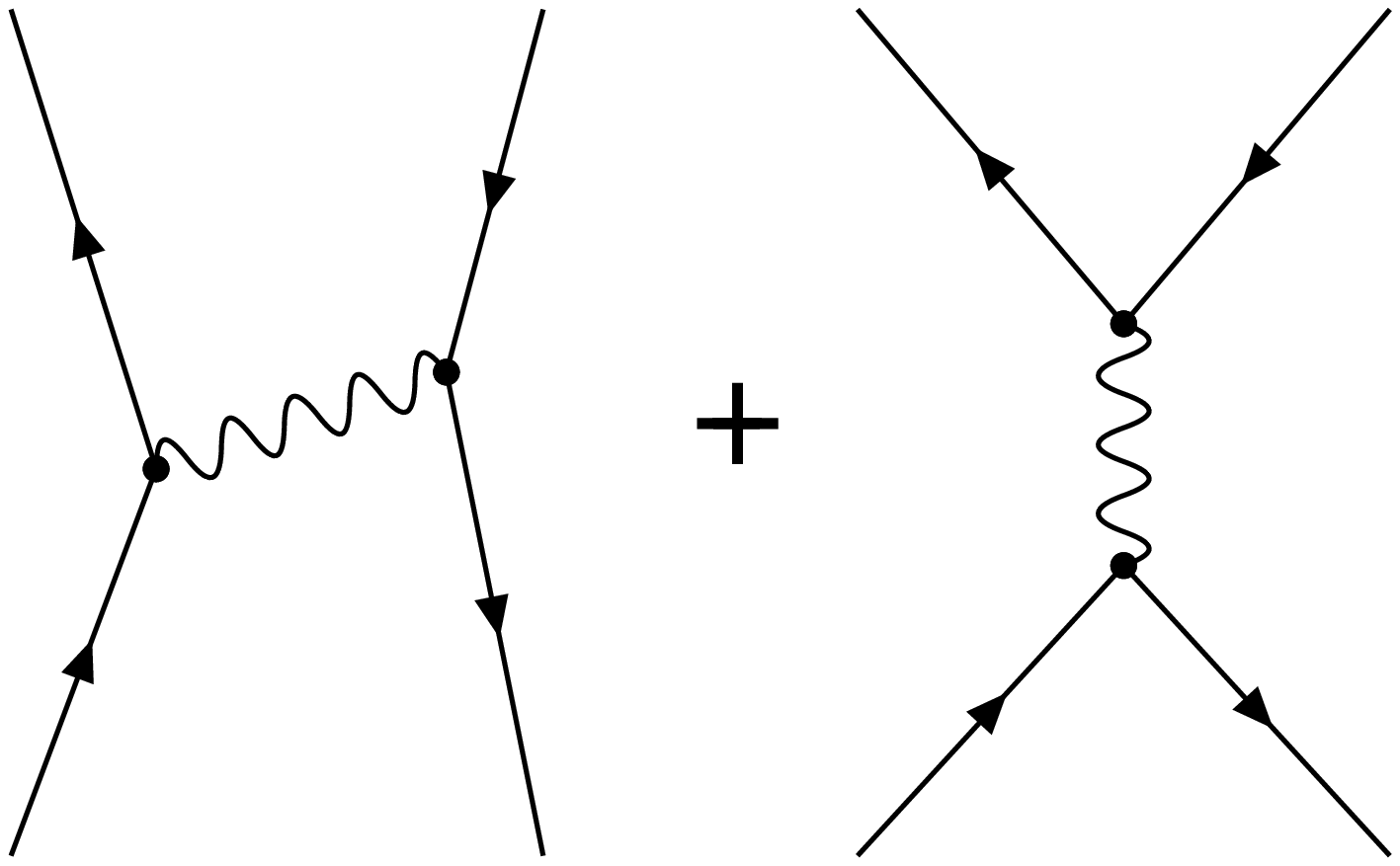}\hspace{0.15\textwidth}
\includegraphics*[width=0.44\textwidth]{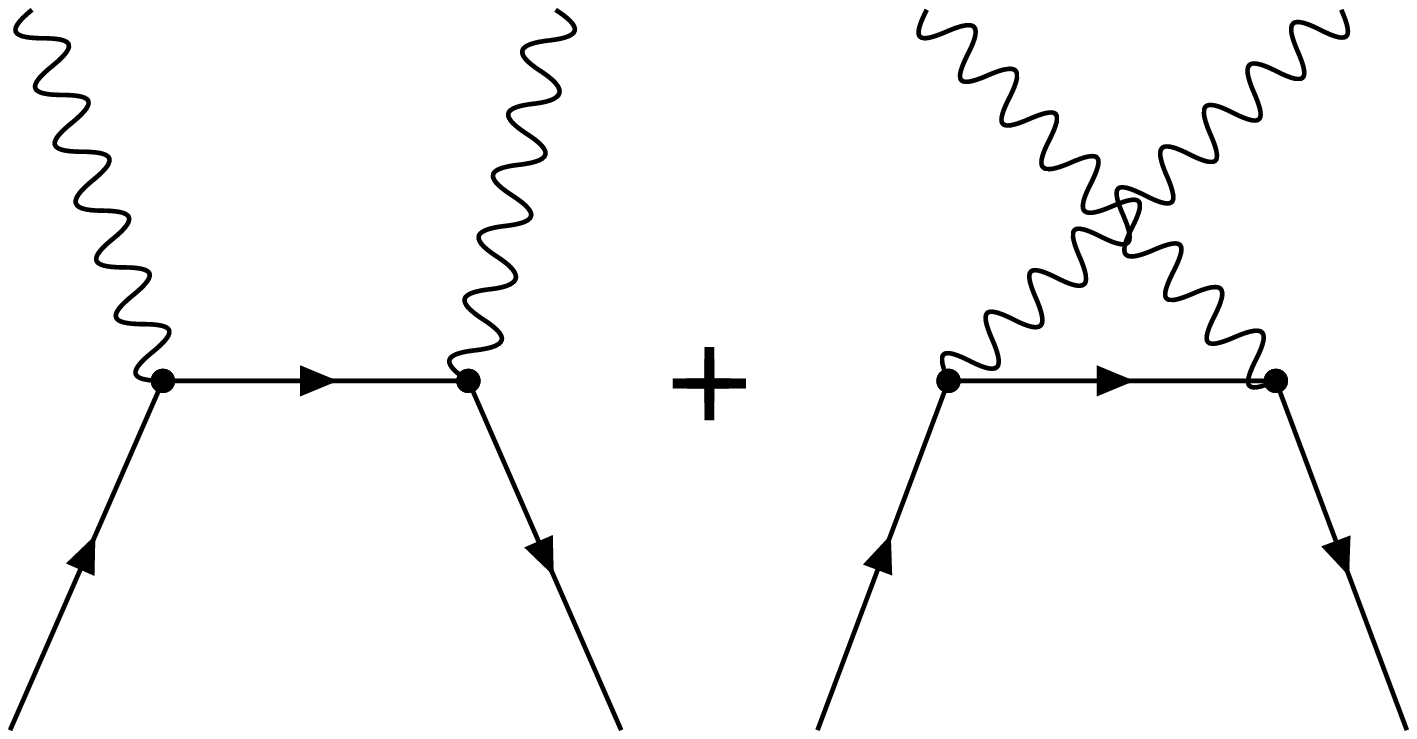}\\[1em]
\includegraphics*[width=0.35\textwidth]{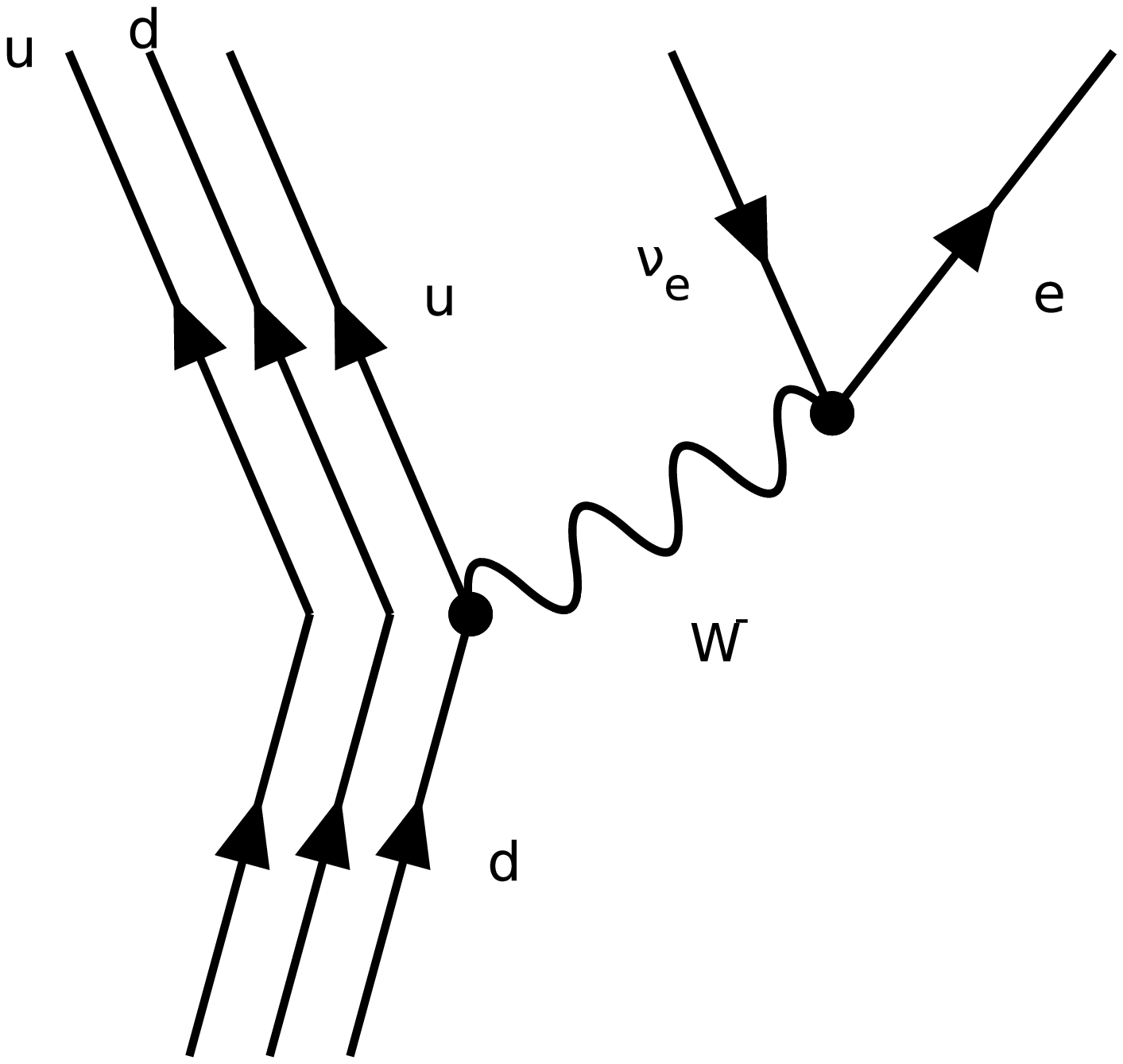}
\end{center}
\caption{Some simple processes. Top left: Bhabha scattering
($\pos\el\to\pos\el$).  Top right: pair annihilation
($\pos\el\to\gam\gam$). Bottom: $\beta$ decay ($n\to
p^+\el\bar{\nu}_e$). \label{processes}}
\end{figure}

Because of energy and momentum conservation, none of the elementary
interactions can take place on its own --- at least two such
interactions are required to have a physical
process.\footnote{\emph{Note added in translation:} This is not
entirely true.  $2\to2$ particle scattering processes mediated by
quartic gauge or scalar interactions are possible.}  Some simple
physical processes are illustrated in figure~\ref{processes}.

\begin{figure}[hbt]
\includegraphics[width=\textwidth]{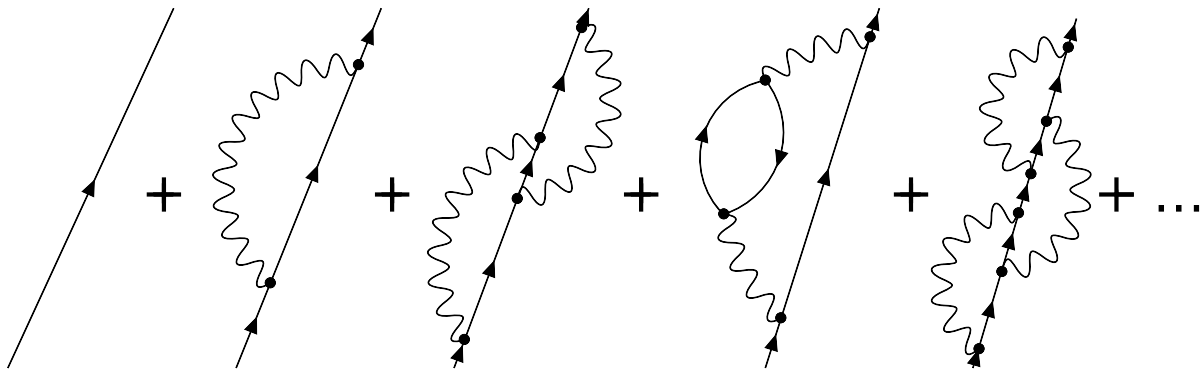}
\caption{The self-interaction of the electron.\label{electron-self-int}}
\end{figure}

A problem that appears with interactions is to define what a free
particle is.  Any particle can at any time interact with itself ---
i.e., we may have processes where one particle comes in and one
particle, of the same type and with the same energy and momentum,
comes out.  An electron will for example from time to time emit a
photon, only to shortly after absorb it again (i.e., we go from a
`pure' electron state to an electron--photon state and back).  In the
same way a neutrino will from time to time be `dissociated' into an
electron and a W-boson, while a photon will be `dissociated' into a
fermion--antifermion pair.  There is no way of seeing from a particle
whether the system has `visited' one or more such states, or how often
this has happened: the particles have no memory.  On the other hand,
it has definite consequences for the propagation and properties of the
particles.  To obtain the properties of the electron we can observe,
we must add up the contributions from all possible self-interaction
processes, as in figure~\ref{electron-self-int},
and replace the mass of the `bare' electron with the mass of the
`dressed' electron in the figure.  Similarly, the elementary charge is
changed by `dressing' the photon.  This is
called \emph{renormalisation}.  We find that the observed charge and
particle masses are different from the quantities appearing in the
Lagrangian, which would have been the charges and masses of the
particles in the absence of interactions (in which case we would not
have any possibility of observing them either).  This difference is in
fact infinite!

\subsubsection{The path integral formalism}
\label{sec:pathint}

This overview would not be complete if it did not mention Feynman's
path integral formalism.  Strictly speaking, this formalism does not
belong conceptually to quantum \emph{field} theory,\footnote{\emph{Note
added in translation:} At the time when I wrote this, I had only been
introduced to variants of canonical quantisation of quantum fields,
and not to the functional integral formalism, which forms the basis
for modern treatment of quantum field theory, and is the field
theoretical extension of Feynman's path integrals.} and I have
therefore postponed it until now, but it is mathematically
equivalent.  I will be brief; Feynman himself has an presentation of
the theory intended for the uninitiated in \cite{Feynman:QED}.  His
articles \cite{Feynman:1948pathint,Feynman:1949qed} where the formalism was first
presented are also readable for an `ordinary' physicist.

The formalism takes as its fundamental starting point the particles
and their paths in time and space, and notes that in quantum mechanics
it is in principle impossible to know the path of a particle in
detail.  Hence, according to Feynman, to compute the probability of a
particle going from one point to another, we must add up the
probability amplitudes for all the possible ways this can happen ---
i.e., all possible paths for the particle.  The probability is then
obtained by taking the absolute square of this sum.  Furthermore, all
paths are equally probable, but they have different phases: the
amplitude for each path is $e^{iS/\hbar}$, where $S$ is obtained from
the classical Lagrangian.  This is the situation when the particle is
alone in the world. moves through empty space and has no
interactions.  The resulting amplitude $K(1,2)$ for going from point 1
to point 2, called the \emph{propagator}, will be dominated by the
contributions from paths close to the classical path, and can be
considered a `primitive' element of the theory.

If we are dealing with two identical particles, which still do not
interact, we have to take into account that we cannot know which one
of the two we observe when each has gone from one point to another.
Hence we have two possibilities (the lines now represent the integral
of all possible paths from one point to the other):\\
\setlength{\unitlength}{.05cm}
\begin{picture}(250,80)
\put(60,10){\line(0,1){50}}
\put(100,10){\line(0,1){50}}
\put(135,30){and}
\put(180,10){\line(4,5){40}}
\put(220,10){\line(-4,5){40}}
\end{picture}

For bosons we must add up the amplitudes for these two processes,
while for fermions they must be subtracted.  In this way we obtain the
effect that bosons like each other, while fermions detest each other.
For example we see that the probability of two fermions with the same
spin ending up at the same point is zero, while for bosons it is
larger than `expected'.

When there are interactions in the system we must also take into
account all the possible ways the interactions can occur.  In
relativistic theory in the Feynman picture, all interactions happen by
emission and absorption (creation and annihilation) of bosons
mediating the interaction.\footnote{It is also possible to include
potential scattering in this formalism, but the potentials are static
and do therefore not strictly speaking belong in a relativistic
theory.}   These emissions and absorptions can occur anywhere in space
and time, and in any order, and we must therefore add up the
amplitudes for all these possibilities.  The result is exactly the
same as in the power series expansion of the S-matrix, and the
diagrammatic representations of the processes are of course intended by
Feynman to represent the motion of the particles in time and space.

For charged particles we will also have to include not only paths
where the particle moves forward in time, but also ones where they go
backward in time.  Such paths can be observed as antiparticles.

\begin{figure}[bht]
\includegraphics[width=0.36\textwidth]{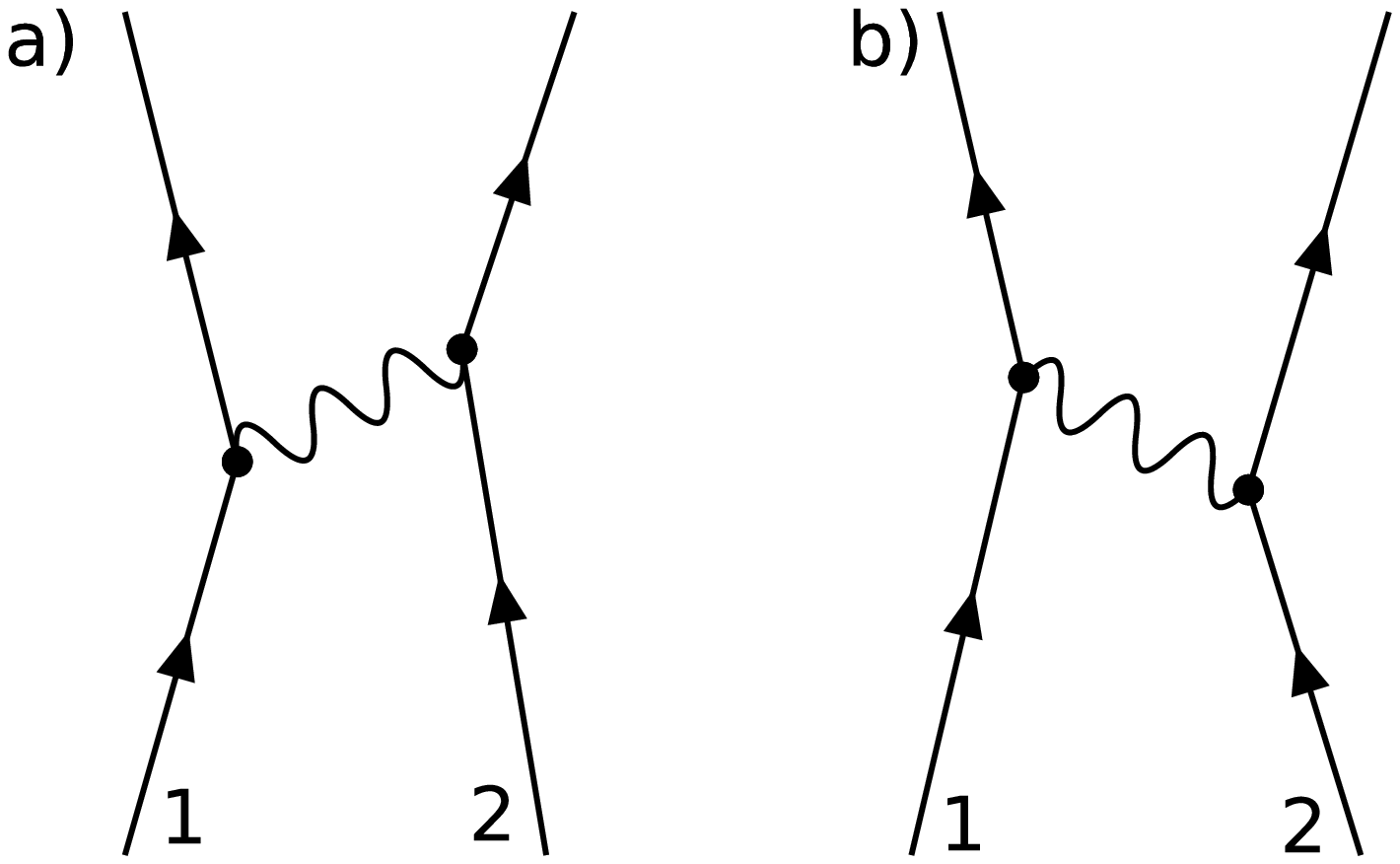}\hspace{0.08\textwidth}
\includegraphics[width=0.54\textwidth]{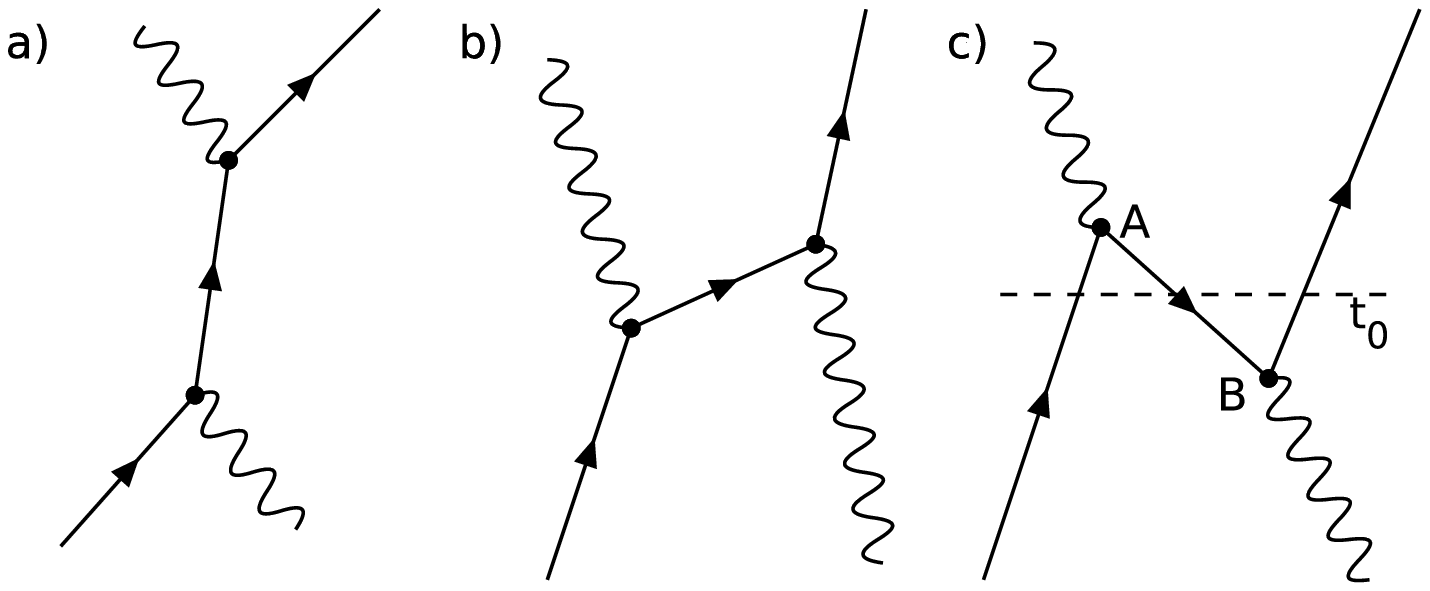}
\caption{Left: M{\o}ller scattering (\el--\el). Right: Compton
  scattering (\el--\gam).}
\label{fig:moeller-compton}
\end{figure}

This is mathematically well-defined and unproblematic,\footnote{The
path is parametrised in terms of a `proper time parameter:
$(x,y,z,t)=(x,y,z,t)(\tau)$, where $t$ does not have to grow
monotonically with $\tau$.} and has clear advantages in that we do
not have to treat separately some processes which are impossible to
distinguish experimentally, and which are only distinguished by the
relative time ordering of spatially separated emissions or absorptions
--- a concept which is not relativistically invariant.  This can be
illustrated by two examples, as shown figure~\ref{fig:moeller-compton}.

In the first example (M{\o}ller scattering) it is not necessary to
treat the cases a) and b)
separately, i.e., it is not necessary to worry about which electron
emits the virtual photon and which one absorbs it.  The formalism
does not distinguish between emission and absorption, and we could
just as well say that in case b) it is electron 1 that emits a photon,
which travels backwards in time and is absorbed by electron 2 (it is
not possible to distinguish between a photon moving forward in time
and one moving backward in time --- the photon is its own
antiparticle).  Indeed, by way of a suitable Lorentz transformation
diagram b) can be transformed into a) (virtual  photons do not have to
move at the speed of light), so the two can be treated as
one.\footnote{On the other hand, fermionic statistics requires that
the diagram where electron 1 and 2 are interchanged in the final state
must be subtracted.}

In example 2 (Compton scattering), diagrams b) and c) can be treated as one, while diagram
a) must be treated separately.  Thus we see that it is not the order
of emission and absorption in time that matters, but rather the order
along the worldline of the electron.  Diagram c) can be interpreted
(in a traditional fashion) as one photon creating an electron--positron
pair at B, whereupon the positron travels to A and annihilates the
electron there --- or (in the Feynman picture) as the electron
travelling to A, emitting a photon, turning and running backwards in time
to absorb the photon at B, whereupon it turns around again and behaves
`properly'.\footnote{If you went in at time $t_0$ and observed, you
would see three particles: two electrons and a positron.  That would,
however, be a different process.  It is essential to integrate over
all possible interaction points.}

  \chapter{Physics and philosophy}
\label{chap:philos}
 
\section{Basic concepts of ordinary life}

\subsection{Things, space and time} 
\label{sec:things}

The starting point for our experience of and thinking about the world,
and for all our activities, is found in our ordinary life.  The primary
conceptual framework of ordinary life is absolutely necessary --- not in
the sense that it cannot be revised (although it is probably the most
robust of all conceptual frameworks), nor in the sense that we have to
use it (explicitly) in all contexts --- but in the sense that in order
to live and function socially we need some minimal set of assumptions
and concepts, which are encoded (explicitly or implicitly) in the
primary conceptual framework of ordinary life.  Since our understanding
of the world is ultimately based on ordinary life, this is also where we
find the original starting point for a more refined understanding, as
expressed for example in physics.  In my opinion, a correct
understanding of basic concepts of physics must therefore be based on
showing their relation to our everyday understanding of reality.

The most important elements of our (human) everyday understanding of
reality are the (macroscopic) \emph{things}.  By a thing, I
provisionally mean something independent, limited, which we can touch
and see, and thereby immediately grasp as one entity.  Starting from
this, we find that almost all our daily tasks and conversations relate
in some way or other to concrete things.  The things form the starting
point for our orientation in and understanding and recognition of the
world.  In this context it is not at all without interest to remark
that we are also things --- admittedly of a special kind compared to
may others, but still things.  This is reflected in the fact that what
we usually recognise as things are and must necessarily be of the same
order of magnitude as ourselves --- with dimensions from about 1\, cm
to about 10\, m.  For us to see something as a thing, it must be such
that we could treat it directly as a single entity, both physically
(we should be able to, at least hypothetically, move it) and
psychologically (we must perceive the whole thing).  Also entities
with larger or smaller dimensions can be considered things to the
extent that they can be observed and (in our imagination, at least)
handled indirectly, but this is only possible within certain limits.
If the entities become too large or too small, the observation and/or
the handling of them will be too unlike the daily observation and
handling of things, so that the parallel is no longer valid.  Entities
larger than a planet can hardly be thought of as being handled in any
sensible way, and can therefore not be considered things.  I will come
back to the limit for how small a `thing' can be; for now, I can
estimate it to be around the size of a macromolecule.

I will claim that the things, with their objective existence, form
the basis for us being able to talk meaningfully and unambiguously
about affairs in the world --- indeed, they form the basis for us
having any idea of a world in the first place.  Two questions then
arise: how do they do this, and which conditions must the things
satisfy in order to form such a basis?

When we talk about, or in some other way relate cognitively to matters
in the world, it is necessary that we have \emph{something} to relate
to: that we are not considering nothing or something that we have no
idea about.  Moreover, we must be able to recognise or identify it as
a \emph{something}, and (if we are to have meaningful relations with
other people) describe it to other people in such a way that they also
know what we are talking about, and hence are also able to identify
it.  And, since we are talking about matters in the world, it must
also in some way or other be unambiguously located in the world. If it
is to have any value to us in our ordinary life, it must also have a
definite, unambiguous relation to our everyday experiences and our
everyday activities.\footnote{We can of course tell tales, stories or
  fables about things and phenomena we have not put in any relation to
  our ordinary world, and have no difficulty in identifying the
  elements and understanding the meaning of such stories, but they do not
  tell us anything directly (only analogously) about our world.  This
  is not an attempt to devalue such stories, fables and myths --- but
  this is not what we are talking about here.}  But how can this
happen?  How can we even say that something is located unambiguously
and objectively in such a way that we and others can relate to it;
that the world and matters in the world are such and such, and not
only appear like this to us?

A description of something gives us no guarantee that what we describe
exists in the world, and much less where or how we should search to
find it.  The location can therefore not happen by way of a
description.  And a location is always relative to something --- how
can we then locate something unambiguously?  We could locate it
relative to ourselves, but we would still not have any measure or rule
for the location --- it will be arbitrary and without any possibility
of recognition.

Let us imagine that I locate all phenomena in the world on a spherical
shell around myself.  (I cannot have any idea of \emph{distance},
since this requires me to be able to move \emph{among} the phenomena
--- i.e., that they are located relative to each other, and not to
me.)  I will not be able to turn my head or move around --- that would
mean that the whole world would move, and I could not say that the
phenomenon that is located in direction $\alpha$ now is the same as the
one that was in direction $\beta$ a little while ago.  This becomes
particularly acute if I close my eyes for a moment: how could I for
example say that the red that was straight ahead of me just then is
the same as the red that is to my right now?  It can even be difficult
to find criteria for calling both red --- to do this, I would have to
be able to compare them objectively, and to decide how different they
can be and still be said to have the same colour.

We also have no guarantee that other people's world (defined by
location relative to them) is the same as ours --- it is even
questionable whether we could form an idea of other subjects who could
have a world.  The world would equal my world, which I could not in
any sensible way distinguish from my own subjective sensory
impressions.  I would have to say with Wittgenstein: `I am my
world.'\footnote{{\em Tractatus} 5.63}

We therefore need something that can operate as a fixed, unambiguous
and person-independent framework for the world.  At the same time, this
framework has to be related to us in some way, since otherwise we
would not be talking about matters in our world.  The things
constitute such a framework: when we locate something in space and
time, this is the same as locating it relative to the things in the
world.  And the things exist objectively --- something which is is
manifested by us all agreeing that they are there.  By way of the
things, we can at least agree on a common relative framework for the
universe --- when two people are in the same place, i.e., near the
same things (which is usually the case when talking with each other),
they can agree by referring to the things they see around themselves,
and locate everything else relative to this.  Alternatively, they can
refer to named things whose location is already taken for granted,
also relative to their `immediate' surroundings.\footnote{This
  argument is given a thorough exposition by P.F.~Strawson in
  \emph{Individuals} \cite{Strawson}.}

We see that there are some conditions that must be satisfied for this
to work.  Firstly, the things must by and large be at rest relative to
each other, otherwise they could not in any way form a fixed framework
for the world, and we would hence have no criteria for how to
reidentify or recognise something as the \emph{same} thing as opposed
to a `doppelg\"anger'.  This is required if it is to make sense to
talk about the things as individual things, and not only as
collections of general properties. That in turn is required in
order to talk about them as being present in the world --- we have no
other way of definitely determining whether something exists in the
world than to show it: \emph{There it is, as an individual phenomenon}
--- we cannot \emph{point} at something other than as an individual
phenomenon.  But then we must in some cases also be able to say that
two things are (numerically) different things even though they have
identical appearance and properties.  The things must also be
relatively stable --- they cannot be destroyed by the tiniest
disturbance or after a very short time (by our measure).  Secondly, we
must ourselves (at least) be able to move relatively freely among
these things, ortherwise we could not have any ideas about what
existed in other places than our own.

The things are also essential in making it possible to talk about and
make sense of the idea of extension: because of the stability of
things, we can use certain things as measuring rods for extension in
general.  It is essential that these measuring rods are things with
`normal' dimensions to enable us to have any kind of control over
them.  Firstly, either the measuring rod or the item that is to be
measured must be moved so that they are brought into contact with each
other.  Secondly, we must be able to read off the measurement, which
means that we must have an overview over the entire measuring rod.
One might object that we also measure events on completely different
scales, with measuring rods on those scales, and that the unit of
length (metre) for a while was defined according to a subatomic
scale.  However, these measurements presuppose the normal measuring
rods that are things, in the sense that a measurement can only be
considered to be completed and understood when it is `translated' to
normal scales.  This is the case physically: the final observation
must take place at `our' level --- or at least, it presupposes the
possibility of such a `normal' observation.\footnote{We may have
  measuring instruments that make observations at the everyday scale
  redundant, but they are always built on theories where such an
  observation enters in some place when the theory is to be
  validated.}  It is also the case mentally: we `scale' the
measurements up or down to form an image of them, and this image is
alway an analogy to an ordinary measurement with a ruler.

As regards duration, this concept clearly requires the existence of
change.  Whether it is the things themselves, some feature of the
things, something beyond or between them or a relation between them
that changes, what forms the basis for a precise concept of duration
must be a fairly clearly delineated transition from one state to
another: the duration is the interval between these transitions.  It
is equally clear that the transition must be viewed against something
(relatively) permanent --- for example the general pattern of things.
And if we are to establish a measure of duration, this requires
processes that can be taken to be regular.  The frequencies of these
`clocks' must be of the same order of magnitude as typical times for
human activity --- i.e., from about a second to a few
days.\footnote{The primary clock is the day (e.g., sunset to sunset),
  and then the oscillations of pendulums, burning candles, sand
  flowing in hour glasses, etc.  Of these, only the day can be defined
  without any explicit dependence on things.}

\subsection{Activity and change}
\label{sec:change}

We all know and see that change occurs --- it makes no sense to deny
this.  Change is also a requirement for life: all processes of life
(including thought) are changes.  The fact that we are alive implies
that we (as things) are changing.  And for us to be alive, we
must be able to handle and treat the things (at least some of them)
for our purposes.  This would not be possible if all things were
eternal and unchangeable.  It would also at least be hard to notice
anything that does not undergo any change --- it becomes an irrelevant
background.
     
This means that the things, and not only phenomena beyond and between
them, must have the possibility of undergoing change and even
destruction.  But how can these two cases, change and destruction, be
distinguished?  In other words: how can we say that a thing is still
not just a thing, but also the \emph{same} thing, when it has
undergone change?  Why do we not say that the thing has been destroyed
and another thing has appeared --- and in which case could we say just
that?  We must be able to distinguish between these two cases if we
are to relate to things in the first place: the things being
relatively constant does not prevent us from perturbing them,
or the things themselves from changing slightly on their own accord.

It is also important that we distinguish between the changes that we
cause with our activities, and those that occur independently of us,
without any contribution from us.  The first type of changes are part
of what makes us acquainted with the things in the first place --- our
handling of the things acquaint us with them and their specific
features, and distinguish between ourselves (as things) and the things
we handle.  At the same time, by handling the things we grasp which
aspects of them are changeable and which are conserved, and we come to
appreciate their independence of us at the same time as we appreciate
our independence of the things.  But as long as I do not have any
conception of the second type of change, the world is still centred
around me, and has no objective existence.\footnote{Piaget writes
  about how small children experience the world in this way.}  I may
distinguish between myself and the world, but the world depends on me
and relates first and foremost to me (the sun follows me in the sky).
To see the world and its changes as fully objective, I must have the
idea of changes that occur because the things are the way they are,
independently of what I do with them.  The things must be able to
change themselves, according to their nature, so to speak.

For us to be in a position to say that the thing is the same thing
even when it is changes, i.e., that the thing changes, there must be
something about the thing that is not changed --- something that we
can identify as the \emph{essence} of the thing.  This is not only
what we can call its essential properties, but the whole web of
relations and possible changes that make the thing what it is to us.
It is part of the essence of the thing to undergo a certain type of
change if it is exposed to certain kinds of external influences ---
e.g., a thing made out of china will break if we drop it on the floor,
while a balloon expands when we blow into it.  There are also changes
that usually happen without any specific trigger --- like a plant
growing --- and changes that do \emph{not} happen according to the
nature of the thing --- like a billiard ball beginning to grow.  All
of this makes the thing what it is , and something we can recognise
and relate to as having an independent existence.  To the extent that
we consider the thing with its essential features, we can think of it
as a \emph{substance}: something (relatively) independent and
(relatively) constant.  This substance can then be in several
different \emph{states} --- a balloon can for example be filled with
air or other gases to a greater or lesser extent --- without changing
its essence or substantive character.\footnote{This distinction
  (between substance and state) is also useful beyond where we can
  talk about subtance in the usual sense.  In physics, the idea of a
  substance is usually replaced by the concept of a \emph{system},
  which is a collection of entities (particles, fields, etc.),
  possibly including boundary conditions.  This system may be in
  different states, and can to a certain extent be considered a
  substance.}  A change in the system of a thing (or a system or
entity) can be labelled an \emph{accidental} change, as opposed to
creation and destruction, which are \emph{substantive} changes.

To summarise: We recognise our world (and hence are able to relate to
it) primarily by recognising (and relating to) the things, with their
(largely) unchanged mutual relations.  We recognise the things in turn
by noticing that (what we perceive as) their essence is unchanged and
that they are located at (more or less) the same place relative to the
other things.  (If a thing has moved, we must have a reasonable
explanation for how this has happened if we are to maintain that it is
the same thing.)

\section{Matter and forces, physics and natural philosophy}
\label{sec:matter}

This is sufficient for us to live in and relate to the world
unreflectively.  That the things are only relatively stable does not
matter much for our ability to relate to them as substances: we are
also only relatively stable.  However, a world that is to such a large
extent `accidental' or `coincidental' will be alien to us; we cannot
feel safe towards it --- there may always be surprises in store, which
will mostly be unpleasant.  Granted, many things may be explained by
pointing out that this is how it \emph{is}.  For example, we reject
many questions from children or fools of the kind
\begin{quote}
    `Hvoffer har en buko ingen vinger?\\
     Hvoffer si'r den bu og ikke vov?    \\
     Hvoffer sitter neglen paa min finger? \\
     Hvis den satt paa nesen, var det sjov.'\footnote{Why has a cow no
       wings?// Why does it say moo and not woof?//
Why is the nail on my finger?//
If it was on the nose, it would be fun.
From \emph{ Sp{\o}rge-J{\o}rgen}, a Danish children's book from 1944.}
\end{quote}
But still: much of what happens is accidental, and we could still ask
the `childish' questions about why things are as they are --- or why
there are so many different things.  The world is still alien.  To
feel at home in nature, it is necessary that we \emph{understand} it
--- that we can grasp its essence.\footnote{Feeling at home in nature
  has (at least) three elements: understanding, control and
  adaptation.  The purpose of basic sciences is, as I will argue
  regarding physics, to provide understanding.  There has however
  often also been a large focus on control, which is reflected in the
  demand that science should have applications, and that it should
  first and foremost provide predictions, and satisfy the
  nomological--deductive `explanation scheme'.  There are also strong
  political and economic forces acting in support of this, although
  this is not the only reason for the focus on control.  This is
  however an attitude which is far from that of the basic researcher.
  The point about predictions there is to satisfy the requirement that
  the theory should agree with reality, and that it should be
  testable.  Predictions are far from being the main aim; the main aim
  is to reach an explanation or understanding that makes us see that
  what happens follows, as one might say, from the `nature' of the
  things (or entities).  This is as much the case for natural sciences
  as for the humanities, although the methods and the character of the
  understanding may be very different.  Control is important, but
  belongs to the technical disciplines, not the theoretical.  I will
  discuss this further in section~\ref{sec:instr}.  As regards
  adaptation, this is most important for behaviour, ethics and
  ecology, which I will not discuss here.}

Religion --- in particular `primitive' religion --- can be one kind of
attempt to understand the world: the essence of the world becomes more
or less human, and hence less alien to us.  And even if we cannot
assume that nature has a humanoid essence, for us to feel at home in
it, it must consist of structures we can conceive, not of horrific
monsters (the terrible unknown).  It must be possible to understand
what happens on the basis of these structures, and we must therefore
extend our concepts beyond the concepts of ordinary things so that
they may encompass such structures.

Another side of the same problem is that the world must necessarily be
conceived and experienced as one world.  If the world, consisting of a
countless collection of things that constantly change or every so
often are obliterated (or new things are created), plus any number of
non-thing phenomena, is to be understandable and possible to relate to
in practice, i.e., not just a chaos, all things must have something in
common.  Moreover, when a thing is not eternal, but has once appeared 
and will once disappear, it cannot be considered completely
independent, but must be considered a manifestation of a more
underlying substance.  We must imagine that there is something that
does \emph{not} undergo substantive change, and that such changes are
accidental changes of this something.  The thing for example also
depends for its existence on `lucky coincidences' having created it
once upon a time.  The things (with their essential features)
thus become merely relative substances compared to the more
fundamental substance, which in the final instance must be the substrate of
everything in the world (also non-thing phenomena such as sounds,
flames, mountains, lightning, water, etc.).\footnote{Here we may see a
  shift in the concept of substance, from denoting a web of
  properties, possible relations and changes which together make up
  the essence of an independently existing thing, to mean either the
  essence of the whole world, or something that is absolutely
  conserved in time.  The first is how Aristotle uses the concept; we
  find the other two meanings in Spinoza and Kant.}
 
We might imagine that this more fundamental substance resides either
\emph{in} the thing itself and in other phenomoena --- as something
that all things and phenomena are made out of, regardless of what kind
of phenomena they are --- or \emph{beyond} (or rather \emph{behind})
the things, as an eternal pattern or origin (which may for example
express the essence of the things).  The latter is however of very
little help in understanding nature as we experience it.  Not only is
this pattern incapable of undergoing substantive changes; it cannot
even undergo accidental change --- and we can hardly see our world of
things as a state of this.  The world as a structure in space and time
will rather become more incomprehensible, and there is no explanation
either of how creation and destruction of things is possible, or why
there are several things of the same kind.  We do not even have any
grounds for saying that what we experience as one world is the same world,
since the respective positioning of the things in space and time
hardly can be part of the eternal pattern.  We are therefore left to
seek the fundamental substance in the phenomena themselves.  This
first (or last) substance is what is called
\emph{matter}.\footnote{\emph{Note added in translation:} In  the original, I
  used the word \emph{urstoff} as well as \emph{materie}, but there is
  no good English translation of \emph{urstoff} (the nearest would be
  original or ultimate stuff), so I have left it out.}

We see that matter appears as the first principle for the unity and
order of the world: by way of it we can explain that it is the same
world we find ourselves in at all times and places, and that this
world is not just a chaos of wildly changeable phenomena and unrelated
things.  In a sense, matter becomes the essence of the world.

In this way, we can understand that the world is one, but changes are
for the most part as incomprehensible and uncertain as ever, even
though we have the security that something is conserved.  To be able
to understand how change happens, we must in addition to the passive
principle (matter) have an active principle which engenders the
changes.  I will call this principle the \emph{ultimate
  force}\footnote{\emph{Urkraften} in Scandinavian.} --- and it must
be \emph{comprehensible},\footnote{Comprehensible as a `natural
  concept': something it is possible to become acquainted with.} in
the same way as matter is.

Both physics and natural philosophy have as their task to find the
fundamental principles (active and passive) for the world we have
around us --- to say something about what is common, what does not
change, what does change and why it does --- i.e., to say something
about (ultimate) matter and ultimate force.  Indeed, the two
disciplines were for a long time identical (Newton called his main
work `Mathematical principles of natural philosophy`) --- but one may
still find a slight distinction from the outset.

Philosophy aims to argue completely \emph{a priori}, i.e., without
taking account of what we at the present time may happen to have
observed or experienced, or what the world happens to look like right
now.  It wants to say something about matter or ultimate force which
must be valid regardless of what we might experience later.  A natural
philosopher will preferably argue purely logically, and hence produce
necessary truths --- or, at least: truths with a far more general
validity than what may be obtained on the basis of current experience.

However, this rarely leads anywhere.  One problem is to find a
starting point which leads to anything other than tautologies --- to
find this, one must assume that certain facts about the world are
known and certain.  Another at least equally important problem is
that there will always be certain hidden assumptions in any chain of
reasoning, assumptions which will depend on our current experience.
This will make it impossible to pursue a strictly logical chain of
reasoning.  It is however possible to produce arguments that others
will subsequently have to take into account --- like Parmenides
setting the agenda for Greek philosophy by denying the possibility of
change.  This may of course be fruitful, and may indirectly yield
greater insights.

A more `appropriate' task for natural philosophy might be to
investigate the preconditions for our current knowledge of the world,
and hence which assumptions about the structure of the world lie
implicit in our current knowledge.  The question is thus: What must we
assume about the world if our current knowledge is mainly correct? or:
What must we assume if we are to exist and have experience of the
world, given how we today understand ourselves and our world?  This
is no guarantee against future surprises, but it will be easier to
explain the surprises, and therefore understand the world better.  It
will also be easier to see which parts of the old knowledge are still
valid.\footnote{This is also how my reflections here are to be
  understood.} 

Physics seeks the concrete --- the concrete (ultimate) matter and
force, and the laws for how they interact.  The physicist always talks
about concrete quantities or \emph{entities} in the world.  By an
\emph{entity} I mean a something that appears in the world, and which
we hence can imagine to be identifiable in some way or other --- it
has (at least) a kind of `quasi-individual status'.  An entity is not
necessarily a substance, but it must be capable of being in different
states.  For example, ``the'' matter (singular definite) is an
entity.\footnote{I am deliberately using an imprecise word, since
  entities can be of many different kinds, and need not have any
  similarity to things, for example.  If I used a word such as object,
  it would be too strongly associated with pure individuals.
  \emph{Note added in translation:} The Norwegian word used was
  \emph{greie}, which would never be used in formal language; however,
  this was in the absence of any more appropriate word such as
  entity.}

To the physicist, ultimate matter must therefore necessarily be a
potential subject of experience, while the philosopher may take it to
be a principle beyond our potential experience.  This contrast may be
illustrated by taking Thales (who said `All is water', i.e., that the
ultimate matter is water) to be (the first known European)
representative of physics, while his student Anaximander (who said
that ultimate matter is `the indefinite') was a typical representative
of philosophy.  But since the two disciplines have such similar aims,
it is no wonder that the main paradigms of fundamental physics tend to
have a counterpart in philosophy, and vice-versa.  And when physics is
pursued to its foundations, it becomes natural philosophy.

More specifically, the research programme of fundamental physics (the
area of physics that has as its aim to explore the most basic laws and
principles of the world) can be sketched as follows.  You start by
postulating one or more specific properties, which are taken to be
common to everything material, i.e., everything in the world
we see around us, to be the essence of matter.  Subsequently, you
explore the dynamical principle(s) that can give rise to the changes
in or transitions between the different appearances of this ultimate
matter.  All other properties and phenomena should then be considered
to be derived from these.\footnote{This is of course an idealised and
  oversimplified description.  The process often happens the other way
  around.}

\paragraph{}
However, what may be claimed to be the essence of matter is not
arbitrary.  It must be something that is or is perceived as common to
everything that is around us and that we call material.  This will be
properties or functions that we almost automatically ascribe to
matter, and if a physical theory does not postulate these as the
essence of matter, it must at least explain where these common
features come from.

\begin{itemize}
\item  Stuff, or matter, is conserved in time.  This is a direct
  consequence of how matter was defined --- as that which is never
  created or destroyed.  Regardless of how much we destroy something,
  the matter will remain.  If it makes sense to talk about quantity of
  matter, this quantity will be conserved.

\item Everything we call material takes up space.  This is the case
  for things, earth (solids), water (liquids), air (gases) and fire
  (plasma).  An important difference between these four states of
  aggregation or states of matter (as we now call them) is their
  hardness or impenetrability: while we cannot penetrate a (solid)
  thing without destroying it to a greater or lesser extent, air
  presents virtually no resistance against being pushed aside.
  Ultimate matter must therefore in some way occupy space (at least as
  one of its secondary properties), while it must on the other side be
  fairly `flexible'.  Matter must at least \emph{refer to} space or
  \emph{be placed in} space, since we can \emph{point at} matter.

\item Matter has an \emph{individuating} function.  Two separate
  things may be distinct because they contain different bits of
  matter.  This was important to Aristotle, since matter could ensure
  that several things that are (qualitatively) identical could still
  exist independently of each other, in contrast to the Platonic
  forms.  We may for example have two completely identical chairs,
  which are both independent things --- although they are
  qualitatively identical, they are not the same chair: they are not
  numerically identical.  Hence it also makes sense, if we were later
  on to spot a chair of the same kind, to ask \emph{which} (if any) of
  the two original chairs we are dealing with.

The last question obviously only makes sense, or can at least only
have a positive answer, if the chair (or our thing or entity) has not
been destroyed in the meantime.  But even then we may ask if the stuff
it is made of is (numerically) the same.\footnote{That something over
  time is the same thing and that it consists of the same matter are
  not equivalent: a thing may undergo a complete replacement of all
  its matter, and remain the same thing.  For example, the atoms in a
  human being are replaced over an average of seven years, but the
  form (structure) is conserved.  Thus we may distinguish between 
  \emph{material} and \emph{thingy} numerical identity over time.  The
  two are of course connected to a certain extent: a certain
  continuity is required --- all the matter cannot be replaced
  instantaneously.}  Since matter is what remains and is conserved
through all destruction and all changes, it follows that matter itself
can never be destroyed; it only enters (as numerically the same
matter) into new things.\footnote{This was Aristotle's explanation for
  how creation and destruction are possible.}  We can thus talk about
how the individual (particular) bits of matter pass through the
things: a bit of matter will always be particular (it is always
\emph{that} particular bit of matter), and it therfore has its own
(numerical) identity.  This identity is then assumed to be conserved
in time.

\item Matter is \emph{separable} and \emph{movable}.  Given the right
  tools we can always divide up and separate a piece of matter, ending
  up with two different pieces of matter.  These may then (if we are
  able to `hold on to' them) be treated separately, as completely
  independent bits, even if they might be identical, or originally
  very tightly bound.  And even if we are not able to hold on to them
  completely, we may make them move --- or matter can move `by
  itself'.  This may be called the \emph{mechanical} character of
  matter. 

\item Since ultimate matter is the stuff that appears in all possible
  forms, it must also have the ability to somehow take on all these
  forms --- what Aristotle calls \emph{potentiality}.  How this
  happens `from the stuff's point of view' is however an open
  question.
\end{itemize}

These properties of matter are not independent, but are closely
intertwined.  This can be clearly seen by for example considering the
necessary conditions for numerical identity to be a meaningful concept.

A clear condition for two entities being considered to each have a
separate numerical identity is that they in fact are (or may be)
separate.  If the entities can only be understood as parts of a larger
whole, if it is not possible to consider each entity as independent,
then the concept of (numerical) identity cannot be applied.  This does
not mean that it must be possible in practice to separate the entities
--- it may well be that it is (in principle) impossible to construct
a tool to carry out this separation --- but it must be possible to
\emph{imagine} them as separate without losing their essence.  In
daily life the separation is effected by \emph{placement in space}:
when two entities are in different places, they are effectively
separated and hence numerically different.  This condition can thus be
related to the material characteristic of being in space.  In daily
life a material entity will always be characterised by a specific
placement in space, while this is not necessarily the case in
experimental science, as quantum mechanics demonstrates.

A condition for considering material entities as separate, and for
talking about material numerical identity over time, is that there are
conserved, extensive (additive) quantities associated with matter.
That some (not necessarily mathematical) quantity must be associated
with matter is obvious, since matter can move in space, and can
therefore not be directly identified with space itself.\footnote{There
  is \emph{something} that can be moved in space --- even Descartes
  would have to admit to that.  To solve this problem, he invoked or
  somehow assumed \emph{two} spaces: matter and space.}  That this
quantity is (these quantities are) conserved follows directly from
the principle that matter is conserved.  The additivity of the
quantities is related to the separability of matter, but goes beyond
the mere question of whether two entities that exist at the same time
are separate.  It is primarily related to the possibility of claiming
that we are looking at \emph{the same} matter before or after an act
of destruction or substantive change.  If we now imagine that two
entities $a$ and $b$ somehow join together in a whole, there must be
some sense in which the whole is \emph{not} greater (or less) than the
sum of its parts: the whole must contain a `quantity of matter' that
\emph{equals} the sum of the `quantities of matter' of the parts it is
made up of.  Hence it will also make sense, if one were subsequently
to destroy the whole and separate off one piece, to ask if this piece
contains \emph{the same} matter as e.g.\ entity $a$.  This is also
gives meaning to any statement that some matter has disappeared out of
a thing, and to questions about where this matter has gone.  It is
only possible to say that you have accounted for all the matter if the
sum of the quantities of matter is the same.\footnote{This is the
  basis for chemistry.  Alternative (non-additive) `formulae' for
  quantity of matter can be imagined, but they do not conserve
  identity.}

Even if we have such an additive conservation of matter, it does still
not make sense to talk about the same matter if during reactions
(substantive changes), the stuff `mixes' in such a way that it is in
principle impossible to identify the stuff from the different
reactants in the reaction product.  It must therefore be possible to
\emph{imagine} the matter as identified and separated (with its
extensive quantities) inside any finite-sized entity.  One (but
probably not the only) way to imagine this is that the material points
are identified with persistent, well-defined worldlines, i.e., that
two worldlines which at some point in time have a finite distance from
each other, will maintain a finite distance at any other time.  The
worldlines may be countable (each worldline belongs to an atom or
elementary particle) or may form a continuum (but with a certain
density of worldlines per unit volume everywhere).  The quantity of
matter inside a volume is directly proportional to the flux of
worldlines through the volume.  Worldlines are not allowed to
intersect, since this would make separation impossible (if it was not
possible to introduce any additional, qualitative criteria for
separation of worldlines), and worldlines cannot appear or disappear.
One may extend the concept of space, so that two lines can be at the
same point in space, but separated in another `dimension' (state
variable).  The criteria for numerical identity being possible on the
microlevel will still be quite restrictive --- it is not a given that
this concept makes sense.

\paragraph{}
\emph{Ultimate force} has considerably fewer restrictions than
ultimate matter, but it also cannot be the subject of arbitrary claims.
It must at least be capable of acting on matter (and matter must be
capable of acting as a source and point of influence for forces).  If
you have said something about matter, you have therefore also said
something about forces.  One should also be able to explain the origin
of the forces that we see acting at the everyday level.  These can
roughly be classified into three kinds.

\begin{itemize}
\item The things act on each other by contact.  The resistance that
  prevents a thing from penetrating another, and that enables a thing
  to push another, is a form of force.  The same is the case for the
  friction or resistance that appears when two things rub against each
  other (as well as the resistance in air and water), and the force
  that makes brittle things break and soft things deform when they hit
  something harder.  All this may be called \emph{mechanical} forces.

\item We also have external forces that change the state of matter
  without this having to occur by direct contact --- \emph{distance
    forces} of various kinds.  The most important of these is gravity,
  but we can also consider phenomena such as light conditions,
  temperature, etc., which act as forces.  Tensions (elastic forces
  such as in springs etc.) may perhaps also be included in this
  category --- if they do not fit better in the next one.

\item The changes that happen with the things `on their own accord'
  must also be considered results of some kind of forces ---
  \emph{internal forces}.  These include all kinds of natural growth
  etc., and generally anything that arises from the internal structure
  of things.
\end{itemize}

\paragraph{}
In our search for ultimate matter there are two competing interests.
On the one hand, there is a programmatic obligation to try to find one
overarching principle, i.e., one and only one ultimate form of matter.
On the other hand, there are so many different kinds of substances (in
the weaker sense of the word) that it seems very unrealistic to be
able to explain all of this from only one or two principles (passive
and active).  A better procedure appears to be to attempt to
\emph{classify} all the different kinds of material appearances we
have in the world, to put the matter into some kind of order.  This
leads naturally to the idea of not one, but several ultimate forms of
matter, or (rather) elements.  The two directions appear to be
mutually exclusive, and they exhibit quite different thought
processes.  While an `ultimate matter thinker' is set on explaining,
and considers the theory of the (multiple) elements as quite \emph{ad
  hoc}, an `element thinker' is more bent on classifying, and may
consider the principles of the ultimate matter thinker as quite
dogmatic and unrealistic.  Both are however necessary, since (to use a
platitude) both the unity and the plurality of the world must be
accounted for --- they represent two different, but equally
fundamental, knowledge interests.  History also has several examples
of the two directions being of mutual assistance --- this is possible
because they normally operate on somewhat different levels, or may
even be associated with two different sciences, like physics and
chemistry.  A successful classification of elements often leads to the
discovery of a new principle of unity, often in form of a new
symmetry.  And a unified scientific theory must always have some
foundation to work on --- a foundation that can only be obtained
through classification.  Hence, the classification of the elements
made quantum mechanics possible, while the classification of the
hadrons in the 1950s prepared the way for SU(3) and the quark model.

This conflict is stronger in philosophy, because we there (for the
most part) operate at the same level (that is at least what we tend to
think).  Here it appears as the conflict between monist and pluralist
directions.

\section{The research programme of fundamental physics}

As I said above, fundamental physics sees it as its task to explain
all phenomena in the world from (preferably) one passive principle
(matter) and one active principle (force), which exist and act in the
world. Usually, to be realistic the task will be limited to try to
completely explain one kind of phenomena or events at one particular
level in the world.  This kind of phenomena or this level is then
considered to be the fundamental, on which everything else dependes,
even if it is not at the moment possible to fully describe this
dependence.  (To view it differently would be to give up the claim to
be considered a fundamental physicist, and request a transfer to a
different science.)  This procedure is fully consistent with what I
sketched in the previous paragraph --- postulating a property or a set
of properties as the essence of matter, followed by exploring the
behaviour of matter in light of this.  The essence of matter is
associated with those properties or quantities (entities) which are
considered the most fundamental in the area of concern.  Hence, in
Newtonian physics the essence of matter will be to have mass, which is
again associated with inertia and gravitation.  If electromagnetism is
taken to be the fundamental theory, the essence of matter will be
charge, and the fundamental entities will be charged particles and
fields. 

Since physics is concrete, it cannot concern itselv exclusively with
general properties, essences and principles, even if it must take
these into considerations (for instance in the form of symmetry
considerations).  All of this must (as previously stated) be
associated with something that occurs in the world; as something
concrete that has these properties or essences or is the carrier of
these principles.  I will call these concrete, primary quantities in
physics \emph{the entities of fundamental physics}.  I will now say
something about how these entities are defined and determined in
physics, and furthermore about how theories are built around these
entities.

\begin{itemize} 
\item We always start from a part of our known conceptual framework
  when we construct a theory.  Some of these concepts are taken to
  correspond to entities in the physical world, and between them there
  are relations that we have specified, and that are taken to
  correspond to essential relations, laws or forces in the world.
  This part of the theory can be denoted the \emph{essential} part,
  and is the primary concern of theoretical and mathematical physics.
  The entities and their relations are here defined purely in therms
  of themselves: the definition is homogeneous.  For example, an
  electron may be defined as a lepton --- i.e., a quantum mechanical
  particle that satisfies the Dirac equation (with spin 1/2), and
  interacts weakly and electromagnetically, but not strongly --- with
  a rest mass of 0.511\,MeV and charge $-1$.

But if we only have this part of the theory, our entities will be
Platonic forms and our world merely a Platonic realm of ideas, without
much of a connection to the sensory world.  We will never be able to
recover our entities in nature --- firstly, because we know nothing
about how or where to look for them, or how they should look to us;
secondly, because the entities are defined purely in terms of
themselves, without any disturbing influence from external factors,
while such factors will always come into play in reality; and thirdly,
because the entities still appear as universalia, and not as
individual entities.  To obtain such a connection, two further parts
of the theory are required:

\item Firstly, a \emph{constructive} part which says where these
  entities appear in the construction of the world (to the extent that
  the world can be said to be a construction).  For an electron, this
  could look something like this:
\begin{quote}
`One or more electrons, together with an atomic nucleus, make up an
atom, when they are bound together by electromagnetic forces in a
quantum mechanical bound state which is electrically neutral.  Several
atoms may be bound with chemical bonds in a molecule, and a quantity
of approximately $10^{23}$ attoms or molecules make up 22 litres of
gas when there are no or only very weak bonds or interactions between
them, while it makes up a few tens of grammes of a solid if there are
stable bonds of different kinds between the atoms or molecules.'
\end{quote} 
We could imagine what a similar exposition would look like for e.g. a
galaxy.\footnote{It will be a bit more problematic for entities that
  only occur in fairly exotic situations, such as muons.  Here we
  could say that they do not play much of a role in the construction
  of the world, but that they appear at the same level as the
  electrons.} 

\item Then, an \emph{operational} part, which gives rules for how we
  can get hold of the entities and register their properties.  This is
  the part that makes experimental physics possible, and which thus
  makes physics to something other or more than pure metaphysics.
  These rules must be heterogeneous, since they should connect our
  entities and the level of observations or instruments of
  observation.  The rules can be very simple, of the kind `If it is
  shining yellow, and you can bite it, it is gold', or very
  complicated, of the kind `If you dig a 27 kilometre long ring, and
  erect some hundred tonnes of instruments of a particular kind (which
  have now luckily been erected), and following this you undertake
  certain operations with some machines, and then look at the screen
  on your computer, which is connected to the instruments, and see a
  pattern of dots or lines out from a centre there, this is an
  indication that in the ring, in the centre of the instrument, a
  $Z^0$ particle occured, and its energy can be measured by adding up
  certain numbers that can also be obtained by the computer.'
\end{itemize}

What happens when a theory concerns levels of reality far from our
everyday level, is often that these three elements (essential,
constructive and operational) diverge more and more, and that the
operational element tends to become very complicated --- although all
three are inextricably linked and mutually supportive.  (It is for
example difficult to justify a measurement if you do not have a theory
for how the measuring apparatus works, and how it is related to the
world being studied.)  The essential element will also tend to be very
abstract for a lay person, but what characterises a scientist with a
good grip on her theory is that she has a good command of all these
three elements and (more or less) immediately can see the connections
both between the different elements and within each part of the
theory.

There is a final element of a physical theory that must obviously be
mentioned.  When the operational element of the theory has been
established, this also gives access to empirical data which may be
used to establish more empirical laws for the behaviour of the
entities (and obviously to test the validity of the already
established laws).  What kind of empirical laws may be established is
however not arbitrary --- it is clearly circumscribed by the essence
of the entities or the theory.  For example, a body in the theory of
relativity cannot interact instantaneously with another body with
which it is not in direct contact.

We may find (and will soon find) that any selection of the properties
we usually are directly acquainted with (the everyday intuitive
concepts) does not have the potential to be basic elements of our
theory, since the laws we can formulate in terms of these concepts are
not sufficient to describe or explain, on the basis of two fundamental
and completely general principles (matter and force), the phenomena we
observe.  Hence we must try to obtain the basic properties from other
parts of our conceptual framework --- and the consequence is they
become, in a sense, not something that can be experienced.  One way of
viewing this is that since we seek to make a theory about nature, and
we cannot assume that this has any human-like characteristics, we must
try to remove the `anthropomorphic' elements that reside in our
ordinary concepts.  We can for example take a mathematical structure
and give it a name (I am not claiming here that this is how concepts
are formed).  We may also find empirical connections between entities
which are operationally defined in analogy with (what is in this
context) well-known entities, but that these relations do not
correspond to our essential definitions.  Then we do not know what we
are talking about before we have established a new concept of the
essence of the entities the theories describe.  This was the
situation in the early years of quantum mechanics.

The positive heuristics of a theory, i.e., the opportunities for
further fertile research, resides among other things in how rich the
fundamental conceptual framework is in possibilites for formulating
new laws or relations.  This is also (I believe) some of the reason
for the success of mathematical formulations.  Mathematics, in
particular following the development of calculus, contains a near
unlimited number of possibilities to formulate both essences and
relations in a precise way, which are hard to find in any other
`language'.

If a theory can be formulated in several logically or mathematically
equivalent, but conceptually different ways, this is a strength rather
than a weakness of the theory --- even though it implies that the
interpretation is ambiguous.  This gives more opportunities for
further development of the theory, by inserting new elements or
replacing some of the elements (entities) with new, analogous ones ---
i.e., several possible directions.  What in one formulation can be a
fairly small and simple change can be a large and complex revision in
the other formulation.  In this context it is important to note that
we never make use of concepts completely beyond the framework of
concepts we already know and work with --- in some sense we have to
choose among our known concepts, or possibly in analogy with these.

Finally, I will illustrate these reflections with two examples of fundamental
physical theories, at quite different levels. One is Anaximenes' theory of air
as ultimate matter, the other is quantum field theory.

\paragraph{}
Anaximenes claimed that everything in the world is air in different forms. The
essential, constructive and operational features of matter are in this case very
evident: invisible, volatile, neutral --- and sustaining life. It appears
concretely in the world as, indeed, air, which is what we breathe (this can
serve as an operational definition); what is usually found above the ground and
water. Furthermore, earth, water and fire are condensed or rarefied forms of
air. Despite the simple construction of this theory, we can see that it contains
everything required of a fundamental physical theory.

The research programme could consist in a further investigation of the processes
of condensation and rarefication. Is condensation and rarefication itself the
fundamental force, or is it gravity or heat? (Anaximenes believed heat was
identical with rarefication of air.) Attempts to e.g. transform air into water
and earth or vice-versa could be paramount. One problem for the programme could
be how to explain the difference between e.g. iron and stone --- forms of matter
which appear to have the same hardness and density. An attempt at an explanation
could be that they have small, but different variations in density, variations
so small that we cannot see them directly. This is however where the programme
would come to a halt as long as there were no means of investigating these
variations. It would also be difficult to explain colours.

Looking back at this theory from our vantage point today, we can see that in
spite of its naivety it contains several lasting insights. If we translate the
theory to statements about the four states of matter (solid, liquid, gas and
plasma), along with the phenomena heat and density, we will find that it chimes
in well with modern statistical physics and thermodynamics: gas is considered
the primary state of matter (i.e., it is at least the easiest in terms of
calculations, which is not necessarily the same thing). When cooled or
compressed (reduced volume, increased pressure), a gas will become liquid and
then solid, while it when heated or rarefied will be ionised (turn into a
plasma). In fact, heat may be seen as identical to the tendency to move to
higher states of matter, while the states of matter on their part are
expressions of the bonds in the system, and hence of its density. However, a
long detour was required before these insights could be formulated in a more
precise language.

\paragraph{}
Without preempting the discussion in sec.~\ref{sec:q-entities} too much, we can
affirm that quantum field theory is centred around the concept of \emph{particle
species}. The particle species can be divided into two main groups: fermions and
bosons. The fermions are characterised among other things by a `natural
repulsion: two fermions of the same species can, if they exist, not be in the
same state at the same time, and they hence have a natural aversion against
being in the same place. Note that this is not due to any interaction (force)
between them, but is part of what it means to be a fermion. Similarly, the
bosons have a `natural attraction': they can be in the same place over a finite
time interval.

Each particle species is characterised by a set of quantum numbers which
determine its behaviour and interactions. The most important quantum numbers are
mass and spin, which determine the propagation of the particles (or the
evolution of a system consisting of one particle species) when left to
themselves, plus charge, colour, weak isospin and weak hypercharge, which
determine how they interact (the shape of the interactions). In addition there
are requirements for invariance and microcausality. This may be considered the
essential part of the theory.

The constructive part of the theory (which I will consider in more detail in
sec.~\ref{sec:qft-constr}) is centred on the concept of bound states. Up quarks,
down quarks and gluons may enter into stable, spatially limited configurations
which are colourless and have electric charge 0 or +1, mass approx.\ 1\,GeV and
spin 1/2: neutrons and protons. By way of residual interactions these may then
form stable configurations with integer charge: atomic nuclei. An atomic nucleus
may combine with photons and electrons to form electrically neutral, stable
configurations: atoms. (The ensemble of bound, electrically neutral states
formed from the same atomic nucleus are denoted as different states of the same
atom.) Atoms may on their hand be bound with electromagnetic residual forces,
giving rise to molecules, crystals, liquids, etc.

The operational part of the theory is primarily based on the possibility of
observing individual particles in particle detectors of various kinds. Most of
these make use of the ionising power of charged particles and photons. That is,
we make use of known theories of ionisation, as well as theories of phase
transitions or drift of charged particles (electrons) in gases with an applied
electric field. A large amount of work on construction and calibration on the
basis of known theories and properties of e.g. cosmic radiation gives rise to
the criterion for claiming that a particle has been observed: a correlated
`track' of e.g. small pulses of currents in wires. Further information about the
particle --- momentum, charge, mass (i.e., particle species) --- may be obtained
by looking at bending in a magnetic field, time spent traversing a certain
distance, etc., while the energy is measured by stopping the particle completely
(and possibly destroying it).

A contrarian may object that what is observed is interactions rather than the
particle itself --- and hence that the existence of the particle is merely
inferred on a more or less flimsy basis. The particle may be considered merely a
construct. In response, it may be asserted that the correlation of the data in
the measuring instruments is \emph{the criterion} for there being a particle
there; this is what makes it possible to talk about observable particles. The
individual particles, thus observed, form the (primary) empirical basis of the
theory, in light of which the rest of the theory must be evaluated.

The masses of the different particle species, the values of the coupling
constants, and various other data, make up purely empirical parts of the theory.
There is also scope for introducing new interactions and particles and revising
the form of the known interactions within certain limits. For example, the Higgs
mechanism is very flexible, there may be a number of ways of grouping particle
species, and one may `invent' new interactions or symmetries with their
corresponding new particle species. I will briefly consider possible research
areas in chapter~\ref{chap:future}.

Certain anomalies are also know (even ignoring the question of whether
renormalisation is valid). These are mostly associated with the relation to
gravity. It is accepted that quantum field theory is valid for all phenomena
except gravity, which is governed by Einstein's theory of general relativity.
These two theories are mutually inconsistence, and it is an open question how
deep this inconsistency is. Part of the problem is that quantum field theory
takes space and time as given, while general relativity is concerned with
modifications of the structure of space and time. This may suggest that the
inconsistency is fundamental. At least, the attempts that so far have been made
to unify the two have not borne fruit. It can therefore not be ruled out that
this problem will have to lead to a fundamental revision of the theory in the
future, just as new discoveries may of course also do.

\section{Reduction, correspondence and complementarity}
\label{sec:reduction}

By \emph{reductionism} I understand an attitude which says that
\emph{everything} (all phenomena, events, entities, properties) is to be
explained by a small number of principles, properties or laws at a more
`fundamental' level, and claims that what occurs at this fundamental level is
strictly speaking all that is real. Everything else is only combinations or
modifications of this, and the concepts expressing these modifications are
strictly speaking superfluous.

In fundamental physics we always seek the most general and overarching
principles possible, from which as much as possible can be derived. The aim is
what in physics circles today (somewhat irreverently) is called a `Theory of
Everything' (TOE). If one set of principles can be replaced by a more general
set (with a wider range of applicability), this is always a big triumph. This is
what I call a \emph{reduction}. Here I am using the concept of reduction both
about what can be considered reduction in the strict sense --- when a theory
(\emph{reducendum}) may be formally explained or derived from another
(\emph{reducant}), which may be considered more fundamental --- and about what
might rather be termed a \emph{scientific revolution}, when an old theory is
replaced by a new one, which after it has broken through may be seen as more
fundamental, and which can explain everything the old theory could.

A reduction has several purposes, and there are several requirements for
considering it to be a success.

The main purpose is to give a deeper, more fundamental explanation and
understanding of the world, and to be able to explain and understand more
phenomena. Often the multitude of theories is reduced because the reducant is
originally created to explain phenomena that are not encompassed by the
reducendum. Hence we get closer to the aim of one fundamental principle on the
basis of which the world may be understood.

It is considered a further success if the reduction contributes to explaining or
predicting additional phenomena that were previously unknown or inexplicable
(e.g., in statistical mechanics, as opposed to classical thermodynamics, it is
possibly to give a proper treatment of energy fluctuations).

The new theory must of course explain everything that was explained in the old
one --- this is implicit in the concept of reduction. But this is not
satisfactory on its own --- if a theory (partly) replaces an old one, this
introduced a new problem requiring a solution: how could the old theory work
that well? Usually it cannot be directly deduced from the new one, since they
operate on different logical levels. And even if parts of the theories can be
shown to yield formally the same results, it may be pointed out that the general
conceptual framework implies that the formalism represents completely different
entities. The theories are thus logically incommensurable, and in light of the
new theory the old one will become meaningless and incomprehensible.

Physicists usually do not consider this a problem, since they are normally more
concerned with empirical evidence than logical niceties. For philosophers of
science this appears more problematic. It appears you have a choice between two
kinds of answer if you believe the new theory is correct. Either you may reject
the old theory completely (as meaningless), or claim that today's science is
talking about something completely different than previously (if you do not wish
to merely give a purely sociological explanation for how people could believe in
such nonsense) --- or you may claim that the old theory still has a certain
(limited) area of validity, where it is approximately correct. The former is
preferred by many philosophers of science, but is not really satisfactory: the
science and thinking of earlier ages is in effect made worthless and
scientifically irrelevant, and our own science is at risk of becoming so too, in
light of the development we must assume will happen. Moreover, even the theories
that are most completely rejected may be shown to have a certain validity.
Ptolemean astronomy gave correct prediction, and Aristotelian physics can
function as a first (and often sufficient) explanation of everyday phenomena:
stones fall down and fire rises up --- that is that. If the second alternative
is chosen, it is necessary to show that there is a \emph{correspondence
relation} between the two theories, to explain that the new one is generally
valid, while the old one still has a certain validity.

The term correspondence relation may denote several different concepts used in
describing the relation between an old and a new theory.\footnote{The positivist
correspondence problem --- regarding the relation between theoretical terms and
observational data --- is beyond the scope of this discussion.}

Firstly, the correspondence principle can be seen as a methodical rule when
working on a new theory: the concepts in the new theory correspond to the
concepts occurring in the old one, and have an analogue or in part formally
equivalent function. In particular, the methods for measurement or observation
are approximately the same. This is \emph{necessary} for us to be able to claim
that this is a new theory within the same science, and not a completely new
science. In addition it functions both as a means to soften opposition to the
new theory, and as a source of ideas for new developments. This was Bohr's main
use of the principle.

Secondly, it can be said to express the requirement that a new theory explain at
least as much as the old one did. This is however also embedded in other
methodological principles.

The third meaning of the term is more interesting and more controversial among
philosophers and logicians. It implies that the old theory is implicit in the
new one, typically as a limiting case, even if the two are formally
incommensurate or contradictory. For example, we say that classical physics
emerges from quantum mechanics when $\hbar\to0$. Taken literally, this is
meaningless. From a physicist's point of view it is on the other hand not very
different from what we always do: make approximations where we ignore what may
be considered irrelevant to the problem at hand. This leads to solutions which
are known to be idealisations to some extent, but which are still a good
approximation of reality --- all theories imply an idealisation. In the same way
we may for example for big systems consider all effects due to Heisenberg's
indeterminacy relation to be so small that they are of no consequence. This
gives a formal solution of the quantum mechanical problem which is identical to
the classical solution (except in those cases where macroscopic quantum effects
occur). Hence, classical mechanics may be considered an approximation to quantum
mechanics within a certain region, and within this region the whole conceptual
framework of classical physics may be used, as long as the limits of its
validity are known. In showing this, we have also explained why classical
mechanics worked as well as it did, by demonstrating a correspondence relation
between it and quantum mechanics. This holds not only for the relation between
these two theories, but for relations between overlapping physical theories in
general, where one is considered more fundamental than the other.\footnote{A
deeper analysis of the importance of correspondence relations in science may be
found in Krajewski \cite{Krajewski}.}

But, again: the theories are also often based on incompatible or incommensurable
conceptual frameworks. The entities of one theory do not fit into the other one
at all. This is not only the case where one theory is often said to contradict
the other (like quantum mechanics and classical mechanics), but also in typical
examples of reduction, like thermodynamics and statistical mechanices. Even in
those cases where a thermodynamical quantity (like temperature) has a direct
analogy in statistical mechanics, the two concepts have very different meanings
and are logically incommensurable. Within certain areas the theories will
overlap; here they (or their conceptual frameworks) will be considered
\emph{complementary}.\footnote{This use of the concept of complementarity does
not quite correspond to Bohr's ideas. His point is that there are pairs of
entities, phenomena or quantities in the world where precise knowledge of one
precludes knowledge of the other. Examples of this are the wave and particle
aspects of light in quantum mechanics, or that detailed and complete knowledge
of the physical and chemical construction of a body precludes knowledge of the
same body as a living organism --- since it would be dead in the course of the
investigation. I would however claim that my concept of correspondence (which in
part is based on Heisenberg's discussions in \cite{Heisenberg:Phil}) retains what is
worth keeping of this. In particular, I would claim that the wave-particle
complementarity is really an aspect of the complementarity between classical and
quantum physics: waves and particles are classical concepts, which do not belong
(at least not in their classical sense) in a quantum mechanics that stands on
its own two feet.}

Another thing must be noted here. As mentioned above, nearly all theories have a
limited region of validity. This becomes clear from noting that a single theory
rarely (never?) can stand completely on its own, and neither can a single
science (like physics). The essential element of a physical theory may do so.
The constructive and in particular the operational elements, on the other hand,
require the occurence of phenomena which cannot be described in purely physical
terms. As far as the operational element is concerned, this is obvious: for it
to make any sense, we must assume ourselves as consciously acting
(experimenting) beings.

The `reverse requirement' in the constructive element is more subtle. The
constructive element is the one that forms a bridge to our everyday world. This
bridge usually has many spans, in particular when the theory is at a level far
from our own. For a theory to work, it is not necessary for all these bridge
spans to be fully constructed, but we must have some idea of where they lead.
Without this the theory is close to worthless as a physical theory, since we
would not know what we were talking about, and what we are talking about would
in any case not be of any relevance as an explanation of the world. Democrit's
atomism, for example, which says nothing about how many atoms of which sizes
make up a body, nor does it give any indcations as to how to find this out, is
merely a metaphysical position.

Only once we have an idea of where we can find the connection between our
fundamental theory and the other levels of reality may we start considering a
reduction. And, to reiterate, a reduction is in general something quite
different from a simple negation. The purpose of the reduction is to explain the
concepts, phenomena or properties that are reduced; this is obviously not done
by rejecting their validity. On the contrary: it is assumed that the concepts to
be reduced are known and given and have their application in their own area. And
if there is a science which covers this area, this science is still as valuable,
and will rather have gained a further dimension and justification from the
reduction --- it constitutes a link between the more fundamental theory and a
level of reality that (normally, at least) is closer to the everyday level.

A one-eyed reductionism ignores these points, and considers what occurs at the
fundamental level as the only true reality, and the concepts that occur at the
other levels as essentially superfluous, if not empty --- and the theories about
them are wrong (period). Only concepts directly (logically and mathematically)
constructed from the primary, fundamental concepts can be accepted --- they can
be part of a kind of `economy of thought'. This is the case for physical
reductionism (everything is really matter or atoms in motion or quantum
mechanical states), which is what I so far have implicitly referred to, and
which cares only about the essential element of physical theories. This position
(which may also be denoted `ultra-realism') implies in effect that we cannot say
anything about the world until we have the fundamental, all-encompassing theory
--- unless we dogmatically assert that we are already in possession of this
theory.

This one-eyed rejection of everything except the fundamental level is also found
in what I will call mental reductionism (that everything is really a construct
from sensory impressions, feelings, thoughts or principles of association ---
i.e., mental quantities), which only are concerned with the operational element
of the theories (if that). I will take a closer look at an important variant of
mental reductionism in the next section. At the end of the day, a reductionist
stance will not be in a position to explain anything. And the world, to the
extent that it exists, becomes completely alien to us.

On the other hand, the opposite of reductionism, a relativism that
does not recognise any fundamental relations, nor any common
principles or criteria, will be just as unable to explain anything.
All theories, worldviews, events, etc. will just be loose fragments
with no value or mutual connections.  This is really a case of being
thrown into an alien, chaotic world.

When working on a physical theory it is entirely legitimate (and
necessary) to consider everything else mere manifestations of what
occurs in the theory --- to consider the theory all-encompassing and
all-explaining.  The aim of physics is to explain everything in the
world, and it cannot accept anything beyond itself, anything not
linked to the concept of matter.  Immaterial phenomena are physically
impossible and oxymoronic.  This does not give any difficulties as
long as it is recognised that what happens at secondary levels,
considered \emph{at and from} these levels, has aspects that are alien
to the primary level.  There must be a formal link (correspondence)
between the fundamental and secondary theories, but at the same time
the independence and necessity of the different theories on their own
levels must be recognised (complementarity).

I can illustrate this with some reflections on what can be seen and
said by an ovserver from different levels.  We often (usually?)
illustrate our theories by imagining an observer at the level of the
quantities we are working with.  (Here we are still working within the
area of physics.)  This observer will see very different things from
what we see --- presumably primarily the entities we assume exist at
this level.  It is however also possible to make the observer a
`physicist' who carries out investigations at our level.  This
requires the observer to be able to consider itself as a `thing', so
that it can handle the entities at its level, but I will not worry
much about this issue here.

A galactic observer will probably see galaxies, radio galaxies,
quasars and similar entities.  There are entities that \emph{we}
really consider more or less accidental collections of stars,
interstellar plasma and dust.  To the extent that we may imagine them
as `things', this is because we have scaled them down in our mind.
Moreover, they consist almost exclusively of plasma (`fire'), which
for the galactic observer will be the primary state of matter, but for
us can never form things.  Solids, which our things mostly consist of,
do not feature much in the world of the galactic observer.  He will
also barely notice the effects of the existence of things (in our
sense) and humans.  Our level risks becoming the most irrelevant one.

At the atomic level, the aggregate states of matter, which are so
important to our perception of things, do not exist. The entities are
also defined more in terms of their relations, and less as independent
`things'. At the subatomic level it is hard to find anything at all that
deserves the label `thing', as I will show in section~\ref{sec:Newton}.
So what does one see at this level? A subatomic observer in our sense of
the word is a contradiction in terms: an elementary particle must be
blind --- observation requires a method and a means of observation, and
the particle cannot have these if this level is the lowest. A method
would require the observer to have a flexible structure and be able to
differentiate between the different parts of itselv. How could for
example an electron have any critera for saying that it was hit by a
photon arriving from a certain direction with a certain energy? At most
it could say that `something happened'. A subatomic observer therefore
requires an even lower level of reality, which we (so far) do not have
the least knowledge of. If we still imagine such an observer, there is
very little we can assume about what this observer can
`see'.\footnote{Henry Margenau has an interesting attemt to describe the
world as it would look to a subatomic observer in \cite{Margenau:Hist}.}
We can however with great certainty claim that it would be a big stretch
for it to `see' (or construct the concept of) us.

\section{Instrumentalism and positivism}
\label{sec:instr}

According to positivism, all we have to go on when it comes to the world
is the impressions that have come to us through the senses. All talk of
a reality `behind' or on top of this is rejected as meaningless --- it
is not possible to give an account of what is meant by such statements.
All our statements about the world must hence be such that they can be
traced back to statements about what can be sensed or (unconditionally)
observed empirically. Those disciplines that try to go beyond this have
no justification; the only discipline with any claim to validity is the
one that deals with relations between observations (observable
quantities), i.e., (natural) science.\footnote{Ernst Mach talks about
three sciences: psychology, which investigates the connections between
our ideas; physics, which investigates the connections between our
sensory perceptions; and psychophysics, which investigates the
connections between sensory perceptions and ideas.} The only valid task
for philosophy is to investigate which statements have meaning and which
are meaningless in light of the criterion above --- i.e., to act as a
kind of servant for science. In this way, positivism claims to represent
a wholly scientific view: only scientific investigations have any real
value.

When it comes to what is considered `primary content of experience',
there are varying views. Phenomenalism, represented primarily by Ernst
Mach, accepted only simple sensory impressions, and considered
everything else (included things) to be constructions and `economy of
thought'. The logical positivists, such as Moritz Schlick and Rudolf
Carnap, eventually found it difficult to maintain such a view, and
introduced a revised requirement that all statements should be analysed
in terms of statements about things, with a given method for verifying
the statements.

Positivism implies an instrumental view of theoretical terms in science.
The theoretical terms only has a value and a meaning in their relation
to observational data --- as a `stenographic' description of relations
between many observational data, as tools for predicting observations,
or as logical constructs (generalisations from which observational
statements can be deduced). The theoretical terms do not denote anything
that can be ascribed any independent existence.\footnote{Mach for
example considered atoms to be merely practical counting variables.}

I will characterise positivism (both the logcial and the phenomenalist
variant) as mental reductionism because it \emph{treats} the world or
science merely as a logical construct from our experiences, whether
these experiences in the final instance be considered pure sensory data
or experiences (concepts) of things. The starting point is that the
primary experiences are given; the subsequent construction of the world
is purely logical.\footnote{Inductive and/or deductive --- here I use
logic in the sense `rational science of thinking`.} The value of
theories is given purely by their logical relation to experience.

According to logical positivism, scientific explanation is nothing but
presenting a general (universal) statement from which singular
(observational) statements may be deduced. The general statement is then
a scientific law, and this view of scientific explanation is called the
\emph{deductive-nomological model of explanation}. Scientific activity
consists in attempting to subsume our (possible) observation under ever
more general laws (with ever greater empirical content).

Positivism appeals to some scientists and others because a positivist or
instrumentalist point of view apparently involves a liberation from all
metaphysical preconditions. All that must be considered is the empirical
data that are immediately given; there is no need to make any
assumptions about the sources (if any) of these data. You are therefore
free to construct scientific theories independently of any philosophical
prejudices, dictated only by `raw facts' --- a `pure' science should be
possible.

A positivist or instrumentalist view of science will always be able to
give a fully consistent `interpretation' of any scientific theory. As
long as there are unambiguous rules for deriving observable consequences
of the theory (which, according to positivism, there must for it to be a
theory), no problems with the interpretation of the mutual relations
between the theoretical terms can ever occur (as could happen in realist
interpretations), since the theoretical terms have no kind of mutual
relations beyond what is determined by their relations to the
observational data.

The problem with a positivist or instrumentalist view is that it in now
way reflects the way science is done. The positivist philosophy of
science undermines itself, since it renders scientific activity
pointless. Science only makes sense if it is taken for granted that it
deals with real entities, and this is also the case for the constructs
(theoretical terms).\footnote{One issue is that no such thing as `pure
observational data' exists: all observations rely on certain theoretical
or metaphysical presuppositions. Logical positivism has partly taken
account of this by taking things as the starting point.} There is no
point in seeking explanations without an implicit assumption that these
explanations reflect something real in the world. Researchers usually
consider their work to be \emph{discovery} of laws, not construction or
\emph{invention} of concepts which may summarise observations in a
concise form. Granted, the concepts and laws we deal with are idealised
constructions --- but this very point, that we can consider our concepts
to be idealisations compared to some real entities, is of utmost
importance for our understanding of science. Because of this, it is also
possible and appropriate to eventually `factualise' the law to make it
fit reality.

If our laws were only convenient descriptionns, there would also be no
point in maintaining a hypothesis if it fitted the observations less
well than a previous hypothesis --- which is something that can be
justified if there is reason to believe that the new hypothesis better
reflects essential features of reality. By maintaining the hypothesis
despite the negative evidence, it may subsequently be developed into a
successful theory. An example of this is the heliocentric view of the
world, which was maintained by severeal people despite Copernicus' model
giving incorrect predictions and all other experience going against it.
Only when the model was developed by Kepler, Galileo and Newton could it
be given the status of a theory. This would be a miracle if you had an
instrumentalist starting point. It would also appear pointless to carry
out experiments to create new phenomena to be explained (unless this is
with an aim to exploit these phenomena). Positivism, which presents
itself as a radical view, could hence result in a rather reactionary
practice.

Positivism (or the deductive-nomological modal in general) is also
unable to distinguish between a scientific theory and a
phenomenological relation or correlation.  Both will have the same
logical form, and they may even predict exactly the same phenomena.
The theory will however be accepted as an explanation, while the
phenomenological correlation will \emph{not}.  The difference is not
at the logical level, but at the level of understanding.  The theory
explains the phenomena by fitting them into a comprehensible or
coherent pattern, as consequences of principles and relations that
appear as natural.  A phenomenological relation or correlation is only
a mathematical expression which does not appear as part of any larger
pattern.  How could the deductive-nomological model explain that
Planck's law of radiation, when it was first proposed, was not a
theory, but that it changed character completely once the quantum
postulate was introduced?  A theory also does not need to give precise
prediction or agree completely with observations.  It is sufficiant
that it is `in principle correct' and provides a qualitative
explanation of the phenomena; one may also choose to adapt the
observational data to the theory rather than the other way around.
For a phenomenological relation this would make no sense.  Norwood
Russell Hanson expresses the inadequacy of the deductive-nomological
model as follows:
 \begin{quote}
`Philosophers sometimes regard physics as a kind of mathematical
photography and its laws as formal pictures of regularities.  But the
physicist often seeks not a general description of what he observes,
but a general pattern of phenomena within which what he observes will
appear intelligible.'\footnote{{\em Patterns of Discovery} \cite{Hanson}, 
p.109}
\end{quote}

Positivism aims to establish the correspondence between theories
at the level of `direct observation' alone --- otherwise, the
correspondence can only be considered a heuristic rule.  This can
however not, as pointed out previously, considered an explanation of
the correspondence between the theories.  There is no explanation for
how the theories `build upon' each other, also conceptually.

Instrumentalism could have worked as a philosophy of science if the
purpose of science had been to control the world, or to give more
precise predictions of what will happen.  A quick glance at (for
example) theoretical physics should be sufficient to convince anyone
that this is not the case.  The research operates (in principle)
completely independently of whether the results will ever have any
application, and experimental situations are constructed which have
nothing to do with everyday reality.  Basic researchers do not carry
out their research for the purpose of being useful, but out of a pure
interest in knowing what the world really is like.

Positivism may, however, have some positive(!) uses:

A theory that conflicts with accepted philosophical ideas may be more
easily accepted if a positivist or instrumentalist view is taken,
since this will make the theory immune to criticism in a consolidation
phase.  One does not make any claims about `reality', but only
describes how certain things are related.  There are a number of
examples of this in the history of science.  We may mention Andreas
Osiander's preface to Copernicus' `De revolutionibus orbium
coelestium', Newton's `Hypotheses non fingo' (on the origin of
action-at-a-distance forces), or Bohr's `There is no quantum world'
(against the classical realism of Einstein and like-minded people).
All this was intended as defences against criticism that was primarily
philosophical, and did not imply that they did \emph{not} take their
theories to be statements about reality.  But by pretending that one
was only describing phenomena in the simplest way possible, the theory
could be given space to develop.  Quantum mechanics is probably also
in debt to positivism --- the positivist climate in the 1920s and
1930s meant that quantum mechanics met with considerably less
resistance than would otherwise have been the case.

It can also be useful to have a `positivist spring clean' in science
once in a while.  If there is a suspicion that a theory is
fundamentally flawed, it can be useful to peel off anything
unnecessary and look only at what can be observed or measured.  There
are several examples of this leading to a theoretical breakthrough.
Einstein's special theory of relativity had as one of its starting
points an analysis of whether it is possible to determine empirically
whether two phenomena are simultaneous, and Heisenberg's matrix
mechanics started from an intention to only study directly observable
quantities.\footnote{It can be mentioned that Einstein in his earlier
works was inspired by Ernst Mach, while Heisenberg in turn was
inspired among others by Einstein.}

Finally, as an example of how far the positivist critique of the
status of `constructs' and `theoretical terms' is from the mentality
of many practicing physicists, I can quote Rutherford's response when
Eddington once at a dinner remarked that electrons were very useful
concepts, but that they did not need have any real existence:
\begin{quote} `Not exist, not exist, --- why I
can see the little beggars there in front of me as plainly as I can
see that spoon.' \end{quote}

  \chapter{Critique of quantum field theory}
\label{chap:critique}

\section{Newtonian physics in the perspective of hindsight}
\label{sec:Newton}
\begin{quote}
     `The classical mechanistic world view was based on the
notion of solid, indestructible particles moving in the void. 
Modern physics has brought about a radical revision of this
picture.  It has led not only to a completely new concept of
``particles'', but has also transformed the classical concept of
void in a profound way.  This transformation took place in the
so-called field theories.  It began with Einstein's idea of
associating the gravitational field with the geometry of space,
and became even more pronounced when quantum theory and
relativity theory were combined to describe the force fields of
subatomic particles.  In these ``quantum field theories'' the
distinction between particles and the space surrounding them
loses its original sharpness and the void is recognized as a
dynamic quantity of paramount importance.'
\emph{Fritjof Capra}\footnote{\emph{The Tao 
of Physics} \cite{Capra}, p.229}
\end{quote}

Newton was an atomist.  But the physics he founded found itself torn
between atomism and continuum theory.  The idea of fields was always
lurking in the background as a threat, even though it was not
developed until the first half of the 19th century and brought to its
conclusion within the framework of classical physics by Einstein.

In Newton's physics the duality between matter and force is clearly
expressed.  Matter --- the passive principle --- was characterised by
mass (inertia and weight) and extension, while the forces --- the
active principle --- acted at a distance, and changed the velocities
(or momenta) of the bodies by an amount depending on the distance
between the interacting bodies.  The paradigmatic example of such a
force was gravity, which acted on all bodies --- but other forces,
such as magnetism, were also known, and Newton believed there would
have to be additional, stronger forces which kept matter together in
our macroscopic bodies (things).  Impact forces could be considered a
special case of forces acting at a distance.  The world consisted of
atoms (or at least matter particles) in motion, under mutual influence
of action-at-a-distance forces.  An important part of the research
programme was to find the forces that were dominant at the atomic
level, and hence perhaps derive the forces at the macroscopic level as
`residual forces'.  An example of this, and one of the most stunning
successes of this programme, is when Boltzmann and others in the last
half of the 19th and early 20th century could explain all of
thermodynamics and all frictional forces (including hydrodynamics) in
principle as a result of microscopic (atomic) collisions or other
forces. 

All of this looked very nice, and it is no wonder that Kant tried to
derive the essence of this progamme (including Newton's laws) purely
\emph{a priori}.  But with hindsight we can point out several issues
that Kant should have seen.

The first, and most obvious one, was a problem for Newton's physics
already at its birth.  17th century physics, brought up on Descartes'
physics (derived purely \emph{a priori}) --- a continuum theory that
only recognised contact forces (impulse, pressure and friction) ---
reacted with horrot to the idea of action at a distance.  Such forces
were considered `occult', and in no way explanatory.  Although Newton
officially refused to construct `hypotheses' about the origin of
action-at-a-distance forces, he could not avoid attempting some
explanations.  The three main types of explanation were as follows.

\begin{itemize}
\item Bodies act directly on each other at a distance.  This
  hypothesis i philosophically the most unacceptable, but at the same
  time it involves the fewest assumptions on top of what can be
  observed, and many people eventually came to terms with this idea.
  This explanation may be more easily acceptable if God is introduced
  as mediating the forces between the bodies.

\item There is a material substance --- the \emph{aether} --- which
  fills all of space and mediates the forces.  This hypothesis
  involves a number of additional assumptions, and leads to some
  paradoxes, but was accepted by many.  Newton made use of this
  hypothesis to some extent.  The reason for its popularity was that
  it gave an apparently intuitive explanation, and the distance forces
  could be reduced to mechanical forces.

\item The points in space have mathematical properties that depend on
  the distribution of matter, and which determine how the forces act
  --- or put differently: there is a field that is defined everywhere
  in space.  Newton linked this to the idea of an absolute space
  (\emph{God's sensorium}) --- an idea which was however in latent
  conflict with Galileo's principle of relativity (which was encoded
  in Newton's first law).  This was not the correct path to take, as
  was made clear by Einstein in his theory of relativity (derived
  almost \emph{a priori}).
\end{itemize}

Kant should have had the requirements to be able to `predict' a
relativistic field theory --- it is almost strange that he did not do
so, in particular since several central points in the theory of
relativity are very close to Kant's arguments.

The connection between the possibility to talk about absolute
simultaniety and the existence of (instantaneous) forces at a
distance, which was Einstein's starting point, was Kant's main
argument for his interaction category: One can only say that two
phenomena are simultaneous if they have an unbroken interaction
(continuous connection) with each other.  As an
example\footnote{\emph{KdRV} \cite{KdrV}, B280} he uses something that
could have been taken straight out of a textbook in relativity: The
light travelling between us and remote celestial bodies forms a
mediate interaction which enables us to determine their simultaneity.
(Kant knew very well that light had a finite velocity.)

It also appears that Kant believed the space between the interacting
bodies could not be empty, since an empty space cannot be
expereinced.  Therefore he would have to advocate either a field
theory or an aether theory, where the field or the aether are
considered to be real entities.\footnote{Kant's opinion on this is
  somewhat unclear.  Some places he appears to advocate an aether
  theory; other places he rejects this idea as absurd.}  Then it will
be natural to assume a finite velocity for the propagation of forces,
based on the principle that a cause never achieves its full effect in
an instant.

All of this leads not only to Einstein's theory of relativity, but
also (and in particular) to a breakdown of the strong dualism between
matter and forces.  The forces are considered fully real, and may well
be carriers of energy (or other measures of quantity of matter).  In
an aether theory, this is almost unavoidable.  The idea of forces at a
distance are also in latent conflict with the idea of extended bodies
on several other points, as follows.

If distance forces (in some form) are acting between the atoms, and
impact forces can be considered merely a special case of distance
forces, then this implies that these forces are everything we notice.
The forces become more primary than matter, which may be reduced to
points: sources of and points of influcence for forces.  In any case
it leads to the breakdown of the analogy between atoms and things
already at this point.  The traditional (Democritian) idea of atoms is
as (infinitely) hard bodies --- as a kind of `super-things',
completely independent of each other, which can never be destroyed ---
and with a clearly delineated extension.  If distance forces
constitute the general case, it will not be possible to distingush
clearly between the atom (its extension) and the surroundings --- or
it is not possible to distinguish clearly between the thing itself and
its effects.  And what kind of \emph{thing} is that?  The idea of a
separate and bounded thing is linked to contact forces, to being
able to say that \emph{there} we are in contact with the thing.  With
distance forces, a clear delineation of the thing will always involve
an arbitrary boundary.

We may expose this paradox even further by asking the questions of where
in the atoms the forces come from, and what they act on --- questions
which were foreseen by Galileo (who refused to attempt to answer them,
since they were metaphysical).  Are they composed of the contributions
from the smallest parts of matter (points), or do they come from a
central point in the atom?  And do they act on the smallest parts, or
on the body as a whole?  If attractive forces act on the smallest
part, why does the body not then collapse to a point?  And external
forces should act differently on the different parts of the body,
which could lead to the atoms being deformed or even crushed.  To
counteract this we would have to assume some kind of `metaphysical'
force inside the atoms, which ensures that they retain their extension
regardless of what happens.

All this has no effect on the macroscopic functions of matter.  In
particular, the forces ensure that `matter' is still extended --- an
`intruding' particle or structure of particles will be more or less
effectively kept at a distance.  The remaing functions are taken over
by the (point) particles.  However, the debate between atomist and
non-atomist conceptions becomes irrelevant when the atoms do not have
to be extended.

There is a second point where the essence of matter had to change in
the Newtonian programme: as more forces were discovered, matter would
have to obtain more essential properties (attributes).

Matter which is only characterised by extension and mass (for now, we
can ignore the argument that extension cannot be an attribute of
microscopic matter), will be insensitive towards all other forces than
those that couple to these attributes, i.e., impact forces and gravity
and similar forces.  It was well known \emph{both} that neither of
these forces could explain stable, microscopic bonds --- impact forces
are always repulsive, while gravity is too weak --- \emph{and} that
other forces existed, which coupled to other attributes.  In
particular it can be noted that if electricity is a fundamental force,
then charge must be one of the attributes of matter.  Moreover, charge
differs from mass in that it occurs in two versions: positive and
negative.  Hence it cannot be reduced directly to one quantity of
matter --- we need (at least) two types of matter, positively and
negatively charged.  Neither can be immediately get rid of mass as an
attribute.  Thus we are for the time being left with (at least) to
mutually irreducible ultimate forces and (at least) two attributes of
matter\footnote{It is possible that the observation that the force
  both in Newton's law of gravity and Coulomb's law behaved like
  $1/r^2$ inspired Kant to attempt to derive this dependence \emph{a
    priori} --- there wer at least a number of attempts by different
  people to find a common origin of the two forces, without success}.

Kant pointed out\footnote{\emph{KdrV}, B249--250. Kant's concept of
substance corresponds more or less to what I call matter.} that the
empirical criterion for being able to identify matter, is its role as a
source of forces. It is thus clear that all measures of source strength
must be considered an attribute of matter. Hence, we should expect that
if more forces are discovered, matter must also acquire more attributes.
This yields a multitude of (apparently) independent, irreducible
matters, or the matter acquires qualities --- which appears odd
considering what was originally required of matter.\footnote{Kant was of
the opinion that qualities are of critical importance for experience,
and that they can be treated within the framework of mathematical
physics. But firstly, these qualities were taken to be sensory
qualities, and secondly, they could in no way replace spatial extension.
Moreover, it seems reasonable to assume that he thought of the qualities
as states of the substance.} In particular we may note that if we assume
several matters (e.g., `charge matter' and `mass matter'), these will
(partly) coincide in space --- the same body will always have components
of several matters. Matter becomes more and more \emph{quality}, and
less and less \emph{extension}.

And even if it once more would be possible to reduce all the different
forces to one ultimate force, this will necessarily have a different
character to the forces we started from. Hence we are left with both
`occult forces' and matter with `occult qualities'. And the forces
effect qualitative changes as much as spatial motion. This, if nothing
else, is a break with the Newtonian programme.

This break also emerges in a different area, where furthermore the
relation between theory of ultimate matter and theory of elements is
illustrated.

Galileo's and Newton's atomic theory was metaphysical. The scientific
atomic theory was developed in chemistry, which contained both
quantitative properties (e.g. mass), taken from Democritean theories,
and qualitative properties (chemical properties), inherited from
Aristotelian and Averroist theory. The interactions were for all intents
and purposes qualitative. The results were around 90 different atoms and
substances which physics could start working on during the 19th
century.\footnote{The historical development is described by van Melsen
\cite{vanMelsen}.} We may also note that extension was in no way a
relevant attribute of chemical matter.

If we now accept that the world exhibits a level structure, and that
matter at deeper levels has different attributes (properties) from those
we experience at the everyday level, it follows that these `deeper'
properties must be observed indirectly. We are forced to construct
experiments in order to observe and define these properties and entities
in the first place, not only to give them a precise value. This has the
following consequences.

\begin{itemize}
\item Observation must necessarily be treated as a physical process.
This brings up for discussion the entire set of questions related to
what observation and sensory perception really are. It is not possible
to claim that sensory perception is unproblematic, nor that it is the
be-all and end-all of experience --- there is a lot more to it.

\item Observation becomes an \emph{active} process to a much larger
extent than before --- we prepare the conditions that make it possible
to `see'. To observe a charge, we \emph{have to} set up instruments such
as conducting plates, wires, batteries etc. (This example is perhaps not
the best, since we can get a shock if the charge is large enough!) This
is in contrast to everyday life, where we do not need to do anything
except turn our heads, i.e., arrange ourselves so that we can lay our
eyes on something; and also in contrast to Galilean experiments, which
serve to increase the precision or \emph{select} what we want to see ---
something that could in principle have happened naturally.\footnote{Kant
saw clearly the importance of an experiment being a far more active
process than an ordinary observation: we force nature to answer
\emph{our} questions (see e.g.\ \emph{KdrV}, Bxiii). This is however
still within the framework of Galilean experiments, where what is
measured is more or less directly observable.} The `deeper' phenomena
cannot be seen by us when they occur naturally --- we can only see them
(identify them) in an experiment.\footnote{Bohr wanted to limit the word
`phenomenon' to only observations or events in a properly defined
experimental setup.} There is therefore no reason why experimental
setups that are necessary for observing one phenomenon, may in fact rule
out the observation of another phenomenon at the same level or a
different level. This is at the root of Bohr's principle of
complementarity.

\item Our observations become to a large extent theory-laden. The higher
level theories form the background for the observations at lower levels,
since the existence of lower-level entities is inferred from experience
and theories at higher levels, and since the experiments assume known
theories. To make the latter point clear: in modern experimental physics
it is essential that the measuring instruments are \emph{calibrated} ---
i.e., parameters are adjusted so that the instruments give correct
values in well-known experimental situations, and so that the sources of
errors are known. This is obviously meaningless unless you have a theory
that says something about expected experimental data and a theory of the
behaviour of the measuring instruments.

This theory dependence is much deeper than what can be claimed at the
everday level. It can be claimed also in that case that to `see' always
assumes a certain pre-established knowledge if we are to see
\emph{something} that we can identify, or if we are to know what to look
for. This knowledge may however to a large extent be reduced to
synthetic \emph{a priori} requirements along Kantian lines, or to
fundamental categories of everyday language --- things that it does not
make any sense to doubt (like things, space and time, as discussed in
section~\ref{sec:things}).
\end{itemize}

This prefigures (in hindsight) many essential features of quantum
mechanics (e.g. in the Copenhagen interpretation) as features that are
implicit in Newtonian mechanics. The concrete appearance of quantum
mechanics, including for instance the discovery of Planck's quantum of
action (as a finite quantity) could of course not have been predicted,
but nor does the argument require this. From an 18th century point of
view, however, this involves too many logical steps into the unknown for
us to expect that anyone --- not even Immanuel Kant --- could have
carried it out without a misstep somewhere. It is pure hindsight --- and
argument that we can carry out now that we know what we can hang it
onto.

However, there is one of the preceding points that Kant may be
criticised for not having raised: the question of observation and
sensory perception as a physical process. This issue had been raised
already by Democritus, and was later discussed by Descartes and
\label{Locke} Locke.\footnote{The reflections of Democritus, Descartes
and Locke were based on a fallacious distinction between `primary' and
`secondary' qualities, and a belief that it is possible to eliminate the
`subjective' and be left with the primary qualities as properties of the
things in themselves. This position was correctly criticised by Berkely,
a criticism accepted by Kant. However, it appears that Kant also (in my
view) accepted too much of Berkeley's other conclusions regarding
sensation. It is clear that sensation must also be considered a physical
process in \emph{Erscheinungswelt}, and if Kant had as his aim to
clarify the preconditions for sensation being used as evidence, he
should also have taken into account the physical limitations of
sensation.} Democritus' reflections lead naturally to a very important
conclusion: the atoms can in no way be regarded as things in our sense
of the word. Let us follow this line of reasoning, which is valid within
a much wider framework than Democritus' atomic theory.\footnote{Although
his opinions on this point are mostly in agreement with quantum
mechanics --- see the next section for a portrayal of Feynman as a
neo-Democritean.}

Democritus' atomic theory is one of the clearest examples of physical
reductionism: everything is really only atoms in motion. Colours,
smells, heat etc.\ do not exist in the things --- they are phenomena
that appear when the `messenger atoms' collide with the `soul atoms' in
the sensory process. Democritus was aware of the problems of such an
attitude --- he lets the senses say to reason, `Poor reason, do you hope to defeat
us while from us you borrow your evidence?  Your victory is
your defeat.' The senses are the only basis we have to speak about
concrete things in the world, and they are based on a signal being
transmitted from the thing to us. We form an image of the thing as a
whole (including extension) by a large number of signals (or a
continuum, to generalise compared to Democritus and quantum mechanics)
are emitted and absorbed by our sensory organs. So far, everyone should
be able to agree. But what happens when the `things' are so small that
they can be compared with the signals --- when they are atoms? Then we
have no possibility to grasp them as separate entities, independent of
the signals through which we observe them. Democritus drew the logical
conclusion of this, and said that the atoms cannot be sensed (\ldots but
he ascribed extension/geometry and number to them, taking these to be
concepts of pure reason\ldots). We may also note that this argument does
not depend on the signals being atomic, only by them being physical.
Democritus' messenger atoms might well be the smallest of all the atoms.

If we add the pragmatic criterion for calling something a thing --- that
we can \emph{handle} it as a single unit --- the point is even clearer.
The atoms (or the smallest constituents of matter) must be handled by
their `peers', which however are themselves the smallest, structureless
constituents of matter. We can thus in no way handle the atoms with
tools that we hold fixed --- the tools will be neither more nor less
fixed than the atoms themselves! Hence it is nonsense to talk about the
atoms as things that can be known or experienced independently of their
relations to other `things'. A matter--force dualism will not solve this
problem, since there we may well be able to move the atom, but cannot at
the same time know that we are doing this.

This was however something neither Kant nor anyone else at that time considered.

\section{What are the entities of quantum field theory?}
\label{sec:q-entities}

One of the most salient features of quantum field theory is that it all
but abolishes the divide between matter (traditionally seen as
\emph{particles} in some sense or other) and forces (\emph{fields} in
classical physics). The divide was already partly erased with Einstein's
demonstration of the equivalence between mass and energy, but is now
abolished in a more subtle way: matter is also fields, while forces are
also particles. Any attempt at an interpretation must take this into
account, although the emphasis may be placed more on one or the other
aspect. Here I will concentrate on four kinds of interpretation that can
be taken as `paradigms' --- a number of other interpretations can be
considered intermediate positions. I will call these four
interpretations the Feynman interpretation, the aether intepretation,
latence interpretations and the S matrix interpretation.

All these interpretations are variants of more or less `moderate
realism'. This means that I am avoiding both the `reductionist ditches',
ultra-realism and positivism. Both those versions of interpretations
have however been present in the debate, so I should say some words
about them. The ultra-realist interpretations or quantum mechanics are
mostly variants of `many-worlds' and `universal wave function'
interpretations: all problems with the theory are solved by
incorporating the observer into the state function, which is seen as the
only true reality. The problem is of course to what extent such an
interpretation says anything at all, since `true' reality becomes
completely unobservable. Margenau's version of the latence
interpretation, which I will consider in section~\ref{sec:latence}, may
be considered a more `moderate' version of these interpretations.
Positivism has played a large and in part constructive role in the
history of quantum mechanics, both in the development of the theory and
in getting it accepted. A positivist attitude to the theory does of
course avoid all problems, as positivism always does, but is subject to
the general criticism of positivism as I have explained in
section~\ref{sec:instr}.

An interpretation should, I believe, deal with both the essential, the
constructive and the operational elements of the theory, although the
essential element usually receives the greatest emphasis. This may be
because it contains the `ontology', and because it usually is in
greatest need of interpretation to be understood. The operational and
constructive elements are typically tacitly assumed. The operational
element receives greater scrutiny in quantum mechanics, and both the
Copenhagen school and positivism places most emphasis on it. This
emphasis is here found in the S matrix interpretation. None of these
interpretations have much to say about the constructive aspect; I will
discuss this separately at a later stage.

Almost as a summary of what I have written so far, I will first say
something about what \emph{can not be} (are not) and what \emph{can
  be} (are) attributes of matter in quantum field theory.  It should
be obvious from the previous section that extenstion and geometry
cannot be attributes.  (It is then of course an important task to
explain why extension is an attribute of all macroscopic matter.)  Nor
can any of sensory qualities (with one exception!) be attributes ---
this is Anaximander's point.  Of attributes which without any problem
(of any note) can be ascribed to matter, we can mention energy,
momentum, charge and colour charge.\footnote{Charge, or electricity,
  which is a fundamental attribute of matter, can be sensed.  Who had
  thought of that?}  These are straighforward, additive quantities.
(Colours obey the law of `colour addition': blue + red + green =
white.)  Mass is a minor problem: whether or not it is considered a
fundamental attribute depends one your point of view.  Mass is not a
directly additive quantity in relativity, but on the other hand it is
a Lorentz invariant.  More significant problems arise when considering
\emph{number} and \emph{being}.\footnote{Here I am considering being
  as an `ontological category', not existence in a logical sense,
  which cannot be any property or attribute.  If matter has being, it
  means that it has the same kind of `full reality' as (or perhaps
  even more reality than) the things --- that it exists completely
  independently, as something definite.  This is in contrast to a view
  that there are several degrees of reality, and that matter has less
  reality than e.g. the things --- a view advocated by Aristotle.} as
attributes of matter --- different ways of solving these problems
lead roughly to the four interpretations mentioned above.  This is
also related to the problem of identity, which I will discuss in
section~\ref{sec:identity}.

These are properties we can attempt to ascribe to matter, and to
assigne a value to.  How this is to be done is an interesting enough
topic in itself ans concerns what we may call the static part of the
theory.  However, it becomes even more interesting when we also
attempt to interpret the dynamical part --- relating to the equations
of motion or the time evolution of the system, which are often to a
larger extent `hidden' in mathematics.

The quantum mechanical equations express the time evolution of
operators and states, and some of these quantities must enter as
entities in any moderate realist interpretation.  As shown in
sections~\ref{sec:state-op} and \ref{sec:transf} there is quite a
significant freedom of choice in how to express this: we may say that
the system is in the same state all the time, but that this state is
characterised by evolving quantities and properties, or that the state
evolves --- or both.  We may also choose which properties to
concentrate on.  This freedom of choice can be considered an essential
feature of quantum mechanics, but can be difficult to carry over into
a philosophical interpretation; we are almost forced to consider one
point of view as more fundamental than the others.

Another essential feature of the theory is that it deals with
objective tendencies in some sense (at least if we choose a moderate
realist interpretation).  These objective tendencies are moreover
associated with subsystems and properties that may be separated to a
certain extent, but not completely.  Examples of this is that neither
individual particles (particle states) nor particle species can be
completely separated from each other, but are to some extent
entangled.  This entanglement will however not be greater than that it
under certain circumstances makes sens to talk about separate
subsystems.  It is also not possible to uniquely identify what is
a system and what is a state of the system; this is related to the
flexibility of expression.

\subsection{The Feynman interpretation}
\label{sec:Feynman}

I will devote a considerable amount of space to this interpretation,
both because it is a quite `popular' interpretation of relativistic
quantum mechanics, and because several of the problems that occur in
all the interpretations appear here.  I do not imply that everything I
present here was Feynman's own views --- I cal it the Feynman
interpretation because it by and large follows the ideas of Feynman,
but I have painted a bit of a `charicature' to make the points
clearare.  Some of what Feynman himself has written is included in the
references \cite{Feynman:QED,Feynman:1948pathint,Feynman:1949qed}.

Feynman considered the \emph{individual particles} to be primary, and
believed that all systems in principle can be described by the motion
and configuration of point particles in space and time.  If we are to
place this within the conceptual framework of quantum mechanics, we
can say that he starts from a particular type of states, \emph{viz}
particle states.  However, since the interpretation is tied to his own
formulation of quantum mechanics, this does not do it full justice.
He distinguishes between fundamental fermions, which can never be
created or destroyed in an absolute sense, and which reprerent
`matter', and (gauge) bosons, which represent the `forces' and which
are emitted and absorbed by the fermions.  The Feynman diagrams give a
space-time representation of this.

So far, this looks like a good, realist and perspicuous interpretation
which does not depart too far from common sense, and the lines back to
Democritus are quite clear.  The elementary particles are, as
Democritus' atoms, equipped with \emph{being} and \emph{position in
  space and time}.  They obviously do not have any extension, so they
cannot interact through impacts, as they do according to Democritus
--- therefore they instead interact by bosons being created and
destroyed, and hence exchanged by fermions.  This can be considered a
`minimal solution' to retain as much as possible of the perspicuous
contact forces.  It is also necessary to satisfy the requirement of
locality in relativity.

Now these particles start behaving oddly, as Feynman says.  They do
not limit themselves to taking one particular path in space, as proper
things do; on the contrary they may decide to take all possible
paths.  This means that the quantum mechanical indeterminacy principle
is satisfied, at the same time as an apparently pressing problem is
solved: it appears incredibly unlikely that a boson should `hit' and
be absorbed by another particle, but when the particles can encounter
each other anywhere, this is not a problem.\footnote{In fact, it is a
  problem.  If there are more than 4 (3+1) space-time dimensions, the
  particles will not hit each other, even if they can take all
  possible paths.}  It can also be considered a kind of `reverse
causality principle': why should the particles choose a particular
path if they do not have a reason to do so?  So this property, which
appears quite strange and incomprehensible when you first encounter
it, can be considered natural and logical once you become used to it.

Just how the principle of `all possible paths' is to be understood is
however not clear.  Feynman emphasised (in line with the Copenhagen
school of thought) that we cannot say that the particle \emph{really}
took one or the other path, although we do not know which one.  This
is equivalent to summing up the amplitudes rather than the
probabilities.  If we imagine that a particle only has to possible
paths to take to any endpoint (as in the double-slit experiment), ther
may be points it can never reach --- not because the paths leading to
these points are themselves excluded, but because their contributions
cancel out.

If we instead try to say that the particle takes all paths at the same
time, we must be careful with what we mean by this.  We can definitely
not take it to mean that the particle `splits' and that each part
takes a separate path --- the particle is always whole and intact.
(We sum up amplitudes for \emph{whole} particles.)  The particle will
moreover never be observed in several places at the same time, so when
we observe it, it is always in a specific location.  The path integral
formalism expresses neither more nor less than the probabilities or
possibilites for going from one place to the other, possibly with
additional constraints for `internal coordinates' such as spin.  This
does of course require that we are able to observe the particle
unambiguously at these points (or, rather: prepare it at the first and
observe it at the last).\footnote{Both an ultra-relativist and a
  positivist `interpretation' will avoid these difficulties.  The
  universal state function interpretation implies that the particle in
  fact \emph{is} everywhere, also when we observe it --- but when we
  observe it, this leads to also us (or our consciousness) being
  `everywhere' --- the consciousness branches into all the possible
  states, but we can only access one at a time.  A positivist will of
  course say that it is meaningless to say anything about the particle
  except at those moments when it is observed.}  One may also imagine
that the particle `sniffs out' the path in front of it, so that it
`knows' what is possible and impossible.  This may appear to bring an
element of teleology into the theory.  A final possible way of viewing
it is a latence interpretation, which we will look at later.

The particles taking `all possible paths' obviously corresponds to
the field aspect of matter.  That there is no difference in principle
between fermions and bosons in this respect is an expression of the
matter-force equivalence.  This equivalence is broken in two ways.

Firstly, bosons are emitted and absorved, while fermions are not.  I
will look closer at this difference below.  Secondly, for gauge bosons
a distinction is made between `real' and `virtual' quanta, a
distinction that is not made for fermions.  For electrons, the same
propagator is used throughout the process, while the propagator of a
photon that is being exchanged differs from the expression used for an
external photon: a plane wave.\footnote{An electromagnetic plane wave
  can only have transverse polarisation, while virtual photons can
  also have both longitudinal and `scalar' polarisations.}  This is
related to the starting point: the movement of electrons from one
place to the other, where the interaction is not viewed as a field,
but as a direct (but delayed) interaction between the elctrons.  This
interaction is represented by the photon propagator, which has a
different character to the free field.  This somehow problematic
distinction can be made less problematic in two ways.  Firstly, the
integration over all possible interaction points can be carried out,
resulting in Feynman diagrams in \emph{momentum space} (the particles
are characterised by their energies and momenta rather than their
positions).  Then we will find virtual electrons as well as virtual
photons.  Secondly, it can be argued that all light has been emitted
from a source at some time, and will at some point in the future be
absorbed, e.g.by a detector, so that the `free' photons really
represent an interaction between the source and/or the detector and
the electrons that enter into the process.  A third way out can be
found in the S-matrix interpretation.  The problem of how virtual
quanta are to be considered is one of the central philosophical
problems in quantum field theory, and will reappear several times (in
various guises).\footnote{More opinions about the problem of virtual
  quanta may be found in \cite{found}.}

In the `ordinary' field theory formulation there is no difference
between fermions and bosons in terms of their constancy --- both
fermions and bosons have variable particle numbers: particles may be
created and destroyed.  In the Feynman interpretation only bosons have
this property.  This realist, Democritean view of matter can be
maintained by giving the particles the ability to move backwards in
time.  With a positron interpreted as an electron moving backwards in
time, all pair creations and pair annihilations may be eliminated by
interpreting them as the electron `turning' in time.  The world-line
of an electron will hence always remain an unbroken, infinite line.

As pointed out in section~\ref{sec:pathint} this is mathematically
unproblematic --- it is a way of parametrising all possible paths ---
and this point of view can have a number of advantages.  We may
however ask: what does it mean that `the electron is moving backwards
in time?  This appears on the face of it to be a contradiction in
terms: motion means change of position with time, and hence requires a
time in which the motion takes place.  Motion backwards in time would
then mean that time decreases with time, or that one at a later time
finds oneself at an earlier time.  It is not quite as paradoxical as
that: the worldline of the particle is parametrised by its
\emph{proper time}, which is a precisely defined concept --- and the
CPT theorem (see section~\ref{sec:transf}) tells us that motion with
reversed proper time is equivalent to the motion of an antiparticle.

We thus have to find out what proper time means, and for that we must
turn to relativity.  Here the concept can be defined in two
(equivalent) ways.  Firstly, the (infinitesimal) proper time interval
can be defined as the Lorentz invariant quantity formed from the
difference between the temporal and spatial coordinates of a particle,
\( d\tau^2 = dt^2 - dr^2/c^2. \) 
This quantity has an important function in the theory of relativity,
e.g. in defining relativistic velocity and in a relativistic formulation
of Lagrangian particle mechanics (not field theory) --- which was the
starting point for Feynman's formalism.  The sign of $d\tau$ is
however not determined by this definition; it can be fixed in a
coordinate system that follows this particle.  However, this
presupposes and observer `attached' to this particle, and this is only
possible if the particle is a macroscopic system.  For an observer to
be attached to a Feynmanian electron is unthinkable.

It is tempting to make the electron (or particle) itself to an
observer, and hence to equip it with consciousness.  Apart from being
rather speculative, this will not be very satisfactory: we can of
course not ask the electron about anything.  The electron can in any
case not be an observer --- apart from being blind, it can certainly
not be equipped with measuring rods and clocks that would make it
possible to conduct objective measurements.  It can in fact not have
anything like an objective concept of time, since it does not
experience anything except possibly sporadic external changes.  Nor
can it have any memory that would be accessible, since this would
violate both the indeterminacy principle in quantum mechanics and the
principle of causality (that an effect cannot come before it cause).
The only option is hence to make the electron a Leibnizian monad,
with the possibility of introspection and monadic consciousness, but
without `windows'.  Despite the speculative nature of this picture, it
can be a possible key to an understanding of Feynman's system: a
monadic particle can in no way be bound by external spatiotemporal
coordinates, and will therfore be able to take all possible paths.  It
will only know its own, subjective `proper time'.  A `pre-established
harmony' will then ensure that this `proper time' corresponds to the
objective proer time --- and perhaps also that the path integral works
as required?  These speculations are however very far from the spirit
of Feynman.

It may get even more speculative if we ask what closed fermion loops
represent.  An electron moves forward in time, turns and goes back and
not only meets, but `swallows' itself.  This appears to suggest an
eternal, rythmic process or circular time, if we still are to take
`motion backwards in time' literally.  It is best, I think, to leave
these speculations at this point.  I will come back to some of these
issues in connection with the problem of identity.\footnote{Other
  problems arising from Feynman's `time reversal' are discussed by
  Margenau \cite{Margenau:1954qed}.}

Another issue with the Feynman interpretation arises (in particular)
from weak interactions.  In weak processes, particles of new kinds are
formed --- e.g., muon decay gives a muon neutrino ($\nu_\mu$, an
electron (\el) and an electron antineutrino ($\bar{\nu_e}$).  This
happens through an exchange of a W boson between the muon and electron
`parts' or the system.  To maintain the conservation of matter
(fermions) and the path integral interpretation, one is forced to
claim that $\mu^-$ and $\nu_\mu$ are `really' just two states of the
same particle, and the same for \el and $\nu_e$, so that the
antineutrino can be interpreted as an electron moving backwards in
time.  This means that the distinction between different particle
species is effectively abolished, and it is a purely empirical
question whether different types of fermions exist.  It also becomes
less and less clear what is really meant by a particle.  Up to now we
have been able to give the particles, mass, charge and some other
`labels'; now this becomes difficult.  We could try to characterise a
particle by its propagator, but it depends on the particle's mass ---
which is not conserved.  This can however be a strength as well as a
weakness.

A final, serious problem with this interpretation is how to explain
bound states.  One issue is the purely calculational procedure ---
process with an unlimited number of gauge boson exchanges must be
included.  This issue is however common to the whole theory, to the
extent that it is applied perturbatively.  There being an
indeterminate number of photons in an atom does for example not
represent any additional calculational problem.  It does however
represent a problem for the interpretation, if one wants to view
photons as primary entities with the attribute of existence.

A more serious problem is that the formalism to a large extent
precludes any consistent definition of a bound state.  We can say that
the formalism deals with \emph{processes} rather than states; to the
extent states occures in the conceptual framework it is as
instantaneous configurations of particles.  The concept of a
stationary state is hard to fit in.  We can try to define it as two
(or more particles) that at a certain time are close to each other
still being close to each other after a long time has passed.  Such a
definition does however not capture the \emph{stationary} aspect of
the state, even if the time is taken to infinity.  It is one thing
that the particles should be in close proximity all the time, and that
a stationary state is essentially time-independent.  In addition, it
at least looks like it assumes the possibility of measuring the
position of the particles \emph{in} a bound state.  This is however
impossible.  In an atom any precise measurement of the position of an
electron will involve such a strong interaction that the electron
immediately escapes from the atom.  When it comes to quarks in
hadrons, it is even theoretically impossible to isolate them and treat
them as individual particles (confinement).  When they are
sufficiently close together, bound in the hadron, they behave as if
they were free (asymptotic freedom), and it should be possible to
identify them.  However, we do not know how many of them there are ---
in addition to the quarks that contribute to the `net' quantum numbers
of the hadron, there may be an arbitrary number of quark--antiquark
pairs.

The conclusion is that it is in any case difficult to view bound
states as something essentially stable in the Feynman intepretation.
They must instead be considered as eternal processes where everything
looks the same at the surface --- like a Heraclitean river.  It
appears that this would contradict the Democritean basis of the
interpretation.  There are however other states of physical interest
that have indeterminate particle numbers, such as states with definite
values for the electromagnetic field strengths.

Despite all these problems, the Feynman interpretation has an obvious
advantage in its realism and perspicuity --- even where it fails.  It
is also the interpretation which best maintains numerical identity on
the microlevel, as we shall see --- although at a considerable cost.
It is also useful in terms of comparing classical physics and quantum
field theory: employing the Feynman interpretation makes it clear just
where the differences reside.  At the same time we are reminded that
in very many cases it is permitted to view the subatomic entities as
particles, but we must then be aware that they are Feynmanian.

Furthermore, information about processes is usually expressed through
Feynman diagrames, and the particle species and the diagrams is
usually what you are first acquainted with --- before you learn the
specific concepts of field theory.  Some qualitative features of
Feynman's formalism or interpretation may be explained to a layperson
without any use of technical terms.  Feynman's own outstanding
physical intuition and ability to present his material in an
understandable way probably also play a role.  He claimed himself not
to understand quantum mechanics.  In that case, we can say that his
presentation of his lack of understanding have made many other people
understand!

\subsection{The aether interpretation}
\label{sec:aether}

I am not aware that anyone has explicitly formulated and advocated
this interpretation.\footnote{Fritjof Capra \cite{Capra} may come
  closest to it.  However, he explicitly bases his arguments on the S
  matrix interpretation and S matrix theory, although he takes a
  considerably more `realistic' point of view of the fields and
  particles.  The result can be considered a blend of the two
  interpretations.}  On the other hand, there appears to a widespread
opinion that the aether is rehabilitated in quantum field theory,
albeit in a very different form to the classical aether, and many
qualitative descriptions of the concept of a quantum field are close
to that of the aether interpretation.\footnote{See e.g., Dyson's
  presentation in \cite{Dyson}.}  The aether interpretation also
present several features which are interesting in themselves and make
many essential aspects of the theory explicit.  For these reaons, and
in order to have a certain completeness in the system of
interpretations, I have included it.

If the Feynman interpretation can be termed Democritean, the aether
interpretation is best termed Heraclitean.  Heraclitus claimed that
\emph{change} is the essential feature of the world.  Everything is in
a state of continuous flux; the world is like an `ever-living fire,
kindling itself and going out by regular measures'  Ultimate matter is
fire, which is at the same time a process or a force: the fire looks
`stable' just because it always changes and is never the same; it
lives by burning and transforming every new pieces of matter.  `Out of
discord comes the fairest harmony,' said Heraclitus, and by this he
meant that discord, conflict and activity are preconditions for the
existence of anything.  Nothing can be in a state of quiet, without
change.  In line with this, he sought the One in the many --- the
difference in the world is just what makes it one.  Difference is a
principle of unity --- a measure of difference and change.

Now we can directly translate Heraclitus' concepts to quantum field
theory.  Fire can be identified with energy, force with fields or
field operators; change consists in the creation and translation of
particles in an eternal dance, while the Lagrangian (or action?) is an 
underlying measure that regulates everything.

To be somewhat clearer, the aether interpretation states that the
\emph{fields} (field operators) are the primary entities.  The fields
are present everywhere in space, not as static but as dynamic
quantities.  This can be related to the field concept originally being
associated with and derived from the concept of force, which is an
expression of changes in nature.  Thus, the nature of the fields is
change.  The `material' aspect of the fields is expressed primarily in
energy, and secondarily in other conserved quantities.  We may note
that energy is also originally associated with dynamics: energy is
what can be transformed into work.

In the world of phenomena the effects of the fields appear as
movement, change and annihilation of particles in a `cosmic dance'.
This term is quite apt --- a dance is a continuous motion, complicated
but rule-bound.  There is a \emph{pattern} in (or behind) the
movements which is just what makes it a dance and not chaos.  In field
theory this pattern resides in the fields and their inherent
properties, and in particular in the Lagrangian constructed from the
fields.

I use the concept of \emph{aether} because the fields in quantum field
theory have a material character which is deeper than just being
energy carriers.  To clarify the peculiarities of quantum field theory
it may be useful to compare it with classical aether theory and
classical field theory.

The classical aether is a very fine or rarefied material substance
that permeates everything, that is present everywhere in the world,
and through which all motion and all forces are mediated.  The
prototype of an aether is found in Anaximenes' \emph{pneuma} (breath
of life) or the Chinese \emph{ch'i}.  It is hence not exclusively a
passive principle, but rather the condition for all motion and all
life; it is (usually) taken to never be at rest, but always in
motion.  It is often thought that the aether can condense into
material things, and dissolve again later.

In classical physics, the aether was primarily taken to be the
material substance through which light is transmitted: light is aether
waves.  Furthermore, distance forces may be taken to be transmitted
through the aether, as pressure or stress.  The function of the aether
is in other words still first and foremost dynamical, as a medium for
transmission of motion and forces.  The aether may also be taken to be
an absolute reference system --- all motion is ultimately related to
the aether.

When the last function is included the aether has lost much of its
original dynamical chahracter, although there may still be flows in
the aether.  But even ignoring this, it is obvious that it is
something material: it can be moved, and has attributes such as
density and pressure, which only apply to matter.  To the extent that
matter is taken to emerge from aether, it is as condensation, i.e.
quantitative, not qualitative changes.  In reality, classical aether
theory can be considered an attempt to completely elimiate forces ---
the forces are reduced to pulsation and flow in the aether, and are
subject (and subordinate) to the law of conservation of the quantity
of aether.

A classical field is on the other hand primarily immaterial.  It is
originally not defined on its own terms, but through its effect on
matter.  The field expresses the dynamics of the system, and no more
--- it does not have any `life of its own'.  The field is there, or
not, with greater or lesser strength, according to the distribution of
matter, and it acts by changing the distribution of matter.

A Newtonian force field (like the gravitational field) can be
considered a compact notation for the sum of the contributions from
several (even infinitely many) material sources to the total force on
a particle at a given position.  What makes it a useful concept is
that the field can be made independent of the particle that might be
at this position --- the field is the same, independently of the mass,
charge, etc. of the particle.  Hence we may say that the field is
there independently of whether there is a particle there or not.
However, no empirical consequences can be drawn from such a
statement.

If we now do not have instantaneous forces at a distance, but finite
propagation speeds, the field acquires a more real status.  It
transmits information about what happened yesterday which is of
importance for what will happen tomorrow, and thus ensures a
continuity between cause and effect.  At the same time it is natural
to make the field an energy carrier --- e.g., the energy that must be
transmitted between two particles in a collision.  It also begins to
acquire a `life of its own' --- we can write down equations describing
the evolution of the field independently of the material sources.

It is however difficult to make the field more material than this.
The field can not be imagined as an aether, where variations in field
strength are condensations and rarefications: there is no law of
conservation of `field quantity' (total field strength).  It is also
very hard to view the material sources as merely regions of space
where the field is particularly strong, as Einstein attempted, or as
regions with a particularly high energy density.\footnote{For example,
  it makes no sense to say that \emph{that} energy is moved in space,
  but it does make sense to talk about moving a piece of matter in
  space.}  Quite apart from any problems arising with the original
definition of the field (from its effect on matter), the behaviour of
the sources is not covered by the field equations.\footnote{This does
  not mean that it is in principle impossible to describe the sources
  field theoretically, but this must involve a certain revision of the
  field concpet.  Either it must include several fields acting
  reciprocally on each other, or certain non-linear structures in the
  field equations.}  For the theory to be Lorentz covariant, the
sources must also be taken to be points, and they are hence not just
regions with large field strengths, but singularities in the field.
We see that from the attempt to create continuity, strong
discontinuities arise, which need special treatment.

Klassical aether theory attempts to reduce force to matter (flow or
vibrations of the aether), and runs into problems with explaining how
matter can be transformed, i.e. what force really is.  The matter may
well become dynamical (have motion as part of its nature), but not
active.  Hence, it is ultimately sterile.  Classical field theory
attempts to reduce matter to force (there is matter where the forces
are strong), and runs into problems with explaining where the forces
originate and what they act on, i.e. what matter really is.  You get
activity without this activity having anything to act on.  Both models
take being as something completely continuous, and runs into problems
with explaining the discontinuous aspect of matter.  How anything can
be delineated and stable remains a mystery.

Quantum field theory abolishes the matter--force dichotomy not by
reducing one to the other, but by going to a higher level of
abstraction.  It operates at the outset with two levels: the fields,
which are underlying, active quantities which are present everywhere,
and the states, on which they act (literally).  The condition for
talking about the fields as active quantities, which have a (possible)
effect everywhere, is that they have something to act on.  But now
matter, which the fields act on, not something which exists
independently, but is on the contrary created by the fields.  Matter
(particles) consists of manifestations of the fields, but these are
not themselves fields.  A particle state (a state that is a particle)
is not identical with a region of space where the field strength or
energy density is large.\footnote{The particle state may however be in
  a state where the energy density is large in a certain region of
  space.}

The aether interpretation emphasises that the fields contain creation
and annihilation operators, and that these act `continuously' in space
and time, and that change is effected by them.  We can thus talk of a
continuous dematerialisation and rematerialisation --- a `dance' of
particles appearing and disappearing.  Taken to the extreme, we may
say that all change in the state of the system consists in creation
and annihilation of particles.  In particular we may note that all
interactions involve creation and annihilation, i.e. materialisation
and dematerialisation.

Another important point, and the reason I have chosen to call this the
aether interpretation, is that this `energy dance' takes place
everywhere in space, although the activity is greatest in the vicinity
of physical particles.  Suddenly, particles (quanta) may appear from
`nothing', only to disappear back into `nothing'.  Vacuum is not
empty; on the contrary it is involved in a continuous process of
materialisation and dematerialisation which form essential parts of
the `vacuum'.  Vacuum is transformed from empty space, via a role as
`container' for the fields to a very living `aether'.\footnote{Vacuum
  fluctuations are physically relevant, and among their consequences
  is that vacuum looks hot to an accelerated particle.  The
  gravitational effect of the vacuum has problematic consequences for
  attempts to construct a theory of quantum gravity.}

Self-interactions can be interpreted as there being around every
physical particle a `cloud' of virtual quanta, which contributes
significantly to the properties of the particle.  The mass of the
particle receives significan contributions from this cloud, while the
interaction properties may be said to be determined by the shape of
the cloud.  Renormalisation asserts that the particle can in no way be
separated from the cloud surrounding it; it is not itself without all
these virtual quanta.  Capra says, `A subatomic particle not only
performs a dance of energy, it \emph{is} also a dance of energy; a
pulsating process of creation and annihilation.'\cite[p. 271]{Capra}
The essential features of the particle are determined by the dynamical
whole it forms part of.  This view is quite distinct from the Feynman
interpretation, where we can either `imagine away' the virtual quanta
or redefine the particle concept to include this `cloud', leaving us
which a fairly clearly delineated entity.

Quantum field theory differs from classical aether theory and
classical field theory (and classical atomism) by the quantum field
being \emph{both} continuous \emph{and} discontinuous.  It is
continuous because it is nicely define everywhere\footnote{The
  derivatives are well defined everywhere.} and discontinuous because
it is quantised.  The manifestations (materialisations) of the fields
are always \emph{quanta} --- we never find half-particles
(half-quanta).\footnote{In the aether interpretation is is probably
  more correct to use the term \emph{quantum} than the term
  \emph{particle}.  A quantum can be considered an excitation of the
  field.}  The discontinuous aspect to a certain extent --- but only
to a certain extent --- makes it easier to account for the particular
(particle-like) features of matter.  Matter as sources and points of
influence for the forces is easy to grasp: all processes involve
quanta.  These quanta play roles both as matter and forces, according
to your point of view.

The aether interpretation denies the existence of matter and forces at
a fundamental level.  These two basic concepts of physics are both
absorbed into the quantum field.  The primary manifestations --- the
quanta --- cannot be identified with either, although they have the
potential to appear as both.  Here we see a difference compared to the
Feynman interpretations, where fermions immediately can be identified
with matter, and gauge bosons with forces.  In the aether
interpretation the Fermi--Bose distinction does not play such a big
role --- all quanta can be considered equally perishable; all changes
of state are creation and annihilation (at the fundamental level),
while movement in space plays an essential role for Feynman.  The
distinction between fermions and bosones is not due to the permanence
of the respective quanta, but what kinds of patterns they enter into.
Similarly, the distinction between matter and forces appears at this
level --- we may talk about `material patterns' (such as particles)
and `force patterns'.  It may be added that the typical patterns are
determined by the form of the Lagrangian --- i.e., how the fields
relate to each other.

In conclusion, we may say that the aether interpretation features
three `layers' or reality at the fundamental level:
\begin{itemize}
\item The fields, which are the most fundamental and underlying, and
  which are typically active quantitites.
\item The quanta, which are the primary manifestations of the fields,
  and which are constantly created and destroyed.
\item The pattern of creation and destruction, where we can glimpse
  the origin of some of the more well-known phenomena: stable
  particles and forces acting between them.
\end{itemize}

The aether interpretation definitely represents an unfamiliar way of
thinking, at least for us Europeans.\footnote{Capra emphasises
  strongly that this way of thought is much more widespread in other
  parts of the world.  Heisenberg also hints at this.}  This is a
strength as well as a weakness.  A strength, because it makes it
possible to a larger extent to be unshackled from old though patterns
which are inadequate for understanding quantum field theory.  A
weakness, because it requires using metaphors to a large extent --- we
do not have access to the appropriate words, and cannot base ourselves
to such an extent on logical rigour.  In particular it is difficult to
relate the interpretation to categories that are fundamental in our
ordinary experience, like the distinction between substance and
attribute or state.  We usually think of being (in the ontological
sense) as \emph{subsistence}, and this is at the root of the Feynman
interpretation, while in the aether interpretation being is primarily
thought of as \emph{activity}.\footnote{Perhaps it would have been
  easier if we had said `The grass greens' instead of `The grass is
  green'?}

It is a problem for the aether interpretation that the world as it
appears to us to a large extent is dissolved, and it is easy to fall
into an existential angst because nothing is solid in the world ---
the angst that physics was supposed to help remove --- or else seek a
more or less mystical insight into what lies behind.  (This is a
general problem in quantum field theory, but is particularly prominent
in the aether interpretation.)  This problem is exacerbated by it
being hard to see that what we consider to be fixed points of
reference or stable structures emerge naturally (although bound states
are more natural here than in the Feynman interpretation --- they are
as natural as individual particles).  Granted, this is a common
feature of all interpretations.  It is more problematic that the
operational part of the theory appears to be absent --- but it should
be possible to add it.

On the positive side, we may note that the interpretation is very
faithful to the essential mathematical framework of the dynamics of
quantum field theory, and thus gives meaning to most of the concepts
or symbols.  Hence, the essential part of the theory is well explained
--- in my opinion, the aether interpretation, correctly understood,
represents the best and most complete understanding of quantum field
theory as a fundamental theory of physics.

\subsection{Latence interpretations}
\label{sec:latence}

Latence interpretations are best characterised as intermediate
interpretations, and various versions have been advocated as general
interpretations of quantum mechanics.  Both versions denying that
quantum mechanics says anything about the behavious of individual
systems (Popper) and versions maintaining a form of realism within the
Copenhagen framework (Heisenberg~\cite{Heisenberg:Phil}) or Kantian
transcendental philosophy
(Strohmeyer~\cite{Strohmeyer}\footnote{Strohmeyer claims that
transcendental philosophy can justify the concept
of \emph{potentiality}, but cannot provide the foundation for a
science of objective probabilities.}) have been proposed.  Margenau's
interpretation in \cite{Margenau:Nature,Margenau:Essay} also belongs
to this group of interpretations.  Some theories of multi-valued logic
also have similarities to latence interpretations.  Here I will not
take sides among these various points of view.  I will however
primarily base my discussion on Heisenberg's and Margenau's views, as
they cast light respectively on two important concepts in quantum
mechanics and quantum field theory.  Heisenberg's interpretation may
serve as a gateway to understanding the concepts of quantum fields and
operators, while Margenau's interpretation provides a good insight
into the concept of quantum states.

The central point of latence interpretations is that quantum states
express \emph{potentialities} or \emph{tendencies} in the system, and
that these potentialities or probabilities in some way or other are
real.  When the system is in a certain state, some properties
are \emph{latent}.  This means that we cannot say that the
system \emph{has} these properties, but neither can we say that the
state is unconnected to the individual state.  This can however still
be viewed in several different ways.

\subsubsection{Heisenberg's version.}

In Heisenberg's view\footnote{I must point out that this section does
not cover Heisenberg's view in its entirety.  For example, he believed
that quantum mecchanics or quantum field theory is not complete, and
that a contradiction between the concepts of quantum mechanics and
relativity (the sharply defined lightcone in relativity vs. the
indeterminacy relations) is what leads to the divergences of quantum
electrodynamics.  It is against this background we can view his work
on S matrix theory and on a universal lengthscale.  The former belongs
naturally in the context of the S matrix interpretation, while the
latter has lost some of its topicality in the light of newer theories,
and belong to a set of issues I will consider in relation to Planck
scale physics in section~\ref{sec:toe}.  These issues are not
necessarily related to the elements of Heisenberg's thinking I will
discuss here.} `the atoms and elementary particles are not as real [as
the phenomenal of everyday life]; they form a world of potentialities
or possibilities rather than of things and
facts.'\ \cite[p.~186]{Heisenberg:Phil}  The state thus describes
propensities, tendencies or potentialities in the system ---
possibilities which all \emph{may} become real, but which \emph{are}
not so in the isolated system.  When a measurement happens, i.e. when
the system is brought into contact with a macroscopic system which
must be described classicaly, a transition occurs from
the \emph{possible} to the \emph{real} (from \emph{potential}
to \emph{acta}) --- the system acquires some property which previously
was only present as a posibility.  Heisenberg makes explicit use of
Aristotle's terminology, but employs it in a new way to establish a
`quantum ontology'.  Strohmeyer~\cite{Strohmeyer} presents his
`quantum ontology' as a kind of synthesis of Kant and Aristotle.

This transition from `potential' to `real' does not only happen during
measurements, although it is most prominent there, since we `force'
the systems into the framework of classical ontology and logic.  For
example, an excited atom will naturally evolve via states containing
the possiblity both for being excited and non-excited, to the pure
ground state.  At the same time the surrounding electromagnetic field
will have possibilites for different excitations.

This language can describe superpositions of states and measurements
of incompatible quantities.  A system may have several values of one
and the same quantity as potentialities (latent values), but only one
value can be real.  Similarly, a particle can have potential position
and potential momentum, but only one of these can be realised at one
time.  It is also possible to talk about greater and lesser degrees of
actuality.

In quantum field theory we may also have latent or real particles:
since the quantum field makes the particle number variable, the very
existence of individual particles can be potential rather than real.
In addition the field can have latent properties both as particle or
particles, or as forces.  These properties are to a large extent
incompatible: for the field to have reality as force, the particle
number must be indeterminate.  The quantum field thus expresses
the \emph{potential} existence of the particles and the forces.  If
the particles are real, they may still be considered as more or less
individual --- i.e., they have a potential individuality which is only
realised when the particles for all practical purposes can be treated
as separated (i.e., when external forces play a much greater role than
statistical interference).

Heisenberg introduces the idea of potential properties at yet another
level (although he does not explicitly use the word potentiality in
this case).  He claims that the system or quantum field not only has a
possibility to manifest itself as particles or forces, but also as
different types of fields (particles or forces of different kinds).
Examples of this can be a proton's potential existence as a neutron,
or a neutrino's potential existence as an electron, which gives the
neutrino electrical properties.  The last 20 years of his life,
Heisenberg devoted himself to an attempt to derive the entire particle
spectrum as excitations of an underlying field which would represent
all of matter --- clearly inspired of Anaximander's and Aristotle's
view of matter, as what is itself devoid of properties, but has the
possibility to take on all possible forms.

Heisenberg puts great emphasis on describing the relation between the
everyday world of fully real things and the `degrees' of reality that
features at the quantum level --- that the potentialities of the
quantum world are realised in the everday world and in measurements.
In this way it is also easier to understand the correspondence between
quantum and classical physics.  Heisenberg also emphasises the
retention of an explicity symmetry between `complementary' classical
descriptions --- pairs of mutually exclusive descriptions, where both
tell us something essential about the phenomena --- a symmetry which
he considered essential in quantum mechanics.  Examples of such
`complementary' descriptions can be the description of the fundamental
as matter or forces, of matter as particles or waves, and of a
particle using position or momentum.

A problem with the Heisenberg interpretation is how to deal with the
transition from potentiality to reality.  Is it at all possible to
describe this transition \emph{within} quantum mechanics?  It is
tempting to treat the measurement process as some kind of `magical'
operation which effects this transition.  This is not very
satisfactory --- there is no proper answer to when (under which
conditions) the transition occurs.  This is at the root of the quantum
mechanical measurement problem.

\subsubsection{Margenau's version.}

In certain respects, Margenau goes a notch further than Heisenberg in
accepting the probabilities denoted by quantum states as `real'.  To
enable him to do this, he first establishes a distinction
between \emph{nature} or \emph{historical reality}, which is the sum
total of all individual events (including all sensory impressions)
and (physical) \emph{constucts} or \emph{physical reality}, which is
enduring and regular.  The quantum mechanical state vector is an
(abstract) construct, which serves a function in physical
explanations, and hence has in principle the same status as other
constructs.  The state vector is part of physical reality, and the
same is the case of the probabilities it encodes.  The individual
observations, on the other hand, belong to historical reality --- and
at the subatomic level there is no direct (one-to-one) correspondence
between physical and historical reality.

Among the constructs we may in turn distinguish between to important
groupings: \emph{systems} (such as crystals, magnetic fields, or
atoms) and \emph{quantities} (energies, wavelength, probability,
etc.).  The quantities can, as the word indicates, have numbers
assigned, while the systems cannot.  Assigning numbers to quantities
can however be complicated, as in quantum mechanics.  Combinations of
a system and a set of quantities can make up
a \emph{state}.\footnote{The system, when it is also characterised by
what kinds of states it can be in, may be called an \emph{entity}.}
The constructs of physical reality are subject to \emph{causal}
laws\footnote{For Margenau, this is if not a necessary condition for
physics, then at least a rule that should be followed} --- i.e., the
changes in the quantities defining a state are determined solely by
time-invariant laws.  When the states are determined by probabilities,
the laws of quantum mechanics are causal.  It is not a problem that
these quantities cannot be determined by a single measurement, but by
a (large) number of measurements of identically prepared systems ---
nothing prevents us from employing such constructs.  `Observables'
such as the position and momentum of a particle is as much constructs
as are probabilities --- there is no reason why these should have any
preferred status.  If we insisted on defining the states in terms of
these quantities, the theoried would be acausal --- but classical
mechanics would also be acausal if we instistend on using colours and
extensions as state variables.

Quantum mechanics is thus a causal discipline.  The states evolve
causally; the states express objective probabilities, which are part
of physical reality.  This does not change because the measurements
always give definite results --- according to Margenau the measurement
results belong to historical, not physical reality.  There is a
correspondence between these, but it is not a one-to-one
correspondence.  The measurement does in no way affect the state of
the physical system (there is no `collapse of the wavefunction').

We see that the entities of quantum mechanics are quantum particles
and fields --- characterised by quantum states (state vectors).
Naturally, in quantum field theory it is the fields that must be
considered entities --- Feynmanian particles, for example, would yield
an acausal theory, since they can move backwards in time.\footnote{He
discusses this in \cite{Margenau:1954qed}.}  If we however consider
the states of the fields at any point in time, the theory is
completely causal.  If we call the states latences, we may furthermore
employ the same `ladder' of levels of latence as Heisenberg uses.

Margenau's interpretation has the advantage that the relations between
quantities, states and entities are made completely clear, and avoids
the problems Heisenberg has in accounting for the transition from
`potentiality' to `reality'.  The disadvantage is that he introduces a
`duplication' of the world that can be problematic.  Where Heisenberg
to a certain extent emphasises the correspondence between classical
and quantum physics, Margenau places a greater emphasis on quantum
mechanics as different from and independent of classical physics.  The
main emphasis is on the essential and the operational aspects;
Heisenberg also considers the constructive aspect to be important.

The latence interpretations are not complete, although they have
greater or lesser degrees of completness.  They are primarily suited
to illuminate and clarify one (important) issue in quantum mechanics:
how to understand \emph{objective probabilities}.  Discussions of the
measurement problem arise easily out of latence interpretations ---
but the different versions give radically diverging answers.  The
interpretations have little to say about essential features of quantum
field theory such as interactions and entanglement of systems.

\subsection{The S matrix interpretation}
\label{sec:S-matrix}

The S matrix theory can, as described (page~\pageref{S-theory}) be
treated as an independent research programme, transcending the limits
of and forming an alternative to quantum field theory.  This was the
aim of both Heisenberg and several of those who got involved with this
theory in the 1950s and 1960s, and who continued to work on it in the
1970s.  It can also be treated as an interpretation of ordinary
quantum mechanics which does not necessarily break with quantum field
theory; this is what I will now consider.  The clearest exposition of
the S matrix interpretation is due to H.P.~Stapp~\cite{Stapp}, who
considers it the best candidate for a pragmatic version of the
Copenhagen interpretation.  My discussion will be based mostly on his
presentation, although I will allow myself some assessments of my own,
which Stapp would probably not accept, to adapt the interpretation to
quantum field theory.  All quotes are, unless otherwise stated taken,
from Stapp's article \cite{Stapp}.

The central concept in the S matrix interpretation is
the \emph{experiment}, consisting of \emph{preparation}
and \emph{measurement}.  In the interpretation of the experiments it
distinguishes between the \emph{observing} and the \emph{observed}
system, and the correlation between preparation and measurement can be
described in terms of the evolution of the observed system.  The
observed system thus forms a link between preparation and measurement.
However, the possibility of such a link, separated from the
experimental apparatus, requires that the processes of preparation and
measurement are physically separated.  Hence, in order to effectively
be able to identify and `isolate' the observed system, it is necessary
to look at the asymptotic form of the correlation between preparation
and measurements --- the limit when the two are infinitely far apart.
This asymptotic form is expressed through the S matrix.

The S matrix interpretation takes over and develops the Copenhagen
school's emphasis on the influence of the experimental apparatus on
what is to be measured,\footnote{Cf. Bohr's insistence that
a \emph{phenomenon} is only defined within the framework of an
experimental setup.} and its insistence that the experimental
apparatus must be described classicaly.  Stapp delineates the latter
point further to being an \emph{operational} or \emph{technical}
description.  We are thus \emph{not} interested in a precise and
detailed description of the apparatus, only in what can be termed
relevant calibration and measurement data.  A precise and detailed
description would not be possible since this would prevent a use of
classical concepts --- since the classical conceptual frameword does
not correspond completely to the nature of the world.  Furthermore,
such a description would require making the observing system an
observed system.

By almost \emph{defining} the observed system by its relation to the
observing system, the S matrix interpretation stresses that a thing or
physical entity can never be understood completely in separation, but
always in relation to other things or entities.  The concept of a
separate physical entity has a precise meaning only when this entity
is infinitely far from the observational instruments; in other cases
we can only talk about its separate, independent existence as a
practical approach to reality.  In general we can say that all
entities are define first and foremost in terms of
their \emph{relations} --- the S matrix interpretation considers the
essential feature of reality (`being') to be relation, as opposed to
the Feynman interpretation, which views it as subsistence, and the
aether interpretation, which considers it to be activity.

To elaborate on the last point, I may first mention the obvious fact
that an entity that has no relation whatsoever to its surroundings,
can never be discovered by or have any effect on these surroundings,
and can hence justifiably be said to not exist in this world (only in
`its own world').  The only criteria the surrounding world has to
decide not only \emph{whether} an entity exists, but also \emph{what
kind of} entity it is, are its (possible) relations to other entities,
i.e. to its surroundings.  It is thus not completely wrong to claim
that an entity is defined by its relations rather than by its essence,
or maybe even that its essence \emph{is} its relations.  Stapp writex,
`An elementary particle is not an independently existing entity.  It
is, in essence, a set of relations stretching out to other things.'

This point also goes beyond the obvious one that we to observe
something (an entity) must be in some relation to it (directly or
indirectly).  It also emphasises that we never perceive something as
existing completely independently; it always has a certain relation to
its surroundings (in space and time).  `The idea of a table existing
alone in the universe has an aura of unreality' --- the table is
located in a certain room among other things, and has its own history
and a future.  This is essential to our perception of the table.  In
other words, the world is perceived more as a network of relations
than as a collection of things.

In line with this, an elementary particle is perceived not as
something that \emph{has} relations, but as itself a relation.  Here
what Stapp calls \emph{macrocausality} plays an important role: energy
and momentum can only be transmitted over large distances by way of
physical particles.  If there is a correlation between preparation and
measurement that implies a transfer of energy and momentum over large
distances, this means that we are dealing with physical particles,
providing in turn an operational definition of these.  An individual
particle is thus not something we `observe', but an objective relation
between the preparing and the measuring systems.  All other
measurements of the observed system can then be calibrated against
this simplest type of relation.  The quantum mechanical measurement
problem is thus considered to be solved.  The same is the case for the
problem of renormalisation.  A `bare' particle is simply an empty
concept, since this implies considering the particle as completely
isolated from it relation to its surroundings --- an idealisation
without any justification at this level.  The \emph{physical}
particles occuring in the S matrix are always `dressed', and those are
the ones that are used for calibration, etc.

The S matrix interpretation strongly emphasises the non-separability
or entanglement in quantum mechanics.  This should be clear from what
I have written above, but requires some further comments.  The
entanglement can be both between preparing and measuring system,
between observing and observed system, and between subsystems of the
observed system.  These all have in commen that the systems or
subsystems can only be considered to be separated in the asymptotic
limit.

If preparation and measurement are entangled, it is not possible to
define any observed system, and the treatment ov the observed system
requires that it is isolated during the process.  If we for example
intervene with a measurement during the process, this must be treated
as two experiments, not one --- the conditions for treating it as a
single experiment (e.g. by use of one wavefuction which evolves
continuously according to the Schr{\"o}dinger equation) are violated.
This is what happens when one intervenes in the double-slit
experiment, or more generally, in any attempt to `follow' the
trajectory of a particle.

The entanglement of subsystems means that during part of the process it
is not possible to describe the observed system in terms of separate
subsystems, even if all preparation and measurement is translated to
properties of such systems. This does not necessarily imply that the
observed system cannot be described all all during these time intervals,
so that a description of the time evolution of the system would be
meaningless. However, during these intervals we must take a strictly
holistic perspective towards the system. If we think in terms of fields
and states, we can for example not talk about electron fields and
electromagnetic fields separately, but must view the system as a
non-separable coupling of these fields, and nor can we talk about
electron states and photon states, but only about a state for the entire
observed system, where electron and photon cannot be separated --- the
relations within the system are at certain times and in certain regions
so intimate that it is meaningless to imagine the subsystems as
separate. But in that case the description cannot be considered a
description of possible measurement results --- the condition for
meaningful observation is violated if the system is not in an asymptotic
state. (If such an entangled state \emph{is} asymptotic, then it is also
observable and localisable --- like a bound state.) In my view it is
however not meaningless to talk about the observed system as such, at
least not if we understand it as what links preparation and measurement,
and assume that this is a controlled and idealised picture of processes
that in fact occur. We may furthermore, if we wish to make use of the
language of the latence interpretation, say that the particles exist
\emph{potentially}. In certain cases it is also possible to divide the
processes into subprocesses --- energy and momentum are transferred
between two subprocesses by an approximately physical particle, which
may be created in one subprocess and destroyed in the other. This means
that in certain cases it \emph{is} possible to describe the time
evolution of the system.

As regards virtual particles, the S matrix interpretation naturally
takes a more dismissive attitude. The virtual particles belong to the
entangled part of the process, but to think of virtual particles is to
think of separate subsystems. Virtual particles can at best be a
practical point of view for calculating the S matrix; a heuristic
viewpoint which according to the S matrix interpretation is highly
misleadinbg. If the concept is to be used at all, the word
\emph{virtual} must be emphasised much more than the word
\emph{particle}, and it must be reiterated that the interaction
everywhere and at all times during the process, so that the particles in
reality cannot be separated. The virtual particle is thus seen as one
(ore more) relation(s) in an entangled web. This is particularly the
case for strong interactions, where it in the low-energy domain would be
in principle wrong to view the system as consisting of a number of
particles. Here it is considerably more fruitful to view the processes
as expressions of the various patterns of relations that the hadrons can
enter into, and the hadrons themselves as entities given primarily by
the relations they have to other hadrons: what kinds of hadrons they can
be formed from and what kinds of hadrons they can help create. Any
hadron has a potential existence in the form of other hadrons --- these
possibilities constitute at least part of the hadron.

As I mentioned in the context of the Feynman interpretation, the
distinction between real (free) and virtual particles can be
problematic. In principle a gauge boson that is emitted and absorbed
during a process is considered virtual, and this can be extended to
cover intermediate states of all particle species. The problem is that
all particle states strictly speaking can be considered intermediate
states, and hence all particles should be virtual. The S matrix
interpretation both agrees and disagrees with this argument.

It agrees, because the idea that the world in essence is a web of
relations implies that no particle can be considered really free, as I
already pointed out. It disagrees, because the idea of free particles
forms a basis for the very concept of the S matrix: even though the
world is fundamentally non-separable, the idea of separate parts is
necessary. When we observe something, we perceive it as localised and
separate from its surroundings --- this is part of what an observation
\emph{consists} of. The idea of a free particle is an idealisation, but
it is useful and perhaps unavoidable. The key is the asymptotic states.
When the system is far from an asymptotic state, it is so fundamentally
entangled that we cannot speak of (real) particles. It can however
evolve asymptotically, and in such cases we may say that the virtual
particles become more real or free --- free particles are defined as the
asymptotic states. Hence, also an intermediate state may be considered a
free particle. The language here is close to that of the latence
interpretation.

The S matrix interpretation accepts as particles only what \emph{can}
occur in an asymptotic (i.e., free) state. Since quarks and gluons
cannot be free, they cannot be considered particles. But this does not
need to imply a rejection of quantum chromodynamics. Quantum fields have
the property that they not only manifest themselves as free or
nearly-free particles, but as often or even more often in fundamentally
entangled states or states with indeterminate particle number. This
fundamental entanglement is exactly what the S matrix interpretation, as
an interpretation of quantum field theory, emphasises so strongly. As
regards the quark and gluon fields, we are here dealing with fields
which are fundamentally entangled all the time, except at those very
high energies where asymptotic freedom takes over. The quark structure
may (and must, according to the S matrix intepretation) be considered an
expression of `internal relations' in the hadrons. Quarks and gluons as
particles can be said to be little more than a misleading heuristic
point of view, but as quantum fields they are a good expression of the
symmetries and relations that occur in and between the hadrons.

The S matrix interpretation has a strength in the strictly pragmatic
approach, where the operational aspect of the theory is emphasised and
explained well, and in the clear presentation of fundamental
entanglement.\footnote{This may be considered a follow-up to (part of)
the `overlooked' part of Kant's project that I sought in the footnote on
page~\pageref{Locke}: an investigation of the physical preconditions for
meaningful, objective observation. This also implies a critique of
measurement theories that attempt to track the path of physical signals
from the observed systems to the brain and into the consciousness, in
order to find the divide between subject and object somewhere along this
path.} The constructive aspect, on the other hand, appears quite absent,
apart from a remark that it is necessary `to deal with representations
of complementary idealisations of parts of the world, rather than a
representation of the whole physical world itself.' It can be even more
difficult to implement this aspect here than in the other
interpretations. The S matrix interpretation (like the Feynman
interpretation\footnote{It has been argued, for example by Dyson
(\PR{75}{1737}{1949}) that the Feynman theory \emph{is} an S matrix
theory}) starts primarily from the scattering problem, which mainly
appears in experimentally constructed situations. The S matrix is itself
explicitly constructed as a collection of measured or measurable
quantities in an experiment. It should now be clear that the scattering
experiment is not fully representative of what actually happens in the
world at the microlevel --- the idea of the S matrix is thus itself an
idealisation with respect to naturally occuring processes.

A claim that the S matrix is the only physically relevant quantity at
the microlevel is therefore dangerously close to a positivist attitude,
even if built on the idea of objectively occurring correlationns. It
must be clear that we are talking about idealisations. An attempt at
directly extending the interpretation to an idea of a `universal S
matrix' will be an invalid generalisation from experiments --- it
assumed a controlled observation without any observer.\footnote{Capra
appears to want to make such a generalisation.}

Another possible extension is to a `relational ontology', noting that
the experimental S matrices select (idealised) parts of the actually
existing web of relations. Here it is taken for granted that we really
are situated within the `universal S matrix'. This is an understanding
that denies the existence of any fundamental entities or any underlying
substance --- everything is but parts of a web. This is an interesting
but problematic point of view. It starts out as clearly
anti-reductionist, but risks ending up in a `new' type of reductionism,
neither physical or mental, but rather `ecological'. Everything is
reduced to relations and connections, which is it ultimately impossible
to get a handle on, since we do not know what they are relations or
connections between.\footnote{The only way out would be some kind of
mystical insight in or unity with everything.} To find our way out of
this tangle, we would have to introduce the notion of something beyond
(or rather, \emph{inside}) the web of connections, so that the web
itself, when we approach the everyday level, must recede in favour of
the components or contents of the web. These components must to a
certain extent be present at all levels. At the level of elementary
particles the quantum fields can play this role, and this will not imply
a major break with the ideas of the S matrix interpretation as long as
they are defined primarily in terms of their relations and asymptotic
states.

\section{What is a particle?}
\label{sec:particle}

Quarks, leptons, photons, W bosons: these are all called elementary
particles, and quantum field theory is often called elementary particle
physics. As I have touched on at several occasions, it can be doubtful
to what extent the particles can be accorded any primary status, or
whether they should be taken to be secondary quantities; for example as
possible manifestations of the fields, or as idealised limiting cases.
There is however no doubt that the particles play a central role in the
theory, and not least in the experiments that are carried out to test it
and gather more information about the phenomena it deals with.

Another question is how to define a particle: what does it mean to be a
particle? We usually understand a particle to be something located at a
specific point in space and following a specific path, and this is the
starting point for the Feynman interpretation, although quantum
mechanics forces us to relax the requirement for specificity. Looking at
the states that are typically denoted as `particles' in the mathematical
formalism (page~\pageref{particles}), we see that their salient feature
is actually that they are not localised, but rather they have definite
values for some quantum numbers. This choice between giving the
particles definite positions and giving them definite energies and
momenta (the particles of the Feynman formalism have no momentum or
energy assigned to them) is a consequence of Heisenberg's indeterminacy
relation. An `alternative definition' of a particle leads to theoretical
condensed matter physics (where non-relativistic quantum field theory is
used) deals with a number of `quasiparticles' which are states
(excitations) of the \emph{entire} system. Other disciplines also employ 
the concept of quasiparticles.

I will leave this question here (and return to it in
section~\ref{sec:qft-constr}), and assume that the particles are at
least somewhat localised. (I will later on consider some consequences of
the localisation not being sharp.) Two other questions concerning the
particles will be discussed more thoroughly: whether a particle can be
taken to be a given entity, essentially separate from its interactions,
and whether a particle can be considered an individual.

\subsection{Dressed and bare particles}
\label{sec:dressed-particles}

With the self-interaction of the particles and the renormalisation
procedure to deal with this phenomenon, the question arises: which
particle is the `real' one --- the `bare' particle which appears before
renormalisation, or the renormalised particle, which includes the whole
`cloud' of virtual quanta which are involved in the self-interaction?
Both the aether interpretation and the S matrix interpretation give a
clear answer to this question: only the `dressed', renormalised particle
deserves to be called a particle. In the aether interpretation this is
because only it represents something `durable', in contrast to the
elementary quanta, which are ever appearing and disappearing (the bare
particle is such an elementary quantum). In the S matrix interpretation
it is because only dressed particles can be measured. I will argue that
also the particles of the Feynman interpretation must be dressed.

I will consider the renormalisation procedure itself to be
unproblematic. This can be, and has been much debated --- it has in fact
been one of the main topics of disput in quantum field theory. The
traditional method of `subtracting infinities' appears mathematically
inconsistent; this is what led Dirac and Schwinger among others to
disavow the theory they had contributed strongly to formulate. It
appears paradoxical that by carrying out such outrageous mathematical
operations we may achieve such an astoundingly precise agreement with
experiments as in quantum electrodynamics --- the highest level of
mathematical accuracy that any scientific theory has exhibited. This
precision is one reason to believe in renormalisation --- another reason
is that the same results can be obtained in several different ways, for
example by varying the number of dimensions of space.\footnote{---
something that sounds terribly unphysical, but which in fact tells us
somethig essential about the conditions for having a consistent and
nontrivial interacting theory.} Furthermore, all the theoretical work
carried out from 1970 on is essentially dependent on renormalisation ---
in particular asymptotic freedom and confinement in quantum
chromodynamics.\footnote{\emph{Note added in translation:} When I wrote this, I
had not yet learned about the Wilsonian renormalisation group which
gives a different perspective on the issue.}

We can imagine two reasons for the problems arising from the
self-interaction. One is that there is an inconsistency in the theory
itself, or that the theory is incomplete. When the `correct' theory at
some point is found, the problem is assumed to be solved. Whether this
will require only minor changes or a complete replacement of the
conceptual framework cannot be known, and the attempts that have been
made to find alternatives have so far not had much success. I will
however claim that even if the theory will have to be changed, the
current conceptual framework has shown its validity at least at the
level of the elementary particles and processes we know today, so that a
qualtitative discussion and description of the phenomena at this level
within the framework of quantum field theory will always be justified.
By discussing renormalisation as it appears today, we can always learn
something essential --- any changes in the theory will not have a
serious impact on these reflections. Any new theory must have quantum
field theory as a limiting case (correspondence), and the approach of
quantum field theory must remain applicable where it is so today
(complementarity). The other reason one could imagine is considerably
more flattering for physicists. It is possible that the mathematics is
incomplete, and that renormalisation can be treated in a fully
consistent manner once a more complete mathematical theory has been
developed (analogously to how the theory of distributions was developed
to deal with Dirac's $\delta$ function).\footnote{A somewhat more
thorough discussion of the problem of renormalisation aimed at
non-experts can be found in a couple of papers by Paul Teller
\cite{PSA,Teller:1989}.}

The main point of renormalisation is that it is in principle impossible
to see from a particle whether or not it has interacted with itself. The
parameters (mass, charge, etc.) which belong to the bare particles are
thus unobservable, and must be replaced with the phenomenolgocial
parameters which are assumed to result from adding up the contributions
from all possible self-interactions. When computing the processes,
everything which arises from the self-interactions of free particles
must then be subtracted, since this is included in the phenomenological
parameters. This would be necessary even if all contributions had been
finite --- the difference is that in that case the bare parameters could
in principle be computed from the measured ones. These values could also
have been checked against values for (bare) masses which might be
determined from a more fundamental theory, if such a theory was found.

What is the implication of this for the Feynman interpretation? Well,
here we pretend to start from a bare particle, and then add the
interactions. However, all the parameters that are put in are (of
course) phenomenological. This means that we have already renormalised
ones. It can also be argued that the `primitive' elements of Feynman's
formalism --- the propagators --- are dressed. The starting point is
that a particle moves from one place to another. It can do so in many
different ways, and the propagator is obtained by summing up all these
ways. It is now obvious that if interactions are possible, then the
particle can not only take all possible paths from one place to the
other; it can also take all possible paths and at the same time interact
with itself in all possible ways an arbitrary number of times. These
self-interactions could only be excluded from the propagator if they
either were very unlikely, and therefore gave only a tiny contribution
to the propagator (which is not the case, since they give an infinite
contribution), or if the probability decreased and went to zero with the
distance over which the interaction, so that we could talk about a bare
propagator at small distances (which is not the case, since the
probability of electromagnetic interactions, including
self-interactions, increase as the distance decreases). A further
argument for Feynman's particles being dressed, is the simplicity and
consistency of the presentation: since all possible processes involving
self-interactions must anyway be included, then starting from bare
particles would imply that absolutely all problems would involve an
infinite number of diagrams and/or states with indefinite particle
number, which goes against the simple and perspicuous starting point.

In order for the concept of a particle to be useful, a \emph{free
particle} must thus be defined as a particle that does not interact with
other particles than itself. A particle consequently becomes (literally)
a wolly concept --- there is always a `cloud' of virtual quanta which
must be included in the particle. This is quite far from Democritus'
ideas of atoms --- it is almost precisely the same particle concept as
appears in the aether interpretation, which we see must be built into
all interpretations. By looking more closely at this concept of a
dressed particle we may gain quite considerable and significant insight
into quantum field theory.

Firsly, we learn the importance of `squinting' at the system. When
looked at `broadly', a particle can be quite sharply delimited. If we
however try to look more closely at it, we do not get greater clarity,
but rather more confusion. Where we thought there was only one particle,
we now encounter a `chaos' of many different particles. Several physical
quantities (such as the strength of the interactions) depend on how
great a resolution we have.\footnote{This is after all not really
incomprehensible. For example, what we consider the lenght of a
coastline obviously depends on whether the ruler we measure it with is a
kilometre or a millimetre long.} We can explain this both by the
particle interacting with itself all the time, and by that we in the
attempt to reach smaller distance must pour in so much energy that we in
fact \emph{create} new particles. Here it is also essential that in
investigating the elementary particles we do not have any other
instruments at our disposal than other elementary particles, and to have
a high resolution in distance we must use particles with small
wavelenghts, i.e., large momentum and energy. All this follows almost
directly from the indeterminacy relations.

Secondly, the point of the S matrix interpretation that the entities are
defined primarily by their relations is reinforced. Interactions are
clearly a form of relations, and self-interactions are linked both to
the coupling between the field representing a particle and other fields,
and with the particle's possible interactions between other
particles.\footnote{One may claim --- as the aether interpretation
possibly does --- that interaction and exchange of particles in fact
consists in quanta escaping from the cloud of virtual quanta that makes
up a particle.} When for example the mass and charge of a particle now
contains significant contributions from the self-interaction, we must
conclude that a particle is not itself without taking its possible
relations to other particles into account. In addition we obviously have
the point I mentioned about how we investigate particles.

\begin{figure}[hbt]
\begin{center}
\includegraphics[height=3cm]{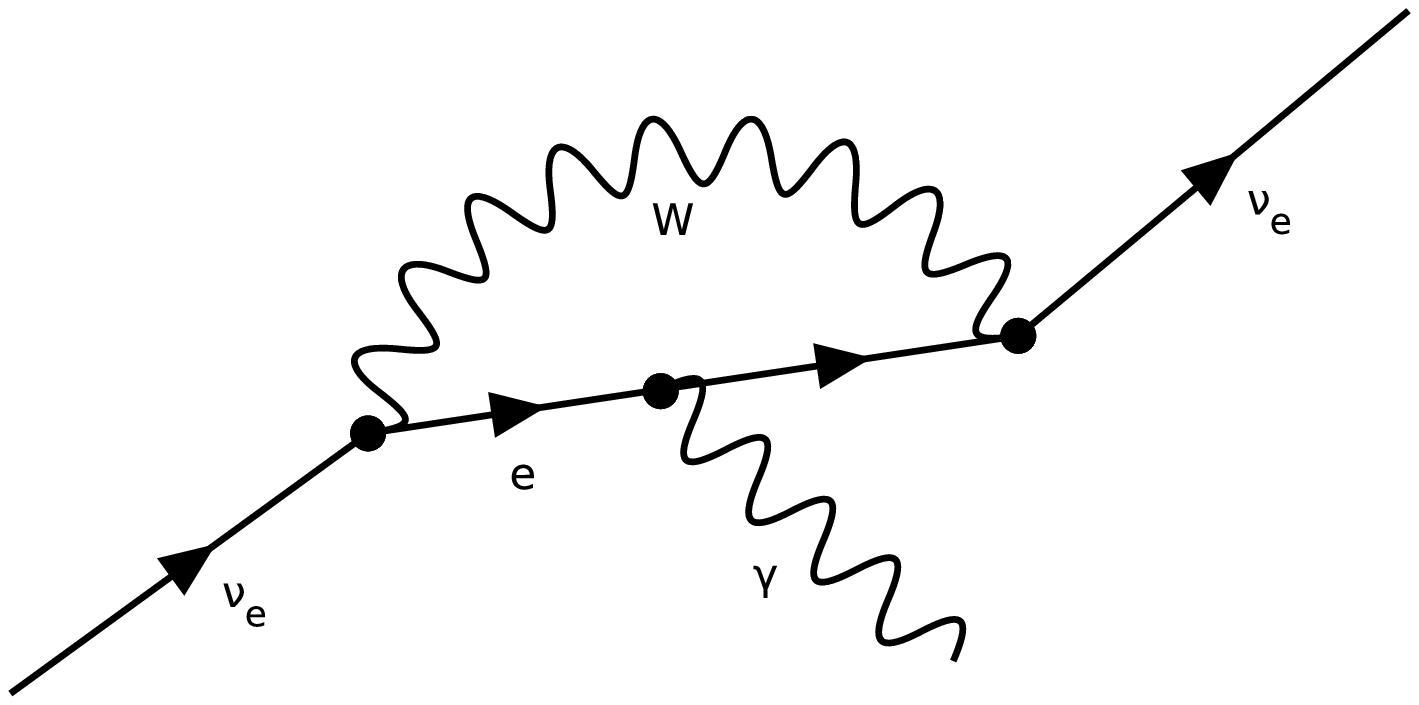}\hspace{3cm}
\includegraphics[height=3cm]{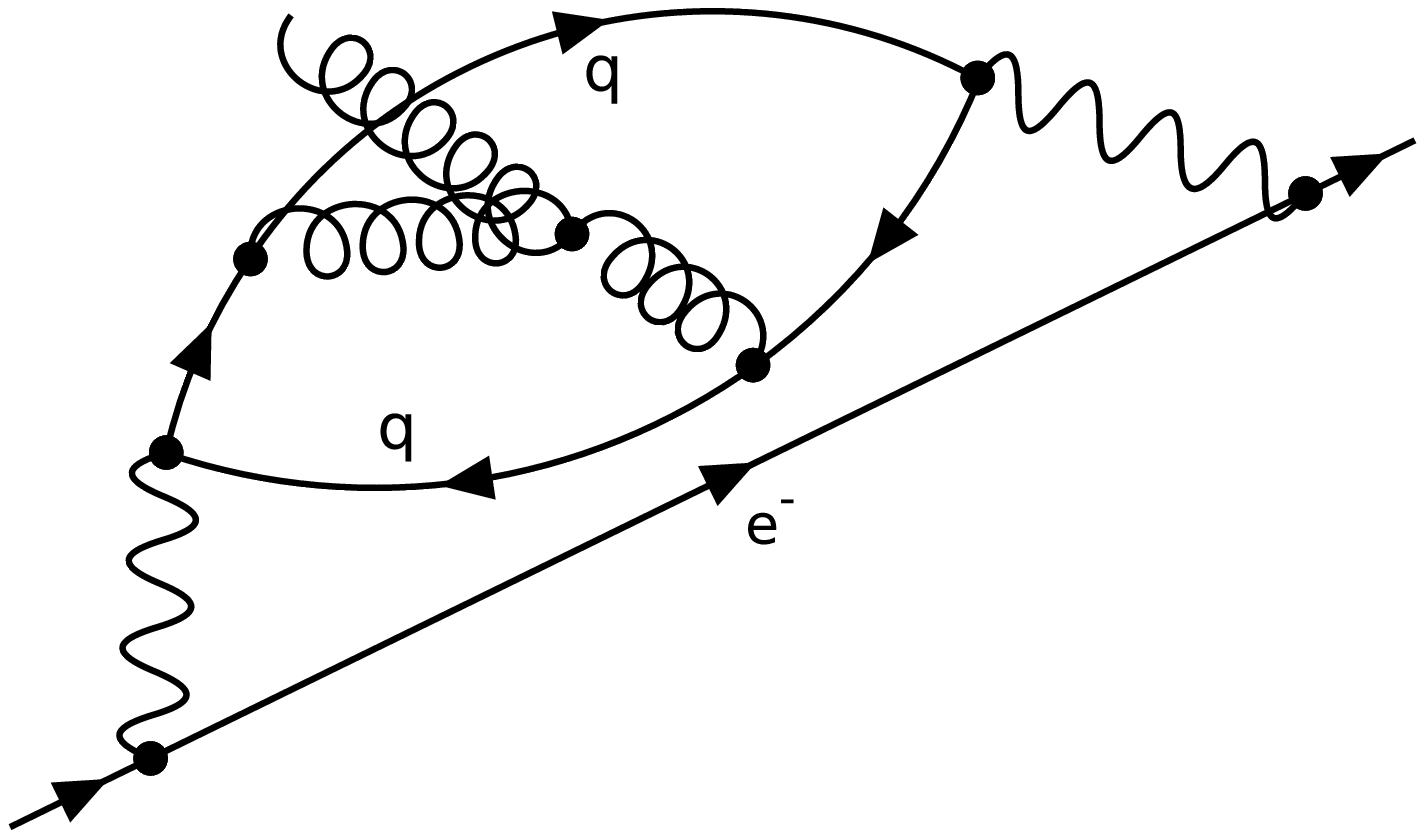}
\end{center}
\caption{The electric properties of the neutrino and the hadronic
contents of the electron.}
\label{fig:nu-electric-el-hadronic}
\end{figure}

Thirdly, we see that any particle will contain all possible kinds of
particles in different quantites. This means that the separation of the
different particle species at most must be considered a limiting case
(or rather, something we can use when squinting). This also has
implications for the properties of the particles, e.g., a neutrino has
electrical properties and an electron has properties with respect to strong
interactions,\footnote{Experimentally, the most precise measurements
have been of the hadronic contents of the photon.} something that can be
seen from the diagrams in figure~\ref{fig:nu-electric-el-hadronic}.
It may thus be said that the claim of the `hadron democracy' or
bootstrap theorists that `all particles contain all particles' (at least
potentially) has a great deal of truth to it, and that it holds not only
for hadrons (although the phenomenon is clearest there), but for all
particles.

\subsection{The problem of identity}
\label{sec:identity}

One of the functions of matter is, as I mentioned in
section~\ref{sec:matter}, individuation. Two things which are otherwise
the same can consist of different matter, and hence be numerically
different, or be considered different individuals. However, it seems
that the particles, which are the carriers of matter in quantum
mechanics, are \emph{not} individuals. If two particles are
qualitatively identical (i.e., they belong to the same species), then
they are if not numerically identical, then at least not independent of
each other. They can in other words not be treated as individuals
(particulars) --- but they are still particular, since there are several
of them. Quantum field theory may well solve this problem by claiming
that the two particles are not two particles, but one state --- but then
some of the `material' aspect disappears, and we still lack an
explanation for why it is practical or correct to classify them by
particle number.

As I explained in section~\ref{sec:matter}, there are certain conditions
that must be fulfilled for the terms particulars and numerical identity
to make sense, and the situation in quantum field theory is related to
some of these conditions being fulfilled while others are not. If matter
is to have a completely individuating function, it must be
comprehensively separated and be linked to extensive quantities. This
may be illustrated by linking material points to persistent,
well-defined world lines, where the quantity of matter is proportional
to the number of world lines.

The separability condition is \emph{not} satisfied in quantum mechanics,
and the idea of separate and unique (well-defined) world lines is not
valid. The latter point is explicitly commented on by Feynman, who
points out that an uncertainty interpretation of quantum mechanics ---
that the particle \emph{really} took a certain path, but that we do not
know which --- cannot yield the results of quantum mechanics. Creation
and annihilation of particles or quanta also invalidate this idea. With
this, we could consider the problem solved: there are no individuals in
quantum mechanics, and the particles can at least not be considered
individuals, and not even as `particles' in the ordinary sense --- the
particles in quantum field theory are nothing but states of the system
of quantum fields, and the particle picture is only one of several
possible representations (alternatively, we may have an indefinite
number of particles).

This, however, is making it too easy for ourselves. Firstly it must be
pointed out that the notion of spatially separated particles or states
with localised energy and other additive quantum numbers plays an
important role also in quantum field theory --- for example, such states
are the basis of measurements in most high energy experiments, as
pointed out in the S matrix interpretation. Secondly, if we only
consider fermions, we will find that a number of the above mentioned
conditions are, in fact, satisfied. This is not irrelevant, and forms an
important basis of the Feynman interpretation.

First it is important to note that only considering fermions does in a
sense not represent any significant restriction in the study of the
concept of matter. The vast majority of what we somewhat imprecisely can
call quantity of matter can be traced back to (approximately) additive
quantities associated with elementary fermions (electrons and quarks or
nucleons). For many practical purposes it is thus possible to identify
`matter' with fermions. That the fermion number is conserved, and can
hence itself be considered a measure of `quantity of matter', is another
reason to study fermions more closely. In practice we will almost always
be able to ascribe to the system a certain net fermion number (which is
more than can be said for energy) --- and even the number of particles
and antiparticles of the various species can often be separately taken
to be numbers which are definite and conserved. Finally, we have what
makes a fermion a fermion: the exclusion principle. This (or the
antisymmetry of the state) implies that fermions `detest' each other,
and if the concept of space is extended to also encompass other state
variables then the conclusion is that the world lines of fermions are
always separated!

This is unfortunately not enough. The properties of fermions makes them
easier to imagine as `real' particles than bosons, but they still have
no individuality. This can be illustrated in two ways within the Feynman
interpretation.

The first of these concerns the question of whether we can say that the
fermions are conserved or not, i.e., whether fermions are created and
destroyed (which would make it impossible to talk about a conserved
numerical identity). Feynman makes is possible to talk about persistent
world lines by giving the particles the ability to go backwards in
time.\footnote{He is also forced to insist that e.g. an electron and a
neutrino are really different states of the same particle, but this is
not of great importance here.} Thus it may also (and does) happen that a
particle that originally behaves `normally' turns and goes backwards in
time, `whereupon' it turns again and continues forward in time. We may
intrude and make an observation at a certain point in time, and find two
particles (moving forward in time) and one antiparticle (a particle on
its way backward in time). If we are to take the interpretation
literally, we are forced to conclude that the two particles we saw in
fact are (numerically) \emph{the same} particle, and that they are also
numerically the same particle as the antiparticle that is observed. In
other words: \emph{According to the Feynman interpretation, two
particles that are observed at different points in space at the same
time, may be one and the same particle.} Two particles which according
to ordinary criteria are understood to be numerically different, can be
numerically identical, and \emph{will be so if} the two particle at a
later time annihilate one another. In this way, whether two particles
are numerically identical now depends on what will happen in the future.
It is also possible that the entire universe consists of only one
particle, whizzing back and forth in time loads of times!

This paradox can only be avoided by admitting that creation and
annihilation of fermions in fact occurs, so that it is correct to say
that the particles are numerically distinct when we are dealing with
macroscopic distances. Only when the distances or time intervals are
very small (and in particular when we are not `watching') is it correct
to say that a positron is in fact an electron moving backward in time.
But that of course removes the basis for talking about an absolute,
material, numerical identity (which does not imply that matter can
disappear --- the energy of the particles, which can be considered a
measure of matter, is transferred to other particles).

My second illustration of the non-individuality of particles is to
consider scattering processes involving two or more identical particles.
In the Feynman formalism, all possible exchanges of the particles in the
final state must be included as equivalent. This is also the case for
particles that may have been created during the process, and particles
that (at the outset) are far apart. Once a particle is born, all other
particles of the same kind must adapt to this and realise that they
cannot occupy the same state as the newborn. We may say that all
particles of the same kind `know about' each other or share the same
history.\footnote{That there is no `seniority' among the particles is
clearly illustrated in the aether interpretation through the claim that
the quanta are created and destroyed all the time, and that there is
therefore no such thing as a quantum that is not `newborn'} So even
though we at the outset can identify each fermion with a continuous
world line, and even though these world lines will always be separate,
we must at the end add together all the possible world lines that link
the initial to the final state in such a way that the result is that the
particles must take each other into consideration (for example, some
processes that would otherwise have been possible, now become
impossible). And we do this not out of ignorance, but because this is
how the particles \emph{are}.

When can we then talk of individual particles? The answer is, with
certain reservations: when the particles are spatially separated. The
reservations are that the particles cannot constitute an isolated system
prepared in a state where they are not separated, and that we should not
consider or study aspects of the system that exclude detailed knowledge
of the system (i.e. to the positions of the particles). The first
condition is violated in EPR type situations such as the Aspect
experiment, where two particles (photons) are created simultaneously in
a total state with a particular symmetry (total spin zero, so that the
spins of the particles are equal and opposite). This system is then kept
free of external influence, so that the symmetry is not broken until the
particles reach the detector (which measures the polarisation of the
particles). The second condition is violated in statistical mechanics,
where we are interested in studying phenomena due to the statistical
behaviour of a large number of particles.

It is crucial to note that the effects of dealing with identical
particles become evident when the system is left alone. If we interfer
with the system, e.g. by performing a measurement, the system is no
longer isolated and an asymmetry can emerge. Firstly we may say that the
particles in the system know about not only each other, but also all
other particles; and secondly the different particles will have
different environments --- the variations in the environment mean that
the state is changed and depends explicitly on the location. Local
effects due to the configuration of nearby particles will dominate the
symmetries of what once was an isolated system.

Here we also see the need to squint at the system, and at the same time
we see that squinting is unproblematic. If we were to follow the
trajectories and states of the particles exactly, we would have great
difficulties (the `trajectory' would consist more and more of points
spread quite chaotically, which would be difficult to interpolate), and
it would be impossible to consider the system as an observed system in a
real sense, since it would not be given the chance to behave `naturally'
--- it would never get a chance to become effectively isolated from the
observing system. The condition for distinguishing between the observing
and the observed system would not be satisfied. But in particle
detectors we do in fact see the trajectories of particles with
sufficient accuracy to give them individuality, although the resolution
is no better than about 0.3--0.5 nm.\footnote{This is the average
distance between molecules or atoms in ordinary matter --- we observe
the ionising effect of the particles.} For high energy particles this is
good enough to separate them, and at the same time energy and momentum
can be determined with great accuracy.

A final question that must be asked in the context of the problem of
identity is, how can particle number be a good quantum number when the
conditions for talking of individual particles are not satisfied? The
answer is one of two. We can observe phenomena which can be uniquely
derived from the existence of a specific number of particles, or we can
say that we would find a specific numerg \emph{if} we observed. This
will be independent of the details of the system and hence does not
depend on the particles being individuals.

One way of understanding the non-individuality and identity of particles
in quantum mechanics is (again) to compare with Leibniz' monads. Leibniz
denies the existence of such a thing as numerical identity --- if two
entities are qualitatively identical then they are also numerically
identica. He sees time and space as an expression of relations between
the monads, and not as something given prior to the individual things.
The monads are identified among other things by their history (not by
their location in space and time). In quantum mechanics we may say that
when the particles are effectively spatially separated, then they are
also qualitatively different, since they have different environments and
the environments in turn influence the definition of the state. When
they are not separated, on the other hand, they have a common history!
The difference is that we can have \emph{multiple} particles, even if
they are not separated and thus have a common history, and we cannot
give them a specific location in space.\footnote{This is a pretty
superficial comparison. There has been a great deal of debate about the
relation between Leibniz and the quantum mechanical problem of
identity.}

It is of course possible to maintain strict numerical identity also at
the micro-level, i.e. to say that when we have a specific number of
particles, they are all individuals. We may for example introduce forces
or potentials of `fermi' and `bose' type which can do the same job, or
which at least reduce the difference between the two alternative
theories to something we currently cannot observe. Such a solution will
however have several disadvantages, since we then are forced to reject
other cherished assumptions. The locality condition will very likely
have to go --- the forces will be a strange kind of action at a
distance.\footnote{David Bohm introduced such an action at a distance in
an attempt to salvage strict realism in quantum mechanics. However, he
eventually saw this as merely a tool to understand the non-locality of
quantum mechanics --- as a ladder we must throw away when we have
climbed up it.} Moreoever, such a model will appear considerably less
`natural' than the one that is now commonly accepted and follows
directly from quantum field theory. To obtain the fermi--bose
distinction as a natural outcome without rejecting individuality would
probably require a radical break with quantum mechanics, and we
currently have not idea of how such an alternative theory would look. It
can also be assumed that the incentive for such a revolution would come
from somewhere quite different.

\section{A fresh look at the functions of matter and forces}
\label{sec:qft-constr}
\begin{quote}
     `Even to a hardened theoretical physicist it remains perpetually
astonishing that our solid world of trees and stones can be built of
quantum fields and nothing else.  The quantum field seems far too
fluid and insubstantial to be the basic stuff of the universe.  Yet
we have learned that the laws of quantum mechanics impose their own
peculiar rigidity upon the fields they govern, a rigidity which is
alien to our intuitive conceptions but which nonetheless effectively
holds the earth in place.'  {\em Freeman
J.~Dyson}\footnote{\cite{Dyson}, p.64.}
\end{quote}

Regardless of how elegant, beautiful and self-consistent a fundamental
physical theory is, and regardless of how well it agrees with the
experiments that are constructed to test it, it can be considered
completely worthless if it does not serve to explain the world we live
in. This means that a theory where the essential and operational aspects
are intact, but which lacks the constructive aspect, cannot serve as a
fundamental physical theory. It should be obvious that if a theory makes
our known world and our everyday perception of reality impossible, then
something must be wrong with this theory. I would also claim that if it
makes the world as we experience it very unlikely, i.e. if `our world'
(of macroscopic objects) cannot broadly follow reasonably naturally from
what may be considered essential features of the theory, then it is at
best incomplete and at worst useless.

This implies a requirement that it should follow as a natural
consequence of quantum field theory (and at least of the Standard Model)
that there are rocks, stars, air etc. Our existence should not be a
`miracle' in light of the theory, and the same is the case for our most
basic physical categories. We must be able to reconstruct this without
too many additional assumptions (e.g., about the parameters of the
theory). It should for example not be the case that our universe could
not exist if the fine structure constant $\alpha$ was equal to 1/138
instead of 1/137.03 --- unless we can give very good reasons for
$\alpha$ having exactly this value.

We may to a certain extent make use of the anthropic principle: the fact
that we exist is sufficient explanation. If the laws had made our
existence impossible, we could not have had any science about it, since
we would not have been there. But the anthropic principle must be used
with caution if it is to have any explanatory power at all. It should
not be applied to muddle up the distinction between a natural outcome
and a miracle, or to suggest that humankind is the purpose of the
universe (that the universe is created so that humankind could live in
it) --- the latter, besides being unscientific, is an expression of an
inordinate lack of humility towards the rest of creation. The anthropic
principle should also not be used to block further investigations of
possible theories. It can be used to show which questions are ill-posed
and what kind of questions shoudl be posed. For example, the question,
`How could it be that everything conspired so that intelligent(?) life
could emerge just here on earth?', can be refuted by pointing out that
this question could have been posed regardless of where in the universe
we had lived. On the other hand it is conceivable that investigations of
the conditions for intelligent life \emph{per se} show that it is very
unlikely that it should emerge anywhere --- and in that case our
existence is a miracle in light of this (although unlikely does not mean
impossible), and if the anthropic principle is now to be employed the
temptation is there to assume the possibility of a vast number of
universes (which do not contain intelligent life).

The constructive aspect is quite absent in the interpretations of
quantum field theory that I presented in section~\ref{sec:q-entities}.
That does however not imply that it is trivial or uninteresting. Taking
the constructive aspect seriously will ultimately mean justifying all
other science (in the sense of grounding the existence of or at least
the possibility of all the objects of the rest of science) starting from
quantum field theory. It would be necessary to study all of our known
world and all our known science to find out on which physical conditions
it is based, and then see if these conditions can be naturally satisfied
by quantum field theory. This task will probably be out of reach for all
the foreseeable future. Studying the conditions for the processes
characterising living organisms being physically possible and natural is
for example a vast project (not to mention the conditions for
\emph{conscious} life, of which we know next to nothing). Here I will
focus on illuminating the physical origin of the fundamental categories
(thing, matter and force) which I discussed in
chapter~\ref{chap:philos}, as well as the conditions for having
chemistry. The latter is intimately connected with the conditions for
having things in our sense. I will also focus on how quantum field
theory explains that things \emph{can} exist \emph{here and now}, and
not how they might have been formed. The latter is described for example
by Weinberg~\cite{Weinberg}.

\subsection{Matter}

The most importan functions of matter are, as I presented in section~\ref{sec:matter},
\begin{itemize}
\item matter is conserved;
\item it is located in space and fills it;
\item it has an individuating function;
\item it is movable;
\item it may take on all possible forms.
\end{itemize}

There should not be any serious problems with the final point. It is
hard to accuse the quantum field of lacking flexibility. The number of
possible states, with different properties with regard to energy,
lifetime, transition amplitudes, etc., appears all but inexhaustible.
They are not even tied to a certain character of being in space.

As regards the first point, we can identify not only one, but several
quantities that are exactly conserved; the most important one for our
understanding of the concept of matter is energy. The exact conservation
of energy is in itself not that interesting --- at the level of
classical physics we think of energy as kinetic energy, potential energy
and heat (internal energy), which are not considered forms of `matter',
but rather as something matter can have. It is of much greater
importance that energy is concentrated in massive particles, whcih at
low energies (in the non-relativistic limit) can be considered stable
--- they are neither created nor destroyed. If the theory ensures the
existence of such particles (preferably massive, stable fermions) then
matter can be considered conserved at ordinary temperatures. At high
temperature or energy we must instead consider the exactly conserved
quantum numbers, but the conservation of matter is in any case
unproblematic.

Filling space, movability and individuality require a lot more. The most
important condition is that at a certain level it is possible to talk
about entities with an internal \emph{structure} --- a structure that
can only be found in bound states. This structure must be evident and
not confined: a confined structure (like the quark structure of hadrons
at normal energies) is equivalent to no structure. The importance of
structure is perhaps best exhibited by considering an entity at the
transition between purely quantum mechanical matter, which does not have
the properties above, and classical matter, which has them: an atom.

An atom can for most practical purposes be considered an individual ---
something with its own `personality'. This is because its structure
allows us to have a continuous knowledge of the position of the atom
separate from other atoms as we can have a persistent interaction with
the atom without destroying it. This requires us to give up the aim of
obtaining a detailed knowledge of the internal structure of the atom and
instead consider it a `woolly object'. The structure of the atom can
absorb the changes in its state that might be induced by position
measurements or interactions. All this also depends on us considering
the atom as an ensemble of states rather than e.g. a single state.

The extension or size of the atom (which is a function of its internal
structure) implies that it makes sense to say that it has a
well-defined, `real' position independently of whether we measure its
position, and that we effectively can indentify the measured position
with the real one. The `indeterminacy' of the atom itself means that the
indeterminacy relations are no hindrance, and we do not need to measure
the position \emph{all the time} to determine the path. A completely
precise determination is not necessary to distinguish it from other
atoms --- the atoms naturally keep distances larger than the atomic size
and the indeterminacy that follows from the Heisenberg relations. In
cases where this does not happen, the atoms are destroyed, either by
entering into molecules, or by strong ionisation occurring. In these
cases two atoms must be described as a single system.

Before taking a closer look at the transformation of matter or the
concept of matter that occurs at the atomic level, I must explain how
such a structure is at all possible (and even natural). First of all I
must take yet another step back, all the way to the field operators and
their states or excitations.

The crucial point in the reconstruction of matter from quantum field
theory is the transition from considering (having to consider) the
fields as entities to considering individual particles or systems of
individual particles which are (potentially) localised. This transition
has already been carried out in the Feynman formulation --- but, as I
have pointed out, this is not adequate for describing or understand all
phenomena within the domain of validity of quantum field theory. The
transition to a space-time description is essential, both for the
operational and the constructive aspect of the theory. This transition
can be said to consist in a shift to consider \emph{ensembles of states}
as entities. This approach can cast light on both `elementary particles'
and `composite' entities like atoms.

If we are to consider the particles entities, we cannot think of them as
states of the system of fields. An entity must itself be able to be in
states, or in other words, it must be capable of undergoing accidental
changes. If a particle is identified with one state, it is impossible to
distinguish between creation or annihilation and other (accidental)
canges. It is also difficult to view multi-particle states as states of
several particles. By defining a particle as an ensemble of states
satisfying certain conditions (e.g., a definite charge, mass and
lifetime, possibly with a certain slack), and noting that multi-particle
states (usually) can be constructed from single-particle states, these
problems are solved. We may also allow ourselves to view the particle as
a continuous existent in space and time. This can be done by allowing it
do be in a `fuzzy' state (with fuzzy values of energy and momentum, but
with the energy and momentum densities reasonably localised), and need
not worry about quantum mechanical dispersion --- it is unneccesary to
keep it in a definite statee. It also does not matter if we perform
measurements. In general all processes that localise a particle
(determine its position relative to surrounding particles or entities)
are allowed, since they (in general) will not destroy it as particle.
Nor does the particle being dressed cause any problems.

The particles are however structureless --- they are described as point
particles. We have essentially ended up in the Feynman description. The
only changes that are possible here are changes in `state of motion' (in
a generalised sense) --- changes that can be taken to be equivalent to a
change of coordinates. This also includes things such as change in spin
orientation or quark colour.

For an entity to have an internal structure it must be in a \emph{bound
state} of several particles (or entities). This means that it must be
possible in certain contexts to consider the entity as composite, and to
somehow (indirectly) identify the `parts'.

When it comes to bound states I must admit that my intuitive
understanding is better than my conceptual one. In quantum field theory
it is not easy to grasp the meaning of the term. Firstly, bound states
must be distinguished from dressed particles, and secondly, it must be
possible to characterise bound states as a certain type of states of the
fields, not just as a combination of particles. Key features are that
the state has a (more or less) definite energy, and that it
(spontaneously or by additional energy being supplied) may dissociate
into a state of two or more free particles. It is also essential that
contributions from different `parts' at different locations in space may
be seen e.g. in scattering experiments. However, the particles will have
undergone a substantial change when they enter into the bound state ---
they have completely lost their identity. A real bound state, as opposed
to a metastable state or a resonance, is energetically favoured over a
state consisting of the free particles. Finally it should be mentioned
that also bound states may be considered ensembles of states, both due
to their possible motion and (not least) due to the possible existence
of excited states.

How naturally bound states emerge from the formalism is also somewhat
obscure. However, it seems reasonable, in view of the non-separability
feature of quantum mechanics, that a system of interacting fields will
have states that cannot (directly) be described in terms of the free
modes (particles). Whether such states can be stable and localised will
presumably not be immediately evident from the field equations but will
depend on the shape of the interaction. We will therefore from now on
take this into consideration. We have three kinds of bound states that
should appear in the standard theory.

\begin{itemize}
\item The first one is the confined (colourless) states of quantum
chromodynamics --- the hadrons. As bound states they are fairly
uninteresting --- all signs that they have a structure are effectively
hidden at low energies, where they behave as point particles. These
states are critically dependent on the group-theoretical aspects of the
gauge theory. It is important for the idea of a gauge theory to be of
any use that these states can be shown to exist; thereafter they can be
considered elementary.\footnote{Note also that excited QCD states are
typically not considered excited states, but as separate particles or
resonances. The reasons for this are in part historical, but are also
related to the large differences between different energy levels in
bound QCD states, and that they only are created in collisions between
particles at high energies, and decay into several particles.}

\item Next we have the atomic nuclei. The existence of nuclear forces
based on exchange of mesons (pions) between nucleons should ideally be
derivable from QCD --- but this is as yet a pipe-dream. We may
qualitatively imagine a mechanism for such meson exchange (as in
figure~\ref{fig:meson-exchange})\footnote{This figure must of course
\emph{not} be taken literally; it is merely a schematic illustration of
a partial process.} but there are no certain indications for this giving
rise to attractive forces (binding of nucleons) which do not at the same
time `break down' the nucleons.
\begin{figure}[hbt]
\begin{center}
\includegraphics*[width=0.9\textwidth]{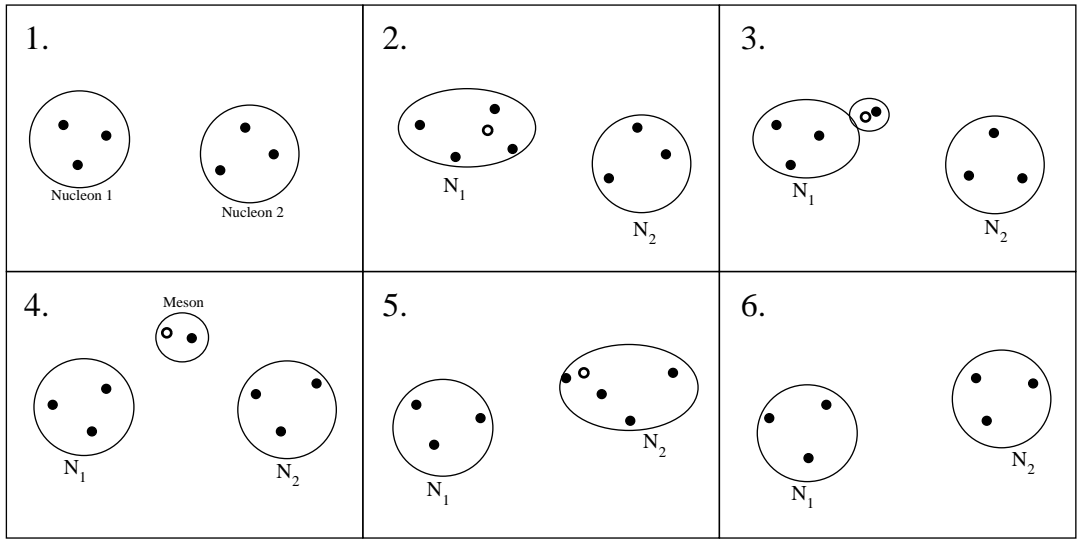}
\end{center}
\caption{Exchange of a meson between two nucleons. The solid dots are
  quarks, the open circles are antiquarks.}
\label{fig:meson-exchange}
\end{figure}
Even if we were to calculate the probability of the meson processes,
this would not be of much help --- we would still be trapped in the
problems of the 1950s. The existence of atomic nuclei is however
(obviously) necessary for the existence of chemistry --- which in turn,
as we shall see, is necessary for the existence of things.

\item Finally we have the electromagnetic bound states, which we have a
much better grip on. By performing a certain gauge transform of the
electromagnetic field we arrive at \emph{Coulomb gauge}, where the field
explicitly splits into an electrostatic part and a part consisting of
free waves (photons). In the first approximation the binding energy is
given by the electrostatic part.\footnote{This was the picture
introduced by Heisenberg and Pauli and used throughout the 1930s.} If we
have a system containing one very heavy charged particle, this can be
considered a static (classical) point charge which hence gives rised to
a real electrostatic potential. Then the Dirac equation can be used to
compute in the lowest approximation bound states for the system
consisting of this heavy particle, matter fields (e.g. electrons) and
the electromagnetic field. We have atoms!
\end{itemize}

A couple of remarks are in place. We have had a certain success in
computing electromagnetically bound states of other systems as well. One
of the early successes of (renormalised) quantum electrodynamics was
positronium --- a bound state of \pos and \el. This means we are not
limited to the highly asymmetric systems of atomic nuclei and electrons,
but the difficulties are greater. It is unclear whether it is possible
to `conjure up' a Coulomb phase also for other gauge fields (e.g.
Coulomb gluons?).\footnote{\emph{Note added in translation:} I am
suprised that at the time of writing this I was not aware of the success
of the potential model for charmonium.} This would however not serve
much other purpose than to make the binding force of the field explicit.
The electromagnetic field is actually different from the others: only
electromagnetism can give us chemistry as we know it.

For our understanding of the atom it is important to view it as an
ensemble of bound states between a nucleus and a number of electrons.
Because the atom is an ensemble we can say that it is the same atom when
it is excited, and partly also when it is ionised or forms part of
molecules. Furthermore we must be aware that the electrons undergo a
substantial change when they are incorporated into the atom. (The
nucleus will however mostly remain `itself'.) In an atom it is in
principle impossible to distinguish between different individual
electrons. What is relevant is the state and the charge density that
they together constitute for the atom as a whole.

Atoms are extended objects. This extension is still `fuzzy' and
ill-defined, but marks a clear transformation of the concept of matter.
This transformation is based on several important requirements, in
particular if we are to account for the existence of other elements than
hydrogen.

The duality between nuclear and atomic forces is crucial. For heavier
atoms than hydrogen to be formed it is necessary to have strong forces
binding the nucleons into compact nuclei. At the same time it is
necessary that the electrons do not feel these forces. Weinberg writes
about this: `If the electrons in atoms and molecules could be
influencedby nuclear forces, there would not have been any chemistry or
crystallography or biology in nature --- only nuclear
physics!'~\cite[p.~153]{Weinberg}. Granted, the nuclei have structure
and extension, and can to a large extent be considered individuals. With
only nuclear forces we could have had a `compact body physics' where
nuclei exhibited behaviour similar to Democritean atoms. However, it is
not the structure and extension of the nuclei that gives matter
structure and extension, and nuclei or Democritean atoms have no
capability to form stable, differentiated and extended matter.

It is also essential that it is in fact a Coulomb potential that is
responsible for the atomic forces, so that the atoms do not collapse or
dissolve. Some slack may be permissible, but a variation of the force
with distance departing too much from $1/r^2$ will not do the job. Other
values of the electron mass and the elementary charge might however be
possible. The chemistry would look somewhat different, and nuclear
physics might have a chance of having a greater impact, but we would
still have atoms with properties largely like those we know. But we need
not worry too much about this as long as we maintain the premise that
the fundamental theory is a gauge quantum field theory --- the simplest
of all gauge groups is the one giving rise to electromagnetism. It is
therefore quite natural that our theory would contain this interaction.

For the structure of heavier atoms than hydrogen, the Pauli principle
--- the fermionic character of electrons --- plays a crucial role. This
is even more important when we come to chemistry, as we will see.

The extension of the atom implies (or is equivalent to) that within a
certain region of space, the structure of the atom is important. Outside
this region the structure is `invisible' --- the atom may be viewed as a
point particle, where some properties (e.g., electric dipole moment) may
reveal that it has a structure, but cannot tell us anything about what
kind of structure this might be. Within the `atomic radius' the
structure reveals itself as a finitely extended charge density
(`electron cloud') and as something that affects scattering or
collisions between atoms. The typical atomic distances in molecules and
lattices may also serve either as a definition of the extension or as an
indicator of the structure.

This diffuse, quantum mechanical extension, and the quasi-individuality
that the atoms acquire at this stage, is sufficient to form a common
denominator for all matter. A more clearly defined extension,
individuality and mobility which is common for all matter does not
exist. From the atomic level on, different kinds of effects are
prominent in the constitution of the various kinds of matter --- in
particular, the aggregate states of matter are constituted in very
different ways.

\subsection{Things} 

Things, and in general everything consisting of solid material, has a
\emph{clearly limited} extension, and can be handled directly
(mechanically). For this to be possible requires chemical bonds between
the atoms. What in turn makes these possible is the quantum mechanical
system of states, the nature of the electromagnetic interaction, and the
Pauli principle. On reflection it is not unreasonable that the discrete,
stationary quantum states are necessary to provide stability --- these
give the system a certain level of `stiffness' (unresponsiveness). A
certain amount of energy must be supplied for any change to happen, and
the system naturally returns to its ground state if it is disturbed. It
has also been shown that electromagnetism, or more precisely the Coulomb
force with its $1/r^2$ dependence, is critical --- a more slowly
decreasing force would lead to collapse, and the same is the case for
purely attractive forces.\footnote{Forces decreasing faster with
distance \emph{may} give rise to stable matter.} Finally it may be
argued that fermionic matter is necessary. Chemistry as we know it has
as perhaps its most fundamental precondition that atomic and molecular
states are `filled'; the ideas of vacant states, filled shells,
valences, etc., are direct effects of the Pauli principle. It can also
be explicity shown as a general result that stable matter is impossible
without fermions. In the words of L{\'e}vy-Leblond,
\begin{quote}
     `In other words, it is the Pauli principle ruling the electrons
which ensures the stability of the world \ldots The specific quantum
nature of the Pauli principle thus is a proof of the need for a
quantum explanation of the most fundamental aspects of the physical
world, namely its consisting of separate pieces of matter with
roughly constant density.'\footnote{J.-M.L{\'e}vy-Leblond: {\em Towards
a Proper Quantum Theory.} Reproduced in \cite{50yr}, pp.\ 197--198. References to
the original articles may be found here.}
\end{quote}
I may add that these conditions for stability are naturally fulfilled as
a consequence of quantum field theory. As soon as the existence of
atomic nuclei is ensured, no further assumptions are required to show
that macroscopic solid matter and things will exist within a certain
temperature interval.

It is however not sufficient that we have solids. A condition for our
experience of reality is also that there is something material filling
the space between the things --- a gas phase. The gas phase is also
necessary for our surroundings to have a certain continuity, so that
there are not for example extreme temperature variations. And of course
we depend on the air around us to breathe. The gas is considerably less
`tangible' than solids --- its extension, separability, etc. are quite
diffuse. The reason we can ascribe such properties to the gas at all is
statistical effects: instead of trying to keep track of every single
atom or molecule, we look at collections of a large number, and these
will in thermal equilibrium and assuming certain external conditions
(such as gravity) take up a certain amount of space. Most molecules will
also stay in more or less the same region (individuality or separability
at large) and on average be moved in the same way by external forces.
All of this is quite independent of the details of the interactions and
the structure of the atoms; it only requires the existence of relatively
stable smallest particles with a certain (but not too strong) degree of
interaction. For our experience of the relation between gas and solid it
is also essential that the gas particles mostly are repelled by solids.

\subsection{Forces}

The concept of force was perhaps Newton's greatest innovation in
physics. In quantum field theory there is not much left of it. We still
often talk about `forces', but they have little in common with those of
Newton, and the language must be considered mostly a relic of the past.
It is really more a shorthand for tendencies or transition probabilities
which are related to which processes are possible. It is indeed possible
to define `potential' and `force', but those concepts are not very
relevant. The relevance increases when dealing with very heavy particles
(which are nonetheless stable) or averaging over a large number of
elementary processes. In the case of the electromagnetic field we can
also take the limit of large field strengths of source strengths
(charges) and large spatial dimensions (wavelengths) and obtain the
classical field, with well-defined field strengths. In this level we
recover the Newtonian force concept. But we do not find the other
forces. Where do they come from?

It could be said that Newton's summarising of a wide range of different
forces in one form (Newton's second law, $\vec{F}=d\vec{p}/dt$)
contributed to muddling the `actual reality', where the forces as a
matter of fact have very different forms and very different origins.
Some forces (in particular chemical forces) were never amenable to the
force law. We might say that almost everything in some way or other
involves electromagnetic forces or process, but it would be grossly
misleading to claim on that basis that the forces can be reduced to
electromagnetism. Many of the macroscopic forces would for example not
even have existed unless a statistical averaging had been performed.

It turns out that macroscopic forces have very little to do with the
`ultimate forces' of quantum field theory. I will attempt to briefly
sketch the origin of these forces. As I tried to explain in
section~\ref{sec:matter}, the forces as they appear to us may
tentatively be divided into three main groupings:

\begin{itemize}
\item Mechanical forces (contact forces between bodies or pieces of
matter). Naturally, these are closely related to the extension of
matter. Here we should strictly speaking distinguish between those
resulting from the stiffness of solids (impacts in the proper sense of
the word) and the others (pressure and friction), which have a more
statistical origin. In particular we may note that viscosity in liquids
and gases is an almost purely statistical effect --- the same is the
case for the elasticity of rubber!

\item External, non-mechanical forces. These may be attractive like
gravity, repulsive like the electrostatic force between two equal
charges, or reorganising like heat. Some of these forces come within the
remit of Newton's force law. These are the ones that most require an
explanation from quantum field theory and are most sensitive to
variations in the theory. In general quantum field theory contains the
possibility of coherent states of bosonic fields, which may give rise to
forces. The only force that survives from quantum field theory up to
large distances is electromagnetism. Howeve, macroscopic
electromagnetism does not give us any fundamental categories, although
modern, industrialised humans are completely dependent on electrical
phenomena. The most important distance force --- gravity --- is one
quantum field theory gives us no clue to understanding.

\item Internal forces or spontaneous changes in state. These forces have
been more or less `banned' since the Aristotelian physics (which was
based almost exclusively on internal, `formal' causes) was replaced by
Galilean or Newtonian physics. However, they find their way back in;
they are typically an expression of the system being in a
non-equilibrium states and relaxes towards equilibrium --- be it
thermodynamic (a state with the largest possible entropy) or mechanical
(lowest possible energy).\footnote{The second law of themodynamics,
stating that entropy always increases, may perhaps be interpreted to
turn Aristotle's concept of change as `realisation of potentiality' on
its head?} Processes within thermodynamics and quantum mechanics or
quantum field theory are perhaps better seen as expressions of `internal
forces' than of the classical, effective, deterministic ones.
\end{itemize}

In general it may be said that `the law of large numbers' plays a huge
role in generating macroscopic forces, and the details of the
microscopic are less important. With a large number of processes at the
microlevel incoherent contributions will cancel each other out or give
rise to friction, while coherent contributions result in a macroscopic
force which looks effective and deterministic. If there is little we can
say \emph{a~priori} about ultimate matter, there is even less we can say
about ultimate force.

Finally I may mention two phenomena which are crucial for our existence
and our experience of the world, and which it is difficult or impossible
to describe adequately without quantum field theory. The first is the
existence of light, which enables us to see. The second is thermal
radiation or emission and absorption of radiation, which contributes to
our ability to see as well as giving a temperature balance which makes
our planet habitable. The thermal radiation from the Sun heats up the
Earth, while the atmosphere absorbs much of the Earth's thermal
radiation --- the famous greenhouse effect.

  \chapter{Future prospects}
\label{chap:future}
\section{Where does quantum field theory stand today?}
\label{sec:status}

\subsection{New theories}

The Standard Model is not complete. It prompts too many unanswered
questions for that to be the case. It contains a large number (17--26,
depending on how you count) arbitrary parameters, including all masses
and coupling constants. The Higgs mechanism has the air of being
introduced to `patch together' everything --- it is remarkable that so
many parameters are pushed on to the Higgs field, without any
explanation for why a field with such properties should exist. It
appears unlikely that this field should be fundamental. We may also ask
why nature has chosen just the symmetry group that appears in the
Standard Model --- could this be merely a manifestation of a deeper,
more fundamental structure?

In an attempt to answer such questions, and possibly also to bring
gravity into the picture, a series of new theories or models have seen
the light of day. However, the idea of quantum field theories with gauge
symmetries has such a strong position that most new theories have taken
this as their starting point. The proposals have then consisted in
introducing new symmetry groups (replacing the `old' ones), new
(alternative) symmetry breaking mechanisms and more particle species
(allowing for a new symmetry principle). The individual proposals are
mostly only of historical and theoretical interest, and it is
unnecessary to discuss them in more detail. Some general features may
however be identified.

The first direction to gain popularity was `grand unified theories'
which sought to describe strong and electroweak forces as manifestations
of one and the same `fundamental force' --- or, in the language of gauge
theories, as subgroups of one symmetry group. The separation of the
strong from the electroweak forces would then be due to symmetry
breaking with a Higgs type mechanism. This also implies that all our
known particles (at least those in the same `family') are `related'. The
first variant --- SU(5) --- had the advantages of using a (relatively)
simple group, keeping the number of arbitrary parameters low (no more
than 22), and all particles in each family were naturally grouped into
two multiplets. The disadvantage was that it predicted a lifetime of the
proton of $10^{30}$ years, while experiments have managed to set a lower
limit of $>\!10^{31}$ years. Later variants became more `artificial',
they often predicted new particles that have not been found, and one
should ask why on earth nature should have chosen such a model --- the
Standard Model appears more natural. Moreover, all these theories are
based on the assumption of `no new physics' between here and the GUT
scale, which is $10^{15}$ GeV, while today's experiments reach
$\sim10^3\,$GeV. That means an extrapolation over 13 orders of
magnitude, which must at least be considered a very bold leap.

Other models featured composite Higgs particles, to make the Higgs
mechanism appear more natural, or with composite quarks and leptons
(preon models), to explain the existence of the three families. And then
we have the attempts at quantum gravity theories, perhaps in part
inspired by the short distance from the GUT scale to the Planck scale
($10^{19}\,$GeV), where gravity plays an important role. I will say
something about these in section~\ref{sec:toe}. None of these theories
have been in a position to be experimentally tested.

On a somewhat less ambitious level we also have investigations of
consequences of alternative Higgs models, massive neutrinos and other
minor revisions of the Standard Model, with effects that should be
observable at `normal' energies.

\subsection{The experimental situation}

So, what have the experiments contributed? In short, the predictions of
the Standard Model are confirmed, and no new physics has been found. The
big accelerator experiments have found W$^{\pm}$ and Z$^0$ with the
predicted masses, and their mass ratio agrees well with what the
Standard Model Higgs mechanism would imply. The search is now on for
evidence for the two missing `pieces' in the `puzzle': the top quark and
the Higgs boson, but so far the result is negative.\footnote{\emph{Note
added in translation:} The top quark was found in 1995, 4 years after
this was written. The Higgs, more famously, was confirmed in 2012.} The
theory gives no direct predictions for the masses of these particles,
but some limits exist. There is however some way to go before these
limits are reached.

Apart from this there has been a painstaking collecting of data about
processes within the Standard Model, with scattering experiments at
different energies. Several parameters have been measured with quit high
accuracy, and a large number of processes have been studied. This has
for the most part not led to any great surprises. The Standard Model has
been confirmed here too.

Some years ago there was quite a fuss created when evidence of a fifth
force was reported, with a range of a few hundred metres and effect
opposite to that of gravity. Several models were proposed to explain
this force: scalar fields with different couplings to baryon number,
isospin and other quantum numbers. Several experiments were also
constructed to to attempt to test this effect. These experiments were of
a quite different kind than the scattering experiments in the
accelerators --- a high degree of inventiveness was shown in
constructing methods for measuring small differences in gravity in
wells, towers, on the sides of mountains, etc. The result was negative:
no reliable effect was found.

\subsection{Non-perturbative QCD etc.}

One area where there could be a fertile interplay between theory and
experiment is those parts of quantum chromodynamics that do not relate
to deep inelastic scattering and hadron showers. Unfortunately,
non-perturbative QCD struggles with finding mathematical methods to
compute anything at all. Some qualitative and very uncertain results may
be found, but there is not much material to be tested.

One phenomenon that should be expected from QCD is bound states of
gluons without any quarks --- so-called \emph{glueballs}. These should
be able to exist as free particles (but with a very short lifetime), and
could give a very direct evidence for the validity of quantum
chromodynamics --- the should leave quite specific `traces'. There are
experiments searching for such effects, but so far no concrete results.

It has also been speculated that there could be a `phase transition' in
systems with extremely large matter or energy density, so that we
instead of nuclear matter were dealing with a `soup' of quarks and
gluons which because of the large density were asymptotically free --- a
\emph{quark--gluon plasma}. There are experiments that attempt to find
traces of such a phase transition by firing heavy ions at each other. It
is also conceivable that this phase transition can occur in neutron
stars --- i.e., that a kind of star even more compact than neutron stars
may exist. Some expected properties of such stars have been calculated.

Non-perturbative QCD is perhaps the greatest challenge the Standard
Model faces today. A breakthrough here would be of great importance for
our understanding of the theory, while the experiments that are required
to test any results need not be insurmountably costly.

\subsection{Particle physics meets cosmology}

One of the most interesting things that have happened the past 15--20
years is the encounter between particle physics or quantum field theory
(the science of the very small) and cosmology (the science of the very
large) in the studies of the `birth' of the universe. This has
traditionally been considered more an area of religion and metaphysical
speculation than serious research. When George Gamow in 1948 proposed
his `Big Bang' theory, based on Einstein's general relativity and the
observation that the universe is expanding, he was for the most part not
taken seriously. However, in 1965 two radio astronomers discovered by
pure chance a completely isotropic `backgound radiation' with properties
corresponding to thermal radiation with a temperature of 3\,K. This
radiation could be explained by Gamow's theory, more precisely from the
assumption that the universe at a very early stage had a very large
energy density and was in thermal equilibriu, and that the cosmic
background radiation was a remnant from this age.

According to this model, by following the evolution of the universe in
`reverse', closer and closer to the starting point, we should find every
higher temperatures and energy density. If we go far enough back the
temperature should be high enough that creation and annihilation of
particles could occur easily, and quantum field theory would be the
correct theory to describe the universe at this time.

If there now are possibilities for observing effects of these early
phases in the history of the universe, this would mean that the early
universe is a unique `laboratory' for particle physics. Early in the
universe there were energies which cannot be reached in any terrestrial
laboratories, and even the most exotic of the new theories should have a
hope of evidence by seeing possible effects of things that happened at
these early times. Particle physicists can thus try to go to cosmology
to have their theories confirmed or refuted.

Similarly, cosmologists may seek the solutions to their problems in
particle physics. Several general features of our universe may depend on
what kinds of particles, fields and interactions existed in the early
stages of the universe. Such a feature may be the relation between the
amount of matter and antimatter in the universe. That space is
homogeneous, isotropic and approximately flat, which is a mystery in the
`classic' Big Bang theory, can also be explained in cosmological models
that rely to a large extent on quantum field theory. More in-depth
presentations of the relation between particle physics and cosmology can
be found for example in \cite{Weinberg,Hawking,Linde}.

\section{The opportunities and limitations of quantum field theory}

Quantum field theory today is based on the concept of quantities
(quantum fields) which are defined everywhere in space-time, and which
are described by a Lagrangian theory. These fields can have (quantised)
excitations, which correspond to physical particles (and more
complicated entities). Introducing interactions by postulating a local
gauge symmetry has proved fertile, and can now be said to be part of
the core of the theory.

In attempting to judge whether future problems of physics can be solved
within this conceptual framework we of course run up against the
difficulty that we know nothing about what the problems of the future
might be. If we did know, we would already be well underway in trying to
solve them. It requires a lot of imagination to envisage all the
problems that may appear --- it often (usually) turns out that the
imagination of nature far exceeds that of humankind. Some implicit
developments and problems may however be read out of the theory.

As a fundamental physics theory, quantum field theory (or the Standard
Model) is quite successful. It encompasses (in principle) all phenomena
in the physical universe except gravity, and where it gives clear
predictions these agree with observations --- sometimes with an
astounding level of precision. There are also no confirmed observations
of phenomena that cannot be accounted for in the Standard Model or some
other gauge field theory. There are still problems with obtaining
sensible results for the strong interaction at low energies and make the
connection with nuclear physics. This may be simply because where are
not yet in possession of the correct mathematical techniques, but it is
not inconceivable that there is something about the theory itself that
makes solving these problems difficult, even impossible. If this were
the case it would be a serious problem --- the theory would then lack an
important constructive aspect.

Quantum field theory also gives rise to consistent world-views. In some
respects these may jar with our everyday understanding of reality, but I
would assume that this problem could be resolved --- both by the
everyday understanding of reality (the content or connotations of the
concepts thing, matter and force) undergoes a certain revision, and by
the two views are considered complementary, since they deal with
different levels of reality. The essence of quantum field theory could
therefore, if it survives as a fundamental physical theory, become an
integral part of the general consciousness.

The very concept of a quantum field appears to be flexible enough to
encompass most of what can be observed at the subatomic level (at least
with the observational methods we know of), although there may quite
conceivably be phenomena that cannot be adequately described using this
concept. I consider it a reasonable assumption that if or when we are
forced to reject quantum field theory, also other principles which have
been considered central will be jettisoned. This can possibly be
illustrated by pointing out that there is a theory or research programme
which views itself as an alternative to quantum field theory, namely S
matrix theory. To a certain extent it exhibits a greater level of
flexibility, which is related to its rejection of the principles of
locality and microcausality as featured in quantum field theory.

There is one inherent inconsistency in the theory as of today, in my
opinion: the quantities `bare particle' and `free field' still appear.
These are entities which are defined such that they strictly speaking do
not exist (since they cannot have any effects), and after
renormalisation they are also replaced by the `dressed' quantities,
which are the relevant ones. It would be an advantage, I believe, if we
could find a way of avoiding these quantites. The conceptual basis for
such a revision are already present in the aether interpretation and the
S matrix intepretation. This would automatically solve the problem of
renormalisation, and could perhaps also help in the development of
non-perturbative methods. I believe the concept of quantum fields would
survive such a revision, even though the formulation would be different.

An open question is whether we really have hit upon something
fundamental in gauge theories. The basis for judging this is not
overwhelming: 2 interactions (electroweak and strong), where our
knowledge of one (strong) is still quite poor. That these can be
described as gauge theories may well be a coincidence; it may well be
that new interactions and particles will be found that can not (or only
with great difficulty) be fitted into a gauge theory. On the other hand
it is conceivable that more gauge interactions will be found at higher
energies. The concept of gauge invariance gives quite clear guidelines
for extending the theory as it stands, both by placing strong
restrictions on the shape of the theory and at the same time by
suggesting possible new interactions and groupings of particles. There
are two important reasons for wanting to keep these concepts. Firstly,
all the properties of the interaction may be derived from a symmetry
principle, and secondly, all gauge theories are renormalisable. If the
gauge principle would turn out not to be generally applicable, it would
result in a crisis for the theory, in particular unless something had
happened with the problem of renormalisation in the meantime.

If new interactions are discovered at higher energies, and these also
can be described by gauge theories, then this will of course give
greater confidence that gauge field theories have hit upon a fundamental
aspect of reality. This will however give rise to two new challenges,
which are already present now, but cannot then be avoided.

The first one is of a more philosophical, not to say purely speculative
nature: What is it that makes gauge symmetries so fundamental? Most
symmetries we know of are fairly intuitive and refer to operations we
can perform (or imagine) --- for example repeating an experiment at a
different time, with the apparatus in motion with constant velocity, or
with particles and antiparticles exchanged. This is even the case for
the abstract coordinate transforms of general relativity: we may imagine
accelerating our system or reparametrising space. Gauge transformations,
however, are in their totality operations on abstract quantities (field
operators) in an abstract mathematical space (the representation space
of a Lie algebra). There is no way we can physically perform or imagine
these transformations. It appears like a glimpse of a hidden structure.
What kind of structure is that? One suggestion (Kaluza--Klein theories)
is that gauge symmetries reflect the symmetries of `hidden' spatial
dimensions (dimensions beyond the four we know).

The second question is much better suited to research and theory
formation. If there are several interactions which originate from
different gauge groups it appears unnatural to postulate all these as
fundamental features of nature. Why should nature have chosen just these
groups, and no other ones? It is possible to try to derive them from one
larger, more fundamental group with a broken symmetry (as in grand
unified theories), but this group will necessarily be more complicated
and less natural, and the Higgs mechanism (or whatever caused the
symmetry breaking) would also be very complicated. The question would
then arise: why has nature chosen just this group and this symmetry
breaking? There is thus no way around the problem of trying to give a
dynamic explanation for the origin of the different gauge groups. This
requires a new level, behind the gauge groups, where these groups emerge
as states or similar. We have now reached a pretty high level of
abstraction: from the individual particles we have gone to quantum
fields describing particle species. From there we have abstracted to the
gauge groups describing the symmetries of the fields, and from there
again to something underlying those\ldots

Another challenge that is known and which is being worked on is quantum
gravity. The theory is at least not complete until it also encompasses
gravity. I assume that this involves a limit to the validity of the
theory --- that a fundamental revision of the conceptual framework (and
hence also of the implicit ontology) will be necessary to include
gravity. We are however far from having any empirical basis for any
statements about this (unless cosmology can give us some hints). This
means it can take a long time before there is any breakthrough. I will
say a bit more about this in the next section.

There may be a certain amount of frustration that nothing much
`happens' in quantum field theory nowadays.  It is not clear that
there is anything wrong with that.  It may be a good thing for quantum
field theory to be `normal science' for a while, so that the main
effort in research is on exploring the consequences of the theory,
collecting new data and (possibly) developing new mathematical
techniques and formulations, rather than on groundbreaking theories
and prestigeous experiments.  Such `normal research' would be a
positive and fertile field within any theory, and will also provide
opportunities for consolidating the theory.  We cannot expect that
there will always be new, revolutionary discoveries --- although we
have been a bit spoilt in this respect in our century.  It is moreover
not unreasonable to assume that any breakthrough will be built on
painstaking work on the theory over a long time --- that it might
for example be more important to have a large number of precise data and
thorough knowledge of various formalisms than experiments at higher
energies.  The breakthrough may also come from a completely unexpected
corner, far from what on might have thought of today.

\section{Theories of everything}
\label{sec:toe}

\subsection{What is a `theory of everything'?}

Physics has as its aim (as I said in section~\ref{sec:reduction}) to
give a unified explanation and description of nature, i.e. a theory
which in principle should be able to explain \emph{all} phenomena in
the world.  This is an aim to strive for (a regulative aim for
physics), regardless of whether it is achieveable.  It is this
ambition for an all-encompassing theory, or a theory that does not
assume any underlying theory or underlying level, which distinguishes
physics from all other disciplines, with the possible exception of
psychology.  A `final' theory, which encompasses all (physical)
phenomena, may be called a \emph{theory of everything} (TOE).  It will
of course be subject to the limitations I described in
section~\ref{sec:reduction} (theories at different levels are
complementary), and will hence in no way replace previous theories or
the other sciences.  It wlll however form a kind of `ontological
basis' for at least all sciences dealing with the inanimate part of
the world.

Postulating something as a TOE requires a good portion of
self-confidence, and is obviously not something that is done
willy-nilly. It is not sufficient that the theory encompasses all
phenomena that are known at the time it is put forward.  The basic
principles of the theory must be simple and natural and should not
give rise to further questions --- the theory must be philosophically
satisfactory (or have the potential to become so).  Moreover there
must be very good reasons for claiming that qualitatively new
phenomena will not be discovred.\footnote{There are several examples
  where it was believed that a TOE was found or around the corner ---
  only for completely new and unexpected phenomena to be discovered
  shortly after.  For example, 100 years ago many people thought that
  the end of physics was near\ldots}  To believe that a TOE can be
found (and even more so to claim that one has already found it) is
really an expression of a rationalist view of nature and science: that
nature (the things in themselves) follow a set of rational principles,
which we are capable of knowing and perhaps discovering by way of
reasoning.  This view goes well beyond what is implied by the
regulative principle of seeking a TOE, and later on I will argue that
it is very doubtful whether such a view is defensible to begin with.

There are some reasons to expect that a TOE may be found, and that it
is possible to guess what it would look like.  Today we have two
theories which together describe (in principle) all known physical
phenomena: the Standard Model of quantum field theory, and general
relativity (the theory of gravity).  It therefore does not seem
unreasonable to assume that if we succeeded in constructing a theory
that unified these, this would be a TOE. There is also a scale which
there are good reasons to believe is fundamental, so that all natural
phenomena can be explained in terms of phenomena at this scale.  If we
combine three constants of nature: Planck's constant $\hbar$ (from
quantum mechanics, the speed of light $c$ (from relativity), and
Newton's gravitational constant $G$ (from general relativity), we get
the \emph{Planck units}:
\begin{align*}
     &\text{The Planck length} & \ell_{Pl} & = \sqrt{G\hbar/c^3} 
                & = 4,05\cdot 10^{-35} {\rm m} \\
     &\text{The Planck time} & t_{Pl} & = \sqrt{G\hbar/c^5} 
                & \sim 10^{-43} {\rm s} \\
     &\text{The Planck mass} & M_{Pl} & = \sqrt{\hbar/Gc} 
                & \sim 10^{-5} {\rm g} \sim 10^{19} {\rm GeV}\!/c^2 \\
\end{align*}
What might occur beyond this scale can be assumed to be irrelevant,
or may be in principle unknowable --- or it is conceivable that there
simply is nothing beyond the Planck scale: it constituetes the
`absolute' dimensions of the world, just as there are no speeds
greater than the speed of light or temperatures below absolute zero.

\subsection{Proposed theories}

\subsubsection{Quantum gravity}

As mentioned in section~\ref{sec:status}, quite soon after quantum
field theory had restored its status people started working on
theories for quantum gravity, which could possibly be candidates for
theories of everything.  A number of models were proposed during the
1970s and 1980s.  Here is a short summary.

First came the idea of supersymmetry --- at about the same time as the
Standard Model was designed. This introduces a symmetry between fermions
and bosons, so that they can be transformed into one another. This would
imply that for every particle there is a `sparticle' with opposite
statistics, i.e., a doubling of the number of particles --- something
that need not be too bad if there is a good reason and it is possible to
`explain away' the unobserved particles. If supersymmetry is made into a
gauge theory we get \emph{supergravity} --- a gravity-like field appears
naturally. Independently of these theories there was a renewed interest
in \emph{Kaluza--Klein theories}. These had been proposed by Kaluza and
Klein in the 1920s, as a way of unifying electromagnetism and gravity.
This could be done by giving space 5 dimensions and identifying the
geometry of the 5th dimension with the electromagnetic field.
Introducing even more space dimension should make it possible to fit in
the `new' gauge fields, at the same time as spontaneous symmetry
breaking at the Planck scale could `curl up' or `hide away' the extra
dimensoins at our energies. Combining supergravity and Kaluza--Klein
theory resulted in \emph{extended supergravity}, where preon ideas could
be recovered and exploited. All this was and remains a paradise for
anyone lover of abstract mathematics --- but the experimental evidence
is pretty absent.

\subsubsection{Superstrings}

According to superstring theory the fundamental physical entities are
not particles, but \emph{strings}, i.e. 1-dimensional objects. String
theory was originally proposed around 1970 as an attempt to describe
hadrons (as vibrational states of the strings), but was abandoned partly
because it predicted some unwanted massless particles. It was discovered
much later by Green and Schwarz that these massless particles were very
similar to photons and gravitons (gravitational quanta), and if string
models were used to describe all elementary particles this naturally
gave rise to a theory of quantum gravity.

If this had been the only advantage of the theory, it would perhaps not
have become so popular --- theories of quantum gravity existed already.
But it also turned out that the theory was nearly inconsistent --- one
ahd very little choice in designing the theory. For example, it
predicted which gauge groups would be present --- the vast majority of
gauge groups would give an inconsistent theory. In general the hope was
that there would only be one consistent string theory --- given the
requirement that the theory of everything would be a string theory,
everything else would follow. This is close to Einstein's aims for the
last 30 years of his working life: a unified theory of all interactions,
based exclusively on principles of consistency.

String theory cannot be consistent in 4-dimensional space. The simplest
superstring model required 25 space dimensions and 1 time dimension; if
interactions were also included the requirement becauce 10 dimensions.
The hope would then be that the 6 unknown dimensions were `curled up' as
in Kaluza--Klein theories, without having any specific mechanism for
this. It has also proved difficult to obtain any results from the theory
with any connection to the energy levels we have reached experimentally
--- to try to derive our known physics (the Standard Model) from string
theory requires quite a lote of arbitrary parameters to be added. It
also turns out that there are several possible string models, destroying
the hope of finding a TOE by only requiring consistency. But
superstrings is still an active area of research --- it is at least a
possible model of reality at the Planck scale, and the only consistent
way of quantising gravity that is known so far.\footnote{--- and
illustrates how exotic physical theories can become at levels far from
our own.}

\subsubsection{Random dynamics}

The random dynamics programme, proposed by Holger Bech Nielsen, has a
completely different starting point from string theory. Where string
theorists claimed that God had no choice regarding the fundamental
theory, random dynamics asserts that God could choose whatever He wanted
--- as long as it was complicated enough. Fundamental physics can be
completely arbitrary or chaotic, or there may simply not be any
fundamental theory. The aim was then to show that our known physics
emerges as a natural consequence --- that almost regardless of what kind
of complicated model is chosen for `fundamental' physics, the Standard
Model and general relativity will emerge at lower energies. In other
words, the physics we know is very insensitive to changes at the Planck
scale. This happens in many other fields. I have pointed out that
macroscopic forces (with two or three exceptions) have very little
connection to the microscopic ones. Chemistry does not depend on the
details of nuclear physics (only that nuclei exist), and very much of
biology can be explained just by the principle of evolution. Similarly,
if all the effects occuring at the Planck scale, only a few will
`survive' to our energies.

In practice it is obviously not possible to deal with random theories.
What is done instead is taking various sufficiently complicated models
and allowing the parameters to be varied stochastically. What can be
shown to be the case for all these models can be said to be a reasonable
outcome of random dynamics. For example it can be shown that gauge
theories can emerge naturally from theories without gauge symmetry ---
and only gauge theories will `survive' to lower energies. Of all
possible gauge theories it is the simplest that will survive the
longest, and these are the ones we see in the Standard Model. One has
therefore `explained' both why nature has a liking for gauge symmetries
and given a dynamical explanation for the gauge groups of the Standard
Model. If random dynamics should be called a TOE or an anti-TOE can be a
matter of taste \ldots

\subsection{Critique of theories of everything}

There are several reasons to be very skeptical towards any proposals for
`theories of everything'. Usually it can be asserted at the outset that
the proposal is premature: it implies an attempt at formulating general
principles for an area we still know very little about. For example,
string theory and all other attempts at putting forward principles for
physics at the Planck level suffer from a serious lack of empirical
evidence --- there is even less to go by than Einstein had when he
proposed general relativity in 1915.\footnote{He had both the perihelion
precession of Mercury and the idea of testing the bending of light
around the Sun.} It turns out time and time again that the imagination
of nature far surpasses ours --- there are far more possibilities for
how matter can behave than we had thought of. Either nature refuses to
obey our principles, or it follows these principles but they are far
from sufficient to cover all that happens in nature.

There is also a case to be made against the very possibility of a TOE.
If we start from the fact that humans are finite beings, it is not
natural to assume that we should have the capability of grasping the
behaviour of all of (infinite) nature. Firstly we have no reason to
believe that the behaviour of nature can be summarised in a finite
number of principles. Secondly, even if it could it does not follow that
these principles are of such a kind that we can grasp them --- that
would imply that our consciousness has a very special status, or that
nature is very `kind'.

There are a couple of things to note regarding the theories we
formulate. We cannot pick concepts without restriction when constructing
the theories --- the concepts must necessarily be ones that we know of
already, concepts that we have constructed and which originate from
something known. All our concepts of the world thus have a certain
character of anthropomorphism --- we are bound by our concepts of
understanding.\footnote{Although I have here chosen a form of words
which is close that of Kant, it is not intended to be taken as a
`correct' interpretation of Kant. Rather, I have used Kant's starting
point and conceptual framework and adapted it to what I wish to express
here.} To adapt the world to our (available) concepts we must perform a
certain idealisation.\footnote{There is nothing wrong with this.
Idealisationns are necessary to have a concept of the world. Some can be
taken to be constitutive for our experience, such as the concept of
things.} When we describe phenomena at a deeper level of nature we try
to rid ourselves as much as possible of these anthropomorphic
idealisations and be guided instead by `pure' observations. But, since
our concepts can never be completely rid of anthropomorphisms, and since
merely distinguishing between the observing and observed system is a
(necessary) idealisation --- and (not least) because of the `reverse
requirements' of our experience discussed in
section~\ref{sec:reduction}, this aim can never be fully achieved. It
may also be said that in physics we seek the things in
themselves\footnote{A TOE can be considered a theory of the things in
themselves, since it gives a complete description of the behaviour of
everything in the world and of `ultimate reality'. There is nothing
beyond the TOE.} --- but they will always evade our concepts of
understanding.

A third reason to be skeptical towards all proposals for TOEs is of a
more ideological kind. If a TOE were to be found this would (eventually)
put an end to basic research in physics. Within at least one area of
human intellectual pursuit we would have reached a final stage where
nothing happened any more. To me, this --- that history (even within one
field) should come to an end --- seems implausible. We may also ask
whether humanity for ever after\footnote{Well, humans are going to die
out some time, so it is in any case a finite period of time.} will be
content with what is claimed to be the final theory, or whether there
will not be speculations about why just this theory holds --- is there
more behind it? A last question can be if it is at all desirable that a
TOE is found? These questions take us into the next and final section,
where much of the content is base on the reflections in Wigner's article
\cite{Wigner:future}.

\section{The future of physics}

There are two important criteria for a physical theory to be good,
besides agreeing with observations. It should be philosophically
satisfactory (simple, beautiful and comprehensible), and it should be
fertile (give a good foundation for further research). These two
criteria are mutually contradictory. If a theory is very satisfactory
(and has a closed form) it gives little inspiration for further research
(and does not give much of an idea for where to investigate), and
conversely, if a theory has many loose ends it feels unsatisfactory ---
an `ugly' theory. Aristotle's physics was very satisfactory, and
obstructed research for 1000 years. The optimally satisfactory theory is
the theory of everything, which bars all further research. I have
presented reasons for doubting that a theory of everything will ever be
found, and I also do not consider it desirable that physics should die
in this way. On the other hand it is hard to imagine that physics will
continue to evolve forever, at the same speed as now or with increasing
speed, always pushing the frontiers, and still keep its interest. What
then will happen to physics?

There is one tendency which is worth noting in this context: physics
(and science in general) is becoming ever larger and less transparent.
It is impossible for one person to have an overview over all areas of
theoretical physics (not to mention experimental physics); it requires a
substantial effort to be sufficiently acquainted with a small field to
conduct research and move the frontiers of knowledge within this field.
And the experiments required to collect qualitatively new data or test
the groundbreaking theories are becoming ever larger and more costly ---
the big accelerators gobble up billions, and the groups working in these
experiments count hundreds of people. If this development continues it
will be suicidal for physics --- few if anyone will be interested in
putting in such a bit effort in order to perhaps shift the frontiers a
tiny bit in a small area.

I can imagin three scenarios for physics in the long term.
\begin{enumerate}
\item A theory of everything is found. This can of course not be
completely excluded. If so, it means the end of physics. For a while
people will work on finding consequences of the theory, and a certain
amount of dissemination will happen, but after a while the theory will
be transformed into a rigid and dogmatic ontology. Only those parts of
physics which are of technological interest will survive. Active
research will be diverted into other fields.
\item A final theory is not found, but the frontiers of physics become
more and more inaccessible, and the interest in physics will gradually
decrease in favour of other sciences and completely different
intellectual pursuits. A long time later physics may experience a
renaissance --- possibley with completely new approaches to the problems
--- or it can go into cold storage until humans become extinct. We may
imagine that the knowledge that has been obtained will be completely
lost, since it requires a big effort just to maintain it. But we may
also imagine that since the interest will only gradually abate there
will be time to process the insights that have been obtained and extract
the philosophical or conceptual essence of the theory that has been
arrived at. Physics will once more be absorbed by philosophy. This is
not the worst that could happen.

\item The most optimistic scenario is that physics remains an active
area of research. Interest may move between different areas of physics,
and there will not always be revolutionary discoveries, but some
interesting things will always be happening. This requires that the
frontiers of physics do not become more inaccessible. One way for this
to happen is that the science is `slimmed down' --- only the knowledge
that is required to reach the frontiers is taught. There may also be
increased specialisation, so that each researcher has a full overview
only over their own small field, and is completely dependent on others
for knowledge beyond this. This may give good collaboration within
research, but will naturally hinder creativity. Another requirement is
that experiments do not become more costly. This is likely to require a
somewhat lower level of ambition than today.
\end{enumerate}

Regardless of what happens, I do not think physics will completely lose
its interest --- the question is whether it takes the form of dogmatics,
philosophical reflection, or active research. After all, fundamental
physics touches on a core existential question, which I assume will be
asked for as long as humans will exist: What kind of world do we really
live in?

\end{document}